%% file: main.tex
\documentclass[12pt,a4paper,twoside]{article}

\usepackage[T1]{fontenc}
\usepackage[utf8]{inputenc}
\usepackage[greek,english]{babel}
\usepackage{alphabeta} 
\usepackage{multirow}
\usepackage[italicdiff]{physics}
\usepackage{verbatim}
\usepackage{amssymb}
\usepackage[pdftex]{graphicx}
\graphicspath{{images/}}
\usepackage[top=2cm, bottom=2cm, left=3cm,right=2.5cm,marginparwidth=1.75cm]{geometry}

\usepackage{eurosym}
\usepackage{macros}
\usepackage{enumitem}
\setlist{nolistsep,noitemsep}
\usepackage{subfig}
\usepackage{gensymb}
\usepackage{lipsum}
\usepackage{indentfirst}
\usepackage{amsmath}
\usepackage{float}
\usepackage[font=footnotesize,labelfont=bf]{caption}

\usepackage[square, numbers, authoryear, round]{natbib}
\newcommand{\sectionbreak}{\clearpage\ifodd\value{page}\else\hbox{}\newpage\fi}

\usepackage[hidelinks,colorlinks=true,linkcolor=blue,filecolor=magenta,urlcolor=cyan,citecolor=blue]{hyperref}

\begin{document}

\include{Titolo}
\newpage
\thispagestyle{empty}
\mbox{}
\newpage

\tableofcontents
\addtocontents{toc}{\protect\thispagestyle{empty}}
\newpage

\include{1-Introduction}
\sectionbreak
\include{2-Metodi}

\sectionbreak
\include{3-Dataset}

\sectionbreak
\include{4-Algorithm}
\sectionbreak
\include{5-Analysis}
\sectionbreak
\include{6-FuturePerspectives}

\sectionbreak

\include{7-Summary_Conclusions}
\sectionbreak
\include{Aknowledgements}

\newpage

\sectionbreak
\bibliography{Tesi_Magistrale}

\end{document}

%% file: Titolo.tex
\newcommand{\spazio}{\hspace{2cm}}

\begin{titlepage}

\begin{figure}[htbp]
\begin{minipage}{0.3\textwidth}
\centering
\includegraphics[scale=1]{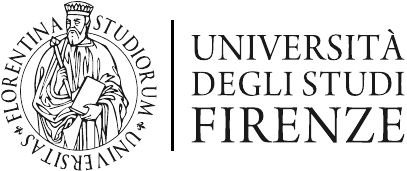}
\end{minipage}
\hspace{0.4\textwidth}
%
\begin{minipage}{0.27\textwidth}
\centering
\begin{flushright}
\large {Scuola di\\
 Scienze Matematiche\\
 Fisiche e Naturali}\break
\hfill\break
Corso di Laurea in\\
Scienze Fisiche e Astrofisiche
\end{flushright}
\end{minipage}

\end{figure}

\vspace{2cm}


	\noindent
	\centering
	\spazio\textbf{\LARGE{Searching for}}\\[1cm]
        \spazio\textbf{\LARGE{chemo-kinematic structures}}\\[1cm]
        \spazio\textbf{\LARGE{ in the Milky Way halo}}\\[1cm]	
        \spazio\textbf{\LARGE{with deep clustering algorithms}}\\[1cm]

	\begin{flushleft}
    \spazio\large{\emph{Candidate:} Leda Berni}\\	[0.5cm]
    \spazio\large{\emph{Thesis Advisor:} Laura Magrini}\\	[0.5cm]
    \spazio\large{\emph{Co-Advisor:} Lorenzo Spina}
	
	\vfill
	\normalsize{
	Academic Year 2023/2024}
	\end{flushleft}

\end{titlepage}

%% file: 1-Introduction.tex
\section{Introduction}
\label{chap1}
\subsection{The Milky Way}

\subsubsection{The structure of our Galaxy}
\begin{figure}[ht]
    \centering
    \includegraphics[width=0.95\linewidth]{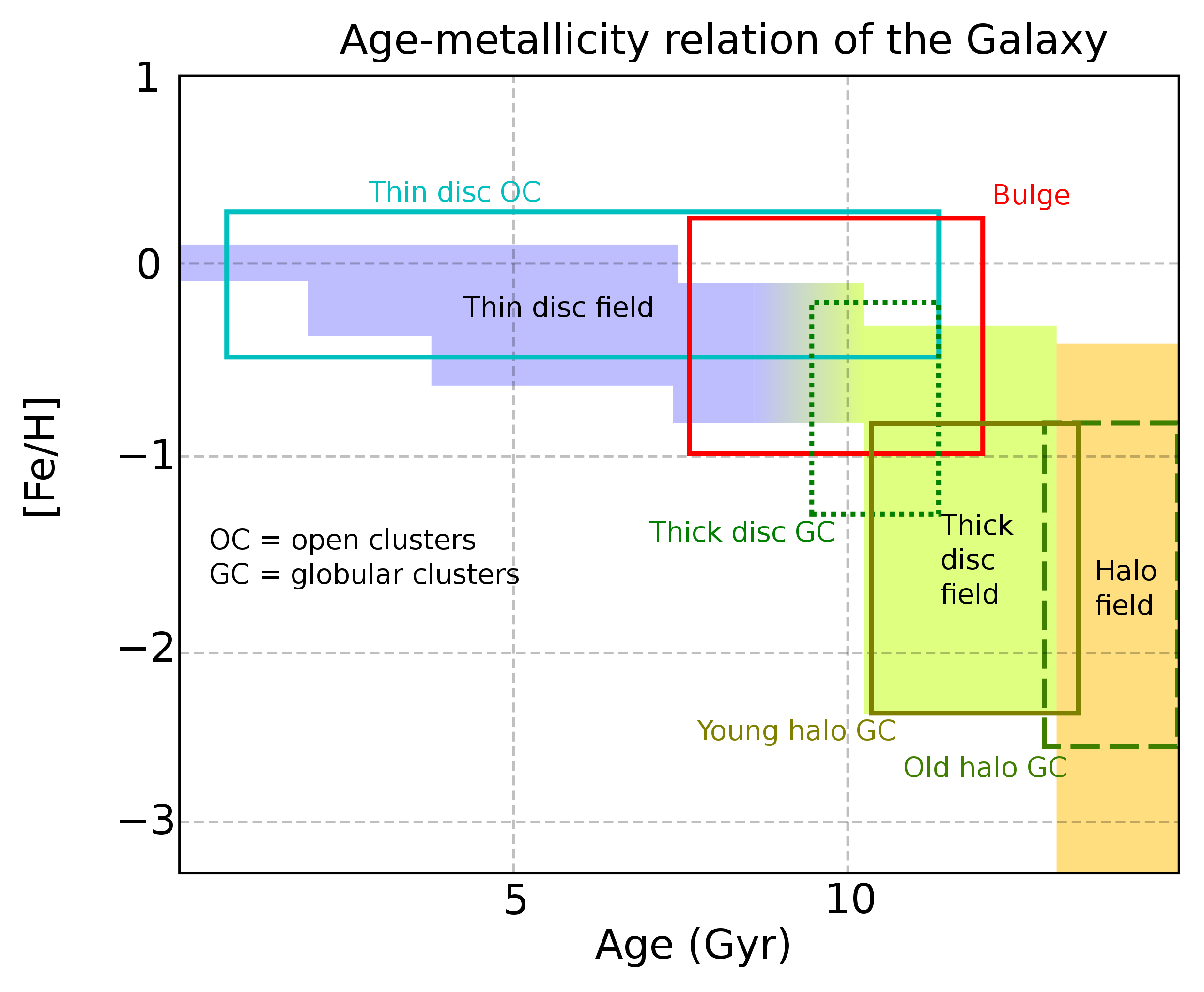}
    \caption{Age-metallicity relation of the Galaxy. Different stellar populations are shown. Filled areas, in order of increasing age are: halo field, thick disc field, and thin disc field stars. Empty frames show: thin disc open clusters (cyan frame), bulge (red frame), thick disc globular clusters (green dotted frame), young halo globular clusters (green frame) and old halo globular clusters (dashed green frame). Figure adapted from \citet{Buser2000Sci...287...69B}}
    \label{fig:agemetall}
\end{figure}

Our Galaxy, the Milky Way (MW), is a complex system of self-gravitating stars, gas, dust and dark matter. Since we reside inside the MW, we can resolve single stars of different masses, from faint dwarfs to supergiants, deriving  their  astrophysical parameters and chemical composition, and inferring their ages. For this reason,  the MW has been subjected to meticulous scrutiny in the latest centuries since the seminal works of \citet{herschel1817xxiv, kapteyn1911milky, curtis1979modern}.  It has been hence considered a benchmark in the study of disc galaxies \citep[e.g.,][]{bland2016galaxy, Guerrette2024RAA....24c5002G}.\newline
In this section, we briefly describe the structure of the main components of the MW: a thin disc and a thick disc, a bulge, which is possibly hosting a bar, and a spherical halo composed of old stars.  These components have different structural characteristics in terms of ages and metallicities of their stellar populations, 
but they can be seen as part of the global evolution of the Galaxy, as shown in the age-metallicity relation in Fig.~\ref{fig:agemetall}, in which we can follow the various evolutionary stages, from the formation of the halo and thick disc, to the formation of the bulge and finally the thin disc. The age-metallicity relationship is not linear, but overall shows the enrichment in `metals\footnote{It is usual in astrophysics to use the term metals to refer to all chemical elements with an atomic number greater than the He one, regardless of their chemical properties. }' of Galactic stellar populations over time.  

The total baryonic mass of the Galaxy, $M_*=6\pm1\times10^{10} M_\odot$ \citep{bland2016galaxy}, is in good agreement with the mass computed using the orbital motion of the Sun around the Galactic centre. This indicates that the Keplerian motion is valid in the innermost part of the Galaxy and that the inner regions are dominated by the baryonic matter.  

\begin{figure}
    \centering
    \includegraphics[width=0.95\linewidth]{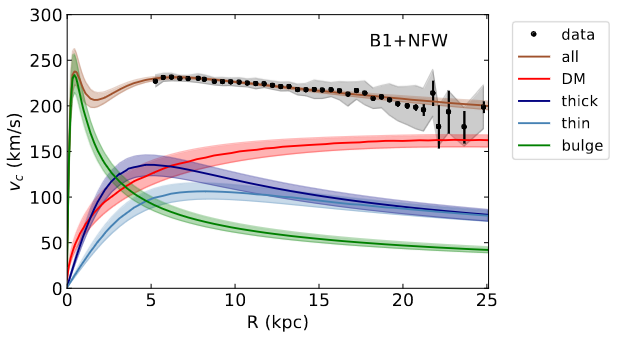}
    \caption{Rotational curve of the Milky Way: the green, blue, cyan and red curves show the theoretical rotation curves for the mass distribution of the bulge, thick disc, thin disc and for the dark matter halo, respectively. The brown curve is the composite rotation curve computed with the baryonic and dark matter mass distribution. The black dots are the observational data of giant stars from \citet{deSalas2019estimation}.}
    \label{fig:rotcurve}
\end{figure}

However, the shape of the rotation curve, shown in Fig.~\ref{fig:rotcurve}, suggests that the outer regions are dominated by dark matter, as postulated in the  $\Lambda -CDM$ theory (Cold Dark Matter) which proposes that the Galaxy is embedded in a \textbf{dark matter halo}, that extends up to at least 100 kpc and that contains most of the mass of the Galaxy. Rotational curves for thin and thick disc and bulge are indeed not sufficient to explain the Galactic rotational curve. The problem might be solved adding the contribution of the  dark matter.\newline
The total  mass of the MW within 200 kpc, including the dark matter component, is  about $10^{12} M_\odot$ \citep{battaglia2005radial}.

\begin{figure}[htbp]
    \centering
    \includegraphics[width=0.75\linewidth]{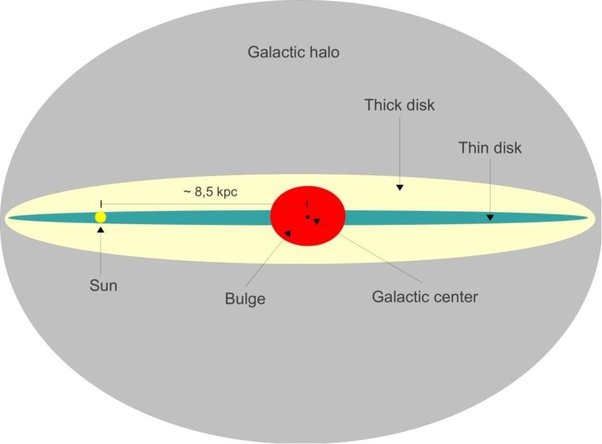}
    \caption{Main components of the Galaxy: in red the Galactic bulge, in green the thin disc, in yellow the thick disc, and in grey the halo. Figure adapted from \href{https://it.wikipedia.org/wiki/File:Milky-way-edge-on.pdf\#file}{\texttt{https://it.wikipedia.org/wiki/File:Milky-way-edge-on.pdf\#file}}.}
    \label{fig:MWComponents}
\end{figure}

The most prominent characteristics of the various components of the MW, as schematically represented in Fig.~\ref{fig:MWComponents}, are the following ones: 

\begin{itemize}
    \item The \textbf{thin disc} is the most noticeable feature of the Milky Way. It has a vertical scale height, i.e. the distance over which the number density decreases by a factor of $e$, of  about 350 pc \citep{Carroll1996ima..book.....C}. It is the site of the ongoing star formation and contains indeed stars that move in an approximately circular orbit. Its current star formation rate is estimated to be about 1.6~M$_{\odot}$/yr. It has been forming stars for at least 8 or 9 Gyr. 
    The Sun is located at a distance from the Galactic centre $R_0$=8.178$\pm$0.035 kpc, as derived by the {\sc gravity} collaboration \citep{GRAVITY2019A&A...625L..10G}. The full diameter of the thin disc, including dust, gas and stars, is about 50 kpc. The shape of the disc is  elliptical, with a ratio of the lengths of the minor and major axis of about 0.9 \citep{Carroll1996ima..book.....C}.
    The disc contains a \textbf{spiral structure} with a number of arms that depends on the tracers used to identify them. For instance, observations of young stars indicate the presence of four spiral arms \citep[e.g.,][]{georgelin1976spiral, lumsden2013red} while near-infrared and mid-infrared observations favour the identification of  two main spiral arms \citep{ drimmel2000evidence, benjamin2005first}.  
    
    According to one of the most widely accepted theories \citep{Shu1970ApJ...160...99S}, the spiral structure originates from density waves, in which stars and molecular clouds are gathered together in higher density regions.
    These overdensities, observed as spiral arms,  are created by a perturbation that modifies the stellar orbits from circularity, adding an  epicyclic component,  as shown in Fig.~\ref{fig:SpiralArms}.

    \begin{figure}
        \centering
        \includegraphics[width=0.5\linewidth]{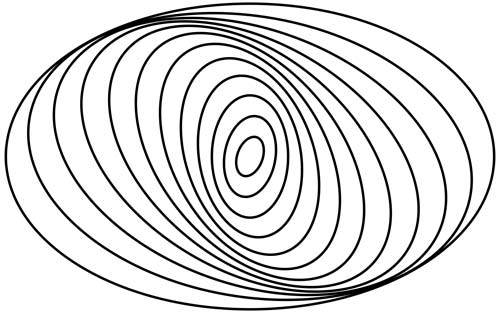}
        \caption{A schematic representation of spiral density waves \citep{kalnajs1973spiral} made up of uniformly filled and perturbed  orbital ellipses. Figure adapted from \href{https://www.daviddarling.info/encyclopedia/D/densitywave.html}{\texttt{https://www.daviddarling.info/encyclopedia/D/densitywave.html}}.}
        \label{fig:SpiralArms}
    \end{figure}
Conditions within the spiral arms favour star formation, therefore arms are the main sites of star formation in the Galaxy (see panel {\em b)}  in Fig.~\ref{fig:SpiralArms}. They can be thus easily traced with the distribution of young stars (particularly O and B type). Fainter stars, instead, live long enough to be able to leave the spiral arms. An artistic  representation of the Galactic spiral arms is depicted in Fig.~\ref{fig:RenditionSpiralMw}.  The Sun is located in one of the spiral arms, called Local arm, which is not one of the major arms of the MW but it is still considered a prominent feature.

    \begin{figure}
        \centering
        \includegraphics[width=0.95\linewidth]{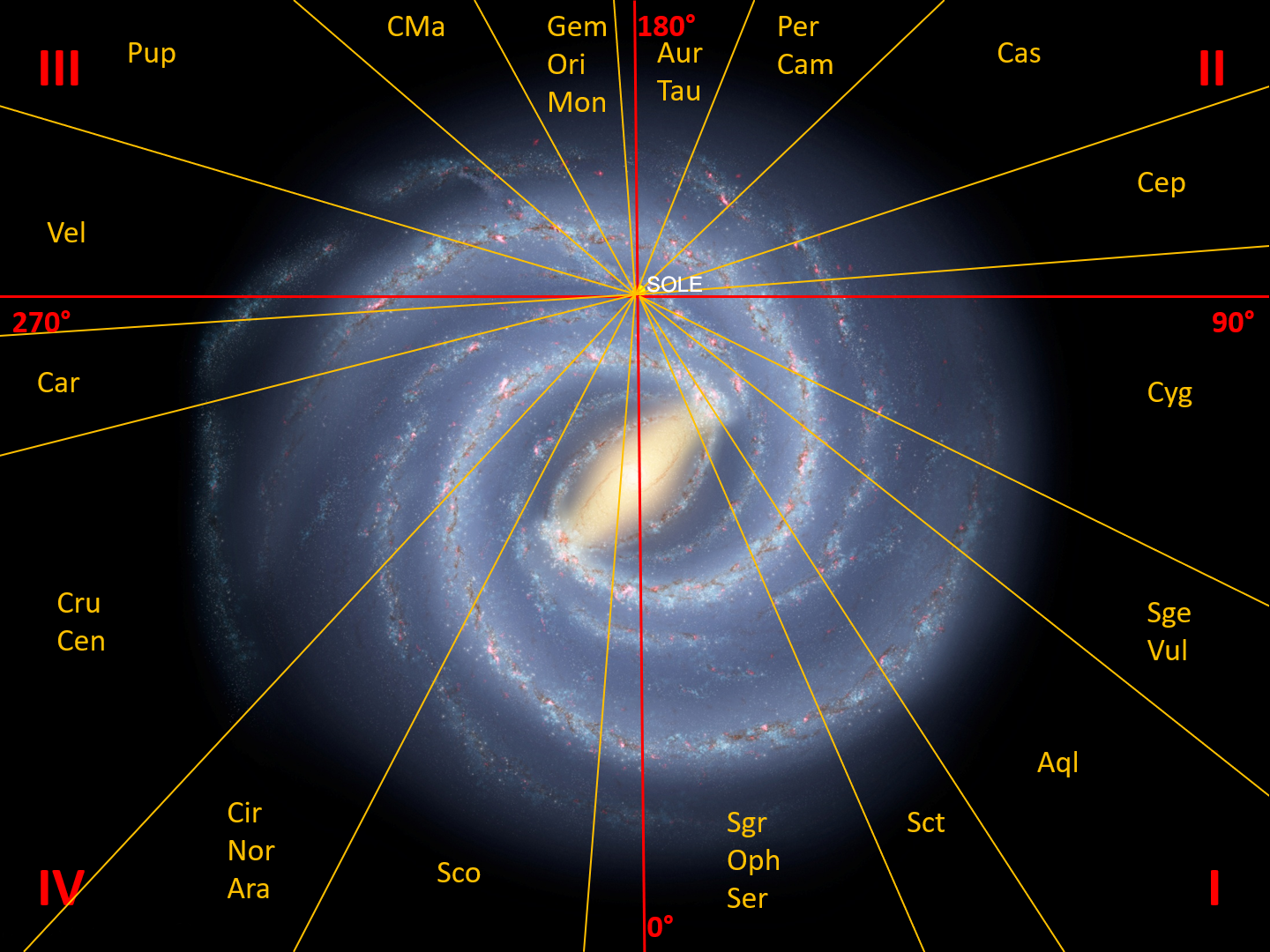}
        \caption{Artistic representation of the spiral arms of the Milky Way, numbers indicate quadrants in the galactocentric system. The location of the Sun is marked with a red circle, while the Galactic Centre with a red star. Figure adapted from \href{https://en.wikipedia.org/wiki/Galactic_quadrant}{\url{https://en.wikipedia.org/wiki/Galactic_quadrant}}.}    \label{fig:RenditionSpiralMw}
    \end{figure}

    The thin disc contains \textbf{gas and dust} as well, primarily located near the midplane and found preferentially in the spiral arms.  The overall distribution of atomic gas within the MW has been measured through the HI absorption line at 21 cm, while the CO molecule is used as a tracer for the molecular gas,  $H_2$. Molecular hydrogen and cool gas are mostly inside the Sun's orbit while atomic hydrogen is more extended. At distances beyond 12 kpc from the Galactic Centre, the disc develops a warp that reaches a maximum angle of deviation from the plane of 15 degrees \citep[e.g.][]{Han2023NatAs...7.1481H, Cabrera2024MNRAS.528.4409C}.
    \begin{figure}
        \centering
        \includegraphics[width=0.75\linewidth]{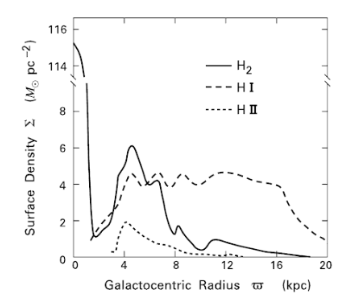}
        \caption{Galactic surface densities of H$_2$, H~{\sc i} and H~{\sc ii} as a function of galactocentric radius. Figure adapted from \citet{palla2004formation}.}
        \label{fig:DistributionHI}
    \end{figure}
    
    \item The \textbf{thick disc} has a scale height of approximately 1.4 kpc \citep{Carroll1996ima..book.....C}. Stars in the thick disc are usually older with respect to the thin disc ones. Their orbits are more perturbed, and their metallicity distribution function peaks at lower metallicities compared to the thin disc. The thin and thick disc cannot only be separated through differences in scale heights and stellar densities, but also from the chemical composition and the kinematic properties of their stellar populations. \newline
    Stars in the thick and thin discs define, indeed, two different chemical sequences in the [$\alpha$/Fe] versus [Fe/H], where the $\alpha$ elements, like O, Mg, Si and Ca, are mainly produced during the evolution of massive stars \citep[M$>$13 M$_{\odot}$, see, e.g.][]{Chieffi2013ApJ...764...21C, Limongi2018ApJS..237...13L}  thus on shorter time scales with respect to iron, produced by stars with lower masses. These differences are a clear indication of  different star formation histories (SFHs) in the two stellar populations that compose the discs.
    
    \item The scale height of stars in the thin disc increases toward the inner regions of the Milky Way, where the disc meets the \textbf{bulge} at a radius of about 2-3 kpc from the Galactic Centre \citep[see, e.g.][]{Zoccali2016PASA...33...25Z}. In this region, the populations of stars are fairly mixed. In the center resides a supermassive black hole, Sagittarius A*.
    The bulge has a composite and complex stellar population \citep{uttenthaler2012constraining}, with more metal poor stars in the latitude interval from $b=-5\degree$ to $b=-10\degree$ and more metal rich stars closer to the plane \citep{babusiaux2010insights}. This vertical abundance gradient in the bulge can be interpreted both as a consequence of the instability that formed the bulge from the thin disc or as generated by a merger.\newline
    Observations show that the bulge resembles a peanut or X-shaped \textbf{bar}, inclined at a substantial angle of $29\pm 2\degree$ to our line of sight (which represents the connection between the Sun and the Galactic Centre) \citep{cao2013new}. This bulge/bar is the most centrally concentrated component and it is heavily obscured. This obscuration is caused by dust, that absorbs the blue part of the spectrum, causing stars to appear dim and reddened. Spectroscopic studies show that the bulge contains a mixture of populations, some very old and metal rich and some that well represent all the Galactic components.
    As of now, the best tracers to study the structure of the bulge are red clump giants (RCGs). They can be easily used as distance indicators thanks to their narrow range of absolute magnitudes and colours. In addition, they are widely spread in the bulge, allowing us to characterize it along different directions \citep{zhao2001high, salaris2002population}.
    Thanks to the study of  RCGs, it was observed that the bulge possesses a double peaked magnitude distribution, known as `split red clumps'. This, together with the fact that bright and faint RCGs are roughly constant at different latitudes, suggests that the actual shape of the bar is a vertical X-shaped structure, as showed in Fig.~\ref{fig:Xshape}.

\begin{figure}
    \centering
    \includegraphics[width=0.75\linewidth]{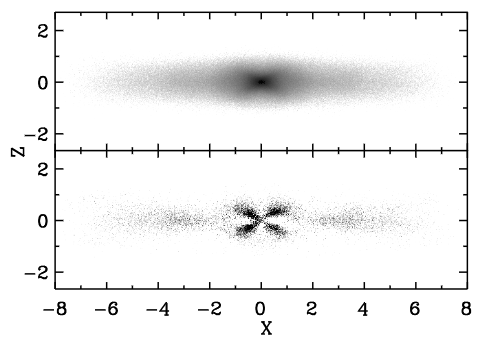}
    \caption{Density map of the stellar populations in the bulge in the X-Z plane. The upper panel shows the side-on view of the stellar density, whereas the lower panel represents the residuals after fitting and subtracting the underlying smooth light contribution. Figure adapted from \citet{li2012vertical}.}
    \label{fig:Xshape}
\end{figure}
    
    \item The \textbf{stellar halo} is the least luminous and most extended component but at the same time it is centrally concentrated, with a half light radius --the radius that contains half of its luminosity-- of about 0.5 kpc. The stellar halo is composed of slow rotating, metal-poor stars and it contains both field stars and globular clusters (GCs). The stellar halo contains the oldest stars in the Milky Way, thus providing us with a picture of the Galaxy in its early stages of evolution.
    Stars in the halo usually have large velocity components perpendicular to the Galactic disc. Stars and GCs in the halo can reach positions far above or  below the plane of the Galaxy. In the stellar halo, we expect to find the remnants of the galaxies that have merged with the Milky Way, i.e. stars that form independent infalling fragments and that are tidally disrupted in the gravitational field, possibly blended together with a component formed `in situ'. `In situ' stars might have been ejected from the disc in their early stages of life, but they could also have been formed in the halo from the primordial gas or from gas stripped from infalling satellites \citep{mccarthy2012global, font2011cosmological}.
    
\end{itemize}

\subsubsection{The current observational landscape}
\label{chap1_landscape}
The last few years have witnessed an epochal revolution, thanks to the launch of the {\em Gaia} satellite, which in its latest release provided distances, proper motions,  magnitudes and colours for more than two billion stars \citep[][]{Gaia2016A&A...595A...1G, Gaia2018A&A...616A...1G, Gaia2021A&A...649A...1G, Gaia2023A&A...674A...1G}. The {\em Gaia} data were enriched and complemented by numerous ground-based spectroscopic surveys (among the main ones, GALAH -GALactic Archaeology with HERMES-
 \citep{desilva2015galah}, {\em Gaia}-ESO \citep{gilmore2022gaia, randich2022gaia}, APOGEE 1-2 -Apache Point Observatory Galactic Evolution Experiment- \citep{majewski2017apache}, LAMOST -Large Sky Area Multi-Object Fiber Spectroscopic Telescope-\citep{yan2022overview}), providing radial velocities, stellar parameters and chemical abundances for million of stars. 
However, this is only the beginning, because we are entering a new era in which massive spectroscopic surveys are being undertaken in both the optical, such as 4MOST -4-metre Multi-Object Spectroscopic Telescope- \citep{deJong2019Msngr.175....3D}, WEAVE -WHT Enhanced Area Velocity Explorer-\citep{Dalton2012SPIE.8446E..0PD}, DESI -Dark Energy Spectroscopic Instrument- \citep{DESI2016arXiv161100036D} and near-infrared, such as MOONS -Multi-Object Optical and Near-IR spectrograph-\citep{Cirasuolo2014SPIE.9147E..0NC}. Spectroscopic surveys conducted with these instruments will cover all Galactic stellar populations, enriching our knowledge of the mechanisms of galaxy formation and evolution. \newline
New space missions and ground-based telescopes enrich the landscape by reaching fainter and fainter stars, as for instance, the {\em Rubin LSST} -Legacy Survey of Space and Time- \citep{ivezic2019lsst},  the {\em Euclid} space mission \citep{laureijs2011euclid}, and The Roman Space Telescope \citep{paladini2023roman}. 
For the next decades, there are plans to build telescopes totally dedicated to spectroscopy, both in the southern and northern hemisphere, such as the   WST (Wide-field spectroscopic telescope) \citep{WST2024arXiv240305398M} and MSE (Maunakea Spectroscopic Explorer) \citep{MSE2019arXiv190404907T}. These telescopes will allow a  spectroscopic follow-up of the new photometric surveys.\newline
In addition to the contribution of the photometric, astrometric, and spectroscopic surveys, the study of stellar pulsations through asteroseismology is proving to be crucial in Galactic and stellar physics. Several space missions are indeed dedicated to monitor of stellar oscillations, such as Kepler \citep{Borucki2010Sci...327..977B}, CoRoT -COnvection ROtation and planetary Transits-\citep{baglin2006scientific},  K2 \citep{howell2014k2}, TESS -Transiting Exoplanet Survey Satellite- \citep{TESS2015JATIS...1a4003R} and in the near future PLATO -PLAnetary Transits and Oscillations of stars- \citep{PLATO2014ExA....38..249R}.  The asteroseismic measurements provide key information on the inner stellar structure, and from it, on the age and mass of the stars. Together with the spectroscopic measurements of the same stars and the direct measurement of the parallax, they  are part of the  revolution in our understanding of the MW \citep{Miglio2017AN....338..644M}. \\
This enormous variety of data at our disposal for the MW has allowed the development of a branch of Astrophysics referred to as {\bf Galactic Archaeology} that aims to study the structure and evolution of our Galaxy by measuring ages, chemical and kinematic properties of stellar populations in different parts of the MW.
In this Thesis, we are mainly focusing on the halo  stellar populations. In the next sections, we provide some details on their characteristics and origin.


\subsection{The chemical composition of stellar photospheres}

The principle guiding Galactic Archaeology is that the photospheric abundances of stars remain, to a first approximation, unchanged during the main sequence and red giant phases. Thus, combined with stellar age estimates, they can be used to reconstruct the chemical composition of the Galaxy over time, somewhat like fossils that retain traces of the past. 
The additional use of kinematic properties also makes it possible to reconstruct their birthplace. In this section, we will briefly describe the methods used to measure abundances from stellar spectra, and some general aspects of Galactic chemical evolution and stellar nucleosynthesis.  

\subsubsection{The determination of stellar abundances}

\begin{figure}
    \centering
    \includegraphics[width=0.5\linewidth]{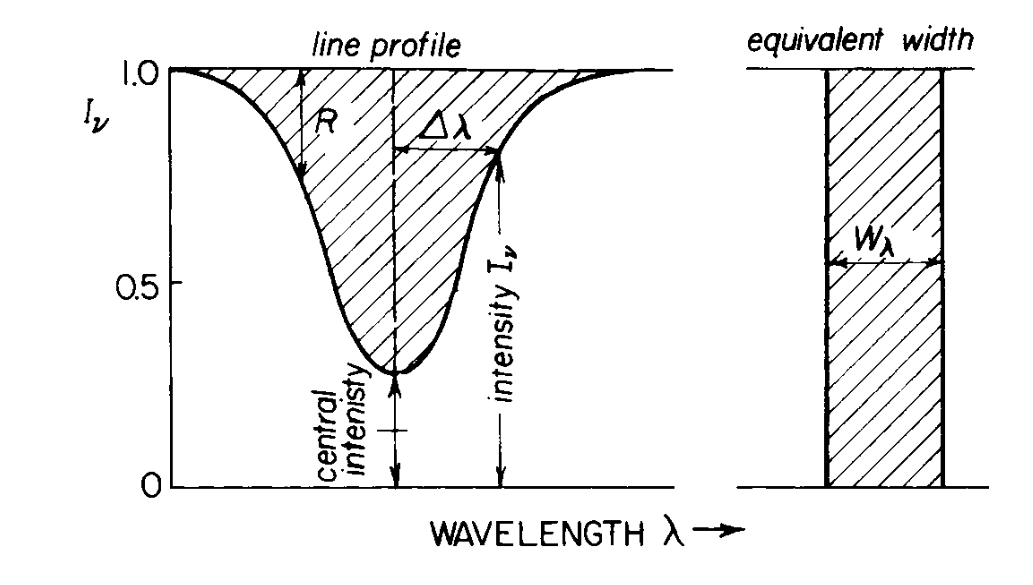}
    \caption{Definition of the equivalent width of an absorption line. Figure adapted from \citep{pagel2009nucleosynthesis}.}
    \label{fig:EW}
\end{figure}

Stellar photospheric composition is measured through spectral analysis, which allows us to derive the abundance ratios of elements versus H. In stars of spectral types from K to F, i.e. stars with effective temperature between 4000 and 7000 K, iron is usually the element with the largest number of absorption lines. For this reason, in stellar spectroscopy, iron abundance is often used as a tracer of metallicity.
The abundance ratio for two generic elements $X$ and $Y$ is defined as:
\begin{equation} \label{eq:AbundanceRatio}
\left[\frac{X}{Y}\right] = \log_{10}{\left(\frac{N_{X}}{N_{Y}}\right)}-\log_{10}{\left(\frac{N_{X}}{{N_Y}}\right)_{\odot}}
\end{equation}
where N$_{X}$ and N$_{Y}$ are the abundances by number of a given element, and N$_{X}{\odot}$ and N$_{Y}{\odot}$ are the same abundances in the solar photosphere. 
When we measure  the iron abundance, we assume  $X$ = Fe and $Y$ = H, while abundances of other elements are usually compared to hydrogen (Y=H, abundances) or iron (Y=Fe, abundance ratios).
According to this equation, stars with the same iron abundance as the Sun, have [Fe/H] = 0, less metal rich stars have negative values and more metal rich stars have positive values. In general, metal rich stars tend to be younger than metal poor stars (see also Fig.~\ref{fig:agemetall}), but there is no linear age-metallicity relationship, and in different stellar populations (and in different galaxies) the same metallicity may be achieved at different times, depending on the star formation history. Stellar spectra, particularly those obtained at high spectral resolution, make it possible to derive both the atmospheric parameters and the chemical composition of  stars. 
The physical basis of spectral analysis is statistical mechanics applied to stellar photospheres, and was for the first time developed in the PhD Thesis of \citet{Payne1925PhDT.........1P}.
The classical method is based on the measurement of the intensity of absorption lines, obtained by measuring the so-called equivalent width (EW), which is the width of a rectangle with the same height of the stellar continuum and with the same area as the absorption line, as shown in Fig.~\ref{fig:EW}. Using the measurement of many absorption lines of the same element (e.g. iron), both in the neutral and ionised state, it is possible, by means of the excitation equilibrium and the ionisation balance, to measure the effective temperature and the surface gravity. 
This requires radiative transport codes, such as {\sc moog} \citep{Sneden2012ascl.soft02009S}, and grids of model atmosphere, e.g. {\sc kurucz} or {\sc marcs} models \citep{Castelli2003IAUS..210P.A20C, Gustafsson2008A&A...486..951G}.   
Once the stellar parameters have been determined, the EWs of the lines of the individual elements allow us to measure their abundance. 
Although these methods have a solid physical basis, they are not easily applicable to very large numbers of spectra. 
Therefore, faster methods have been implemented, based for example on comparing observed spectra with pre-compiled or generated on-the-fly grids of synthetic spectra to extract parameters and abundances \citep[see. e.g.,][]{Recio2006MNRAS.370..141R, Worley2012A&A...542A..48W}. 
Recently, machine learning techniques have been applied to stellar spectroscopy. Starting from {\em training sets} analysed with classical methods, their parameters and abundances, called {\em labels}, are transferred to larger samples of spectra. These methods are usually very fast, and can be applied to large amount of data. Some examples are {\sc the cannon} \citep{Ness2015ApJ...808...16N} and  {\sc the payne} \citep{Ting2019ApJ...879...69T}, widely adopted in the analysis of large spectroscopic surveys. 

\subsubsection{The Galactic chemical evolution}

\begin{figure}
    \centering
    \includegraphics[width=0.5\linewidth]{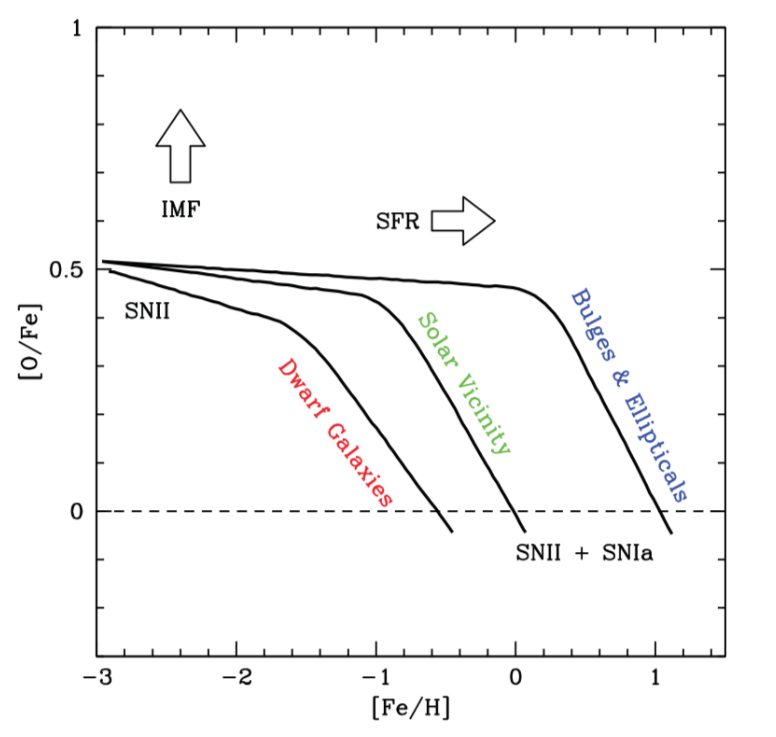}
    \caption{[O/Fe] as a function of  [Fe/H]. The dependence on SFR and IMF is shown with arrows. Theoretical curves  for different kinds of galaxies, or populations in the Galaxy are shown with continuous lines.  Figure adapted from \citep{mcwilliam2016chemical}.}
    \label{fig:alphael}
\end{figure}

Galactic Chemical Evolution (GCE) models aim to explain the formation and evolution of chemical elements in our Galaxy over time and space. 
Initially, only light elements (from H to Li) are formed during the Big Bang, while all heavier elements are produced inside stars, with different processes. GCE models track the evolution of gas, stars and their chemical composition across time and space, taking into account several aspects, such as  the initial mass function (IMF), the star formation history (SFH), the stellar evolution and nucleosynthesis, and the gas flows.
The Simple Model \citep[see, e.g.][]{Tinsley1980FCPh....5..287T}, despite its basic assumptions, provided a first starting point to understand the evolution of our Galaxy. In the Simple Model, the Galaxy was assumed to be closed box, and a first analytical solution relating global gas metallicity to gas fraction was found.
\begin{equation}
    Z\propto ln(1/\mu)
\end{equation}
where Z is the metallicity (expressed in mass) and $\mu$ is the fraction of gas.  This simple relationship tells us that as the fraction of gas decreases --being converted into stars--, the metallicity increases. 
The G-dwarf problem \citep[see, e.g.,][]{Holmberg2007A&A...475..519H} led to the understanding that the evolution of the MW cannot be approximated as a closed-box but involved gas flows, particularly gas infall, which is needed to explain the distribution of the metallicity of old stars in the Solar neighborhood. Subsequent models, like those of \citet{Tosi1982ApJ...254..699T}, relaxed some assumptions like the instantaneous recycling approximation (IRA), enabling the tracking of the evolution of individual elements, with their own time scales.
The inclusion of Type Ia supernovae (SNe Ia) in chemical evolution models by \citet{Matteucci1986A&A...154..279M} provided insights into the evolution of $\alpha$-element\footnote{$\alpha$ elements include O, Mg, Ca, Ti, Si} to iron ratios ([$\alpha$/Fe]). Their time-delay model suggested that early galactic phases were dominated by core-collapse supernovae, with SNe Ia contributing later. An example of the behaviour of [O/Fe] vs [Fe/H] in different galaxies and galactic components  is shown in Fig.~\ref{fig:alphael}. The position of the plateau and the inflection point (the so-called knee) are a function of the star formation rate (SFR) and the IMF, and thus of the evolutionary history of each galaxy or Galactic component.  
This relationship is a fundamental tool to identify stars belonging to different Galactic populations, and in the halo to distinguish between `in situ' or accreted populations. 
One of the most successful model to reproduce the features of the MW is the two-infall model proposed by \citet{Chiappini1997ApJ...477..765C}. This model make the hypothesis of  two major gas infall episodes: one forming the stellar halo and thick disc, and another forming the thin disc. Similar conclusions were drawn from cosmological simulations.
The formation and evolution of the Galactic bulge, with its multiple stellar populations, remain topics of interest. Chemical evolution models and observational data suggest a rapid formation scenario for the bulk of bulge stars, likely involving intense star formation episodes.
For a complete overview of chemical evolution models and their ingredients, see \citet{Matteucci2021A&ARv..29....5M}. 
In summary, chemical evolution models, coupled with observational data from large surveys, continue to improve  our understanding of the formation and evolution, shedding light on its complex history and structural components.

\subsubsection{The origin of chemical elements}

\begin{figure}[htbp]
    \centering
    \includegraphics[width=0.8\linewidth]{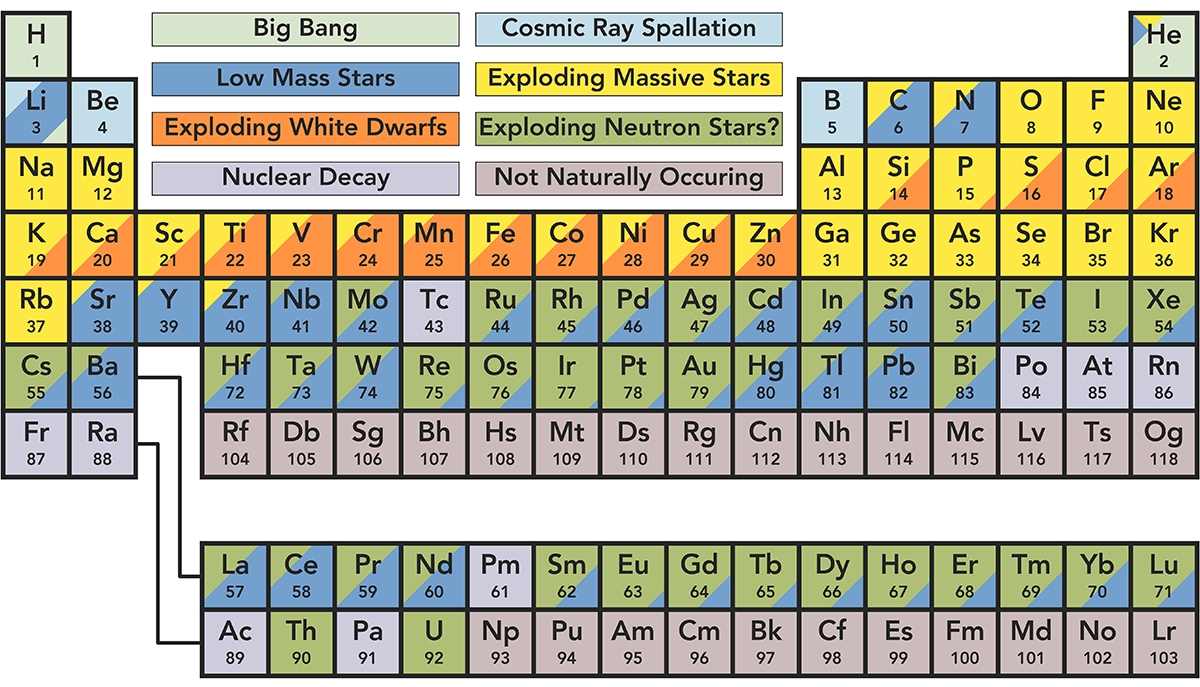}
    \caption{Periodic table of chemical elements in its Astrophysical version: the box corresponding to each element is coloured according to the main process producing it. Figure adapted from \href{https://blog.sdss.org/2017/01/09/origin-of-the-elements-in-the-solar-system/}{\texttt{https://blog.sdss.org/2017/01/09/origin-of-the-elements-in-the-solar-system/}}.}
    \label{fig:ElementsCreation}
\end{figure}

Cosmic abundances are produced through three main mechanisms: Big Bang nucleosynthesis, stellar nucleosynthesis --both hydrostatic and explosive -- and neutron capture.
The kind of nucleosynthesis reactions depends on temperature and density. Usually, the production of heavier elements require higher temperatures and density.\newline
\begin{figure}[ht]
    \centering
    \includegraphics[width=0.8\linewidth]{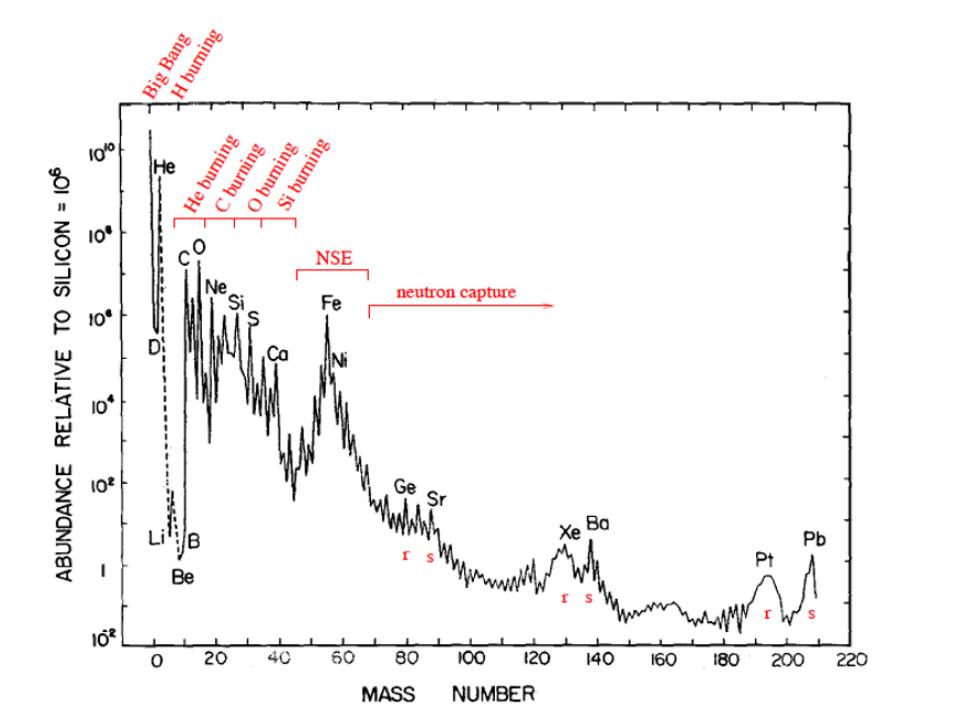}
    \caption{Cosmic abundances: relative abundances versus mass numbers.  The  nucleosynthetic processes are marked in red. Figure adapted from \citet{obertelli2021neutron}.}
    \label{fig:AbbCosmiche}
\end{figure}
The Big Bang theory posits that going back into time, the universe was smaller, denser and hotter. Thus, if we go back in time to the first seconds of the Universe, we get to a moment where temperatures where high enough for nucleosynthesis to happen. This kind of nucleosynthesis was extremely fast, lasting about 15 minutes and managed to produce only hydrogen, helium and traces of beryllium and lithium. No atom with atomic mass number over 7 was produced.
The fundamental process that created most of the heavier elements is stellar nucleosynthesis. Stars, indeed, maintain hydrogen burning in their cores as they are in the main sequence phase. Hydrogen burning can follow different paths depending on the temperature in the core of the stars: the pp cycle is dominant at lower temperatures and can differentiate in slightly different reactions depending on the temperature. When temperatures are much higher, in higher mass stars, the hydrogen burning mechanism is the CNO cycle. This mechanism requires the presence of C, N and O before starting, even in small quantity.\newline
When the star runs out of hydrogen to burn, it contracts as nuclear reactions cannot hold against the gravitational pressure and temperature rises. The contraction stops when temperature rises enough to start helium burning. Helium burns through a process called triple-$\alpha$. In this phase stars manage to create also O and Ne. Moreover, stars create also a strong flux of neutrons that is fundamental in the creation of stronger chemical elements.
The same process of contraction and burning of increasingly heavier elements repeats for C-burning, Ne-burning, O-burning until the star stops with the production of iron-peak elements. These last steps happen only in the most massive stars \citep[see, e.g.][]{Chieffi2013ApJ...764...21C}.\newline
At the end of their lives, most massive stars explode as supernovae. In these explosions they release materials in the interstellar medium and they also create elements like S, Ar, Ca and Fe.\newline
Iron and iron-peak elements are also produced by SN Ia \citep{Leung2020ApJ...888...80L}. SNe type Ia take place in binary systems of stars where one of the two stars is a white dwarf with a time-scale of about $10^9yr$. The star accretes mass from its companion, usually a red giant, and when it gets impossible for it to maintain equilibrium against pressure (i.e. it reaches the Chandrasekhar limit mass), it explodes as a SNe type Ia, enriching the surrounding interstellar medium with heavier elements, in particular Fe and Ni \citep{bland2014origin}.
Elements heavier than Fe are produced through neutron capture \citep{busso1999nucleosynthesis, argast2004neutron, wanajo2013r}, because interactions between charged particles gets inefficient despite the tunneling effect.
There are two types of neutron capture processes: rapid and slow, depending on whether their timescale is shorter or longer than the timescale of $\beta$-decay. Slow neutron capture takes place only in low- and intermediate-mass stars in the thermal pulse during the asymptotic giant branch (AGB) phase. Rapid neutron capture takes place in the explosive phase of the life of massive stars or during mergers of neutron stars, managing to create  highly unstable isotopes too \citep{Thielemann2017ARNPS..67..253T}.\newline
Another important process consists in spallation from cosmic rays, that manages to explain the abundances of Li, Be and B that cannot be explained through the Big Bang theory \citep[see, e.g.][]{Fields1999ApJ...516..797F}.
The astrophysical origin of the chemical elements is shown in Fig.~\ref{fig:ElementsCreation}.

\subsection{Stellar kinematics and dynamics}

\subsubsection{The Galactic reference frame}
\label{chap1_reference}

\begin{figure}[H]
    \centering
    \includegraphics[width=0.5\linewidth]{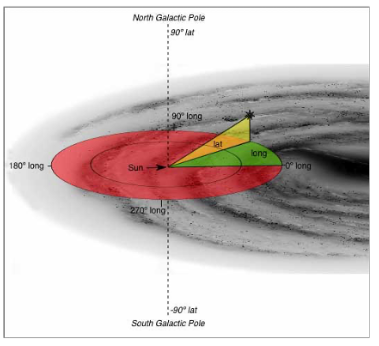}
    \caption{Schematic representation of the Galactic Coordinate System: the location of the Sun is indicated with an arrow. The position of a star is given by its latitude (yellow triangle) and longitude (green triangle). Figure adapted from \href{https://www.universetoday.com/88985/astronomy-without-a-telescope-the-edge-of-significance/}{\texttt{https://www.universetoday.com/88985/astronomy-without-a-telescope-the-edge-of-significance/}}}
    \label{fig:SDR}
\end{figure}

\begin{figure}[H]
    \centering
    \includegraphics[width=0.75\linewidth]{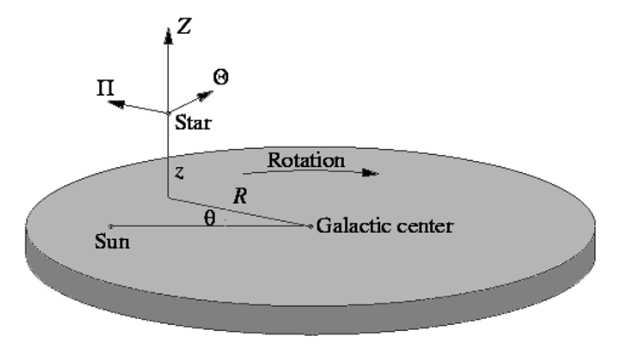}
    \caption{Representation of cylindrical coordinate system:  the center of the reference system is located on the Galactic Centre, the position of a star is given with $\theta$, R, and z. The three arrows indicate the three velocity components in a cylindrical coordinate system (\Theta, \Pi, Z). Figure adapted from \citet{Carroll1996ima..book.....C}}
    \label{fig:SDRcil}
\end{figure}

Observations are always made from Earth or relatively near it, hence the Sun is generally approximated as the site of all observations. Since the distance between the Earth and the Sun is much smaller than distances between the Earth and observed stars in the Galaxy, the difference in positions between the Earth and the Sun is usually neglected. Instead, differences in velocities can be considered relevant.\newline
In order to analyze the kinematics and dynamics of the Milky Way, it is convenient to adopt a coordinate system instantaneously centered on the Sun as it rotates around the Galactic Centre. This system is known as the Galactic coordinate system.\newline
The intersection of the Galactic midplane with the celestial sphere forms the Galactic equator. Technically, the Galactic midplane does not trace exactly a circle because the Sun is not precisely on the plane of the Galaxy, however the deviation is very small as the Sun is located just 30 pc above the galactic plane. Galactic latitude ($b$) and longitude ($l$) are measured in degrees from the line connecting the Sun to the galactic Centre, as depicted in Fig.~\ref{fig:SDR}.
While the Galactic coordinate system is useful for describing celestial positions as seen from Earth, it proves to be a less convenient choice to study kinematics and dynamics because the Sun itself is moving around the Galactic center. We therefore define the cylindrical coordinates system that places the Galactic Centre at its origin, as shown in Fig.~\ref{fig:SDRcil}. In this system R increases outward, the angular coordinate $\theta$ is pointed in the direction of rotation if the Galaxy and the vertical component increases to the north.\newline
It is important to notice that this set of directional choices results in a left-handed coordinate system instead of a right-handed one, because when seen from the north pole, the Galaxy appears to rotate clockwise rather than counterclockwise.\newline
The velocity components in this system are:
\[
\Pi=\frac{dR}{dt}
\]
\[
\Theta=R\frac{d\theta}{dt}\newline
\]
\[
Z=\frac{dz}{dt}
\]
To transform from the Sun to the center of the Galaxy, however, we need to consider the Sun's motion: as it moves around the Galactic Centre in the Galactic Plane, the Sun also moves slowly inward and farther north. We then define the Local Standard of Rest (LSR) to be a point that is instantaneously centered on the Sun and moving in a perfectly circular orbit about the Galactic Centre. The Sun's peculiar velocity relative to the LSR is generally referred to as the solar motion.
Thus, the Local Standard of Rest will be described by the velocities:
\[
\Pi_{LSR}=0,\; \Theta_{LSR} =\Theta_0,\;Z_{LSR} = 0
\]
The velocity of a star relative to the LSR is known as the star's peculiar velocity and it is given by:
\[
u=\Pi-\Pi_{LSR}
\]
\[
v=\Theta-\Theta_{LSR}\newline
\]
\[
w = Z-Z_{LSR}
\]
where $u$ is the radial component, $v$ is the tangential component of the motion and $w$ is the vertical component.

\subsubsection{Motions in the MW}
\label{chap1_motions}

Motions of stars are regulated by the law of universal gravitation: 
\begin{equation} \label{eq:FirstNewton}
    F=-\frac{GMm}{r^2}
\end{equation}
where M is the mass of the Galaxy which affects the star  of mass m, $r$ is the distance between them, and $G$ is the gravitational constant. 
Using the  gravitational potential $\Phi(x)$, the Newton's law can be replaced with the Poisson equation:
\begin{equation} \label{eq:Poisson}
    \nabla^2 \Phi(x) = 4\pi G\rho(x)
\end{equation}
where $\rho$ is the mass density and $\nabla^2$ the Laplace operator. 
A pair of functions $(\Phi, \rho)$ that solves the Poisson equation is known as a potential-density pair. Moreover, since the Laplacian is a linear differential operator, a linear combination of solutions is itself a solution. Consequently, if we consider a set of N masses, the potential is simply the sum of the potentials of the individual masses.\newline
Even though the mass of galaxies is contained in discrete portions, their overall distribution is quite smooth. This allows to consider both the gravitational potential and the gravitational force as smooth functions.\newline
Orbits are the trajectories of bodies in a gravitational field and they can be computed using the second law of dynamics. Stars orbiting inside galaxies usually have a mass which is small enough to allow us to consider the whole orbit independent of mass, as they do not affect the gravitational field. This is usually true for stars and dark matter particles in galaxies but not for massive bodies such as massive black holes and large satellite galaxies orbiting a more massive galaxy.\newline
If we can neglect the mass of the object, the relevant quantities that describe the orbit are:  energy, momentum, angular momentum, and the Lagrangian or Hamiltonian per unit mass. Since the orbit is defined by a second order differential equation, it is fully determined by its initial phase-space coordinates $(x_0, v_0)$.\newline
The second equation of Newton cannot usually be solved analytically, even in the simpler potentials. The full solution is usually computed numerically.
Bound orbits perform oscillations in radius between a pericenter $r_p$ and an apocenter $r_a$ that can be found solving $\dot{r} = 0$ as they are the minima and maxima of $r(t)$.\newline
A convenient quantity to describe the shape of an orbit is the orbital eccentricity $e$, defined as 
\begin{equation}
e = \frac{r_a-r_p}{r_a+r_p}.
\end{equation}
Another important parameter for this effect is the orbital circularity:
\[
\eta=\frac{L_Z}{L_{Z, circ}}
\]
Where $L_{Z, circ}$ is the angular momentum of a circular orbit with the same energy as the star, which thus takes extreme values +1 or -1 for co-planar circular prograde or retrograde orbits, respectively.\newline
When the distribution of matter is flattened in a disc, the situation gets even more complicated. The Poisson equation has to be solved in cylindrical coordinates and we need to resort to Fourier transform and Bessel functions.
Nevertheless, we can compute the solution and obtain the orbits of stars in such a potential.

\subsubsection{Computing stellar orbits}
\label{chap1_orbits}
\begin{figure}[H]
    \centering
    \includegraphics[width=0.7\linewidth]{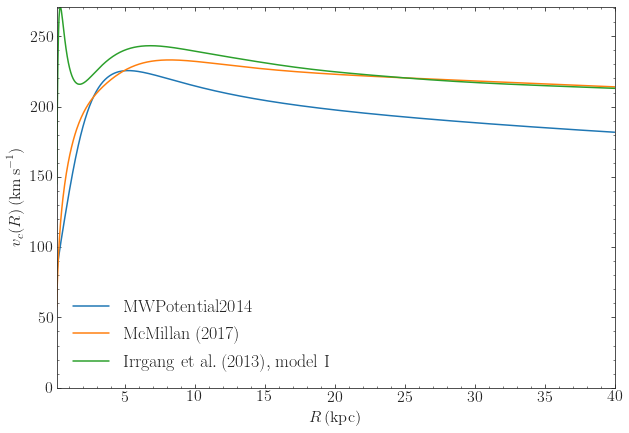}
    \caption{Tangential velocity as a function of galactocentric radius. Rotation curves computed with the Galactic potential models of \citet{mcmillan2016mass}, \citet{Irrgang2013A&A...549A.137I}, and \citet{Bovy2015ApJS..216...29B}. Figure adapted from \href{https://docs.galpy.org/en/v1.9.2/reference/potential.html}{\texttt{https://docs.galpy.org/en/v1.9.2/reference/potential.html}}.}
    \label{fig:gal_pot}
\end{figure}

There are many analytical models to describe  the shape of the galactic potential, as for instance the  models of \citet{Dehnen1998MNRAS.294..429D}, \citet{Irrgang2013A&A...549A.137I}, \citet{Bovy2015ApJS..216...29B},  \citet{mcmillan2016mass}, \citet{Cautun2020MNRAS.494.4291C}. 
They differ in the total mass distribution, and, e.g., in the predicted rotation curves. An example of the rotation curves obtained with three different Galactic potentials is shown in Fig.~\ref{fig:gal_pot}.
Here we describe  the potential  used in this thesis that is the  "McMillan potential" of 2017 \citep{mcmillan2016mass}. This model decomposes the galaxy into six axisymmetric components: bulge, thin and thick stellar discs, H~{\sc i} disc and molecular gas disc.
The model of the bulge is axisymmetric for ease of calculation, it therefore cannot accurately represent the inner few kpc of the Galaxy. This potential is based on the model fit by \citet{bissantz2002spiral}, with an  axisymmetric approximation.
The density profile used is the following.
\begin{equation} \label{DensBulge}
 \rho_b = \frac{\rho_{0,b}}{(1+r\prime/r_0)^\alpha}exp\bigg[-\bigg(\frac{r\prime}{r_{cut}}\bigg)^2\bigg]
\end{equation}
\begin{equation} 
 r\prime = \sqrt{R^2+(z/q)^2}
\end{equation}
with $\alpha = 1.8$, $r_0 = 0.075 kpc$, $r_{cut} = 2.1 kpc$ and the axis ratio $q = 0.5$. \newline
The stellar disc is commonly decomposed in a thin and a thick disc modelled as exponentials.
\begin{equation} \label{Densdisc}
 \rho_d(R,z) = \frac{\Sigma_{0}}{2z_d}exp\bigg(-\frac{\abs{z}}{z_d}-\frac{R}{R_d}\bigg)
\end{equation}
with scaleheight $z_d$ , scalelength $R_d$ and central surface density $\Sigma_0$.\newline
The gas disc is a fundamental component of the Milky Way potential as near the Sun it shows a much smaller scale-height than the stellar component and its presence significantly deepens the potential well near the Sun. 
Both the molecular gas disc and the H~{\sc i} disc are described as follows:
\begin{equation} \label{DensGas}
 \rho_d(R,z) = \frac{\Sigma_{0}}{4z_d}exp\bigg(-\frac{R_m}{R}-\frac{R}{R_d}\bigg) sech^2(z/2z_d)
\end{equation}
Their density profile declines exponentially at large R, but also has a hole in the centre of scale $R_m$
The dark matter halo is described by the simple density profile often used in dark matter simulations:
\begin{equation} \label{DensDMhalo}
 \rho_h = \frac{\rho_{0,h}}{x^\gamma(1+x)^{3-\gamma}}
\end{equation}
In this model $\gamma$ is considered 1, corresponding to a Navarro-Frenk-White profile \citep{navarro1996structure}.
Once the position, the distance (from parallax), the radial velocity and the proper motion  of a star have been defined, it is possible to calculate their orbits  by means of integration methods and using the galactic potential. 
The main parameters that define a stellar orbit are: 
the apogalacticon and perigalacticon radii, the eccentricity, the integrals of motions such as energy E, total angular momentum $L$ for a spherical system or $L_Z$ and the actions $J_r$, $J_Z$ and $J_{\phi}$ for axisymmetric systems.\newline

\subsubsection{The separation of Galactic components with kinematics and dynamics}

\begin{figure}
    \centering
    \includegraphics[width=0.5\linewidth]{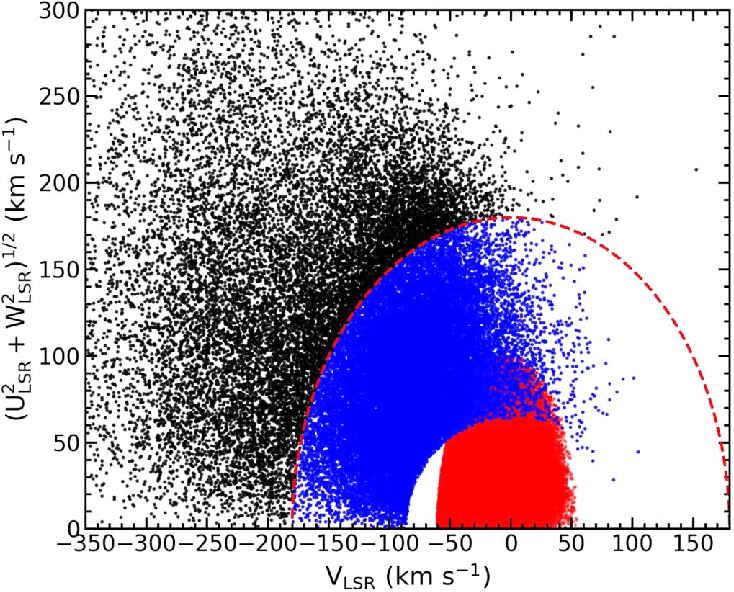}
    \caption{Toomre diagram for stars from LAMOST and {\em Gaia} {\sc dr2}. Black dots are halo stars. Blue and red dots are thick and thin disc stars respectively. The cut in halo selection is at $v_{tot} = 180\;km\;s^{-1}$. Figure adaprted from \citep{yan2019chemical}.}
    \label{fig:ToomreSepar}
\end{figure}

\begin{figure}
    \centering
    \includegraphics[width=0.9\linewidth]{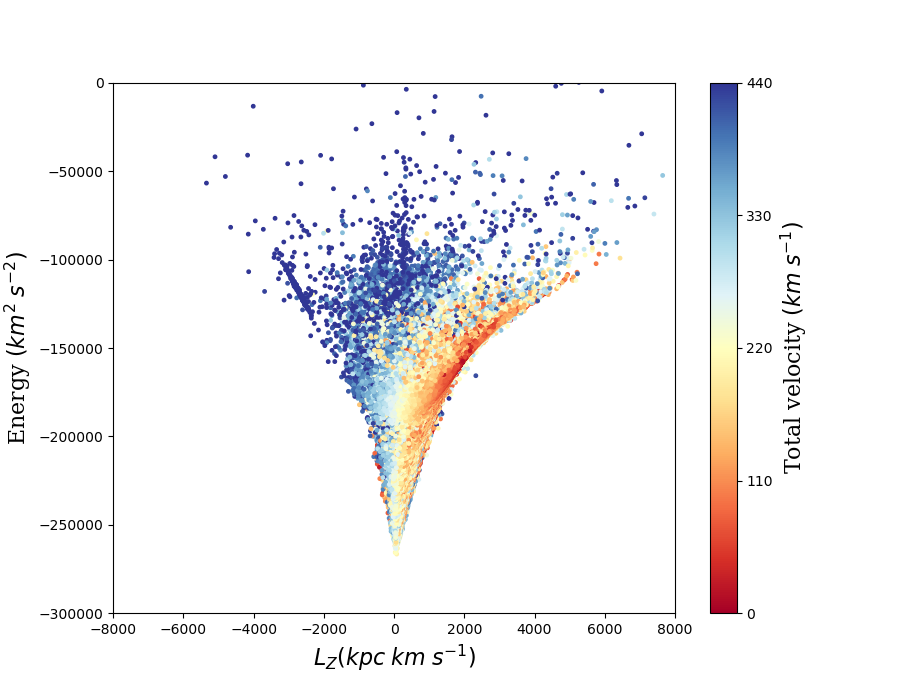}
    \caption{Example of the Lindblad diagram $(E-L_z)$. Stars are coloured based on their velocity: halo stars are in blue $(v_{tot}\geq220 km\;s^{-1})$, disc stars are in red $(v_{tot}<220 km\;s^{-1})$.}
    \label{fig:ELzex}
\end{figure}

The kinematic properties of the populations belonging to the various parts of the Galaxy can be used to separate them. 
One of the classic diagrams used for this purpose is the Toomre diagram. 
An example is shown in Fig.~\ref{fig:ToomreSepar}, in which 
$\sqrt{U^2+W^2}$ is plotted against $V$.
This diagram can  be effectively used to  distinguish disc stars (inside the dashed red line in Fig.~\ref{fig:ToomreSepar}) from halo stars from their kinematics properties.\newline
Indeed, in the LSR, i.e. the system of coordinates centered on the Sun and moving with the Solar angular velocity, 
the total velocity $v_{tot} = \sqrt{V^2+U^2+W^2}$ is expected to be  below $\sim$ 200 km s$^{-1}$ for  stars in the thin and thick disc. Halo stars, instead, are those with a total velocity over $\sim$ 200 km s$^{-1}$. This happens because disc stars in the LSR appear as the `fixed' ones, i.e. to move with velocities close to the Solar one,  whereas halo stars apparently move faster.
Another interesting diagram to study the separation between populations is the Lindblad diagram in which the total  energy is plotted versus the angular momentum $(E - L_z)$.  An example is shown  in Fig.~\ref{fig:ELzex}, where $L_z$ is the z-component of the angular momentum computed in the Galactocentric coordinate system. This plot distinguishes prograde stars from retrograde stars with the boundary set at $L_z = 0$. Depending on the coordinate system chosen, prograde stars can either be at $L_z > 0$ or $L_z<0$. In Fig.~\ref{fig:ELzex}, for instance, prograde stars are at $L_z>0$, where we observe an overdensity in red that corresponds to stars from the disc.
This plot is particularly useful because it has been shown through simulations that structures accreted in the Milky Way, with similar energy and angular momentum at the beginning of their accretion, maintain their original clumps in this plot even after many Gyr of evolution \citep[e.g.][]{helmi2000mapping}.

\subsection{The characteristics of the Galactic Halo: in situ and accreted populations}

\begin{figure}
    \centering
    \includegraphics[width=0.75\linewidth]{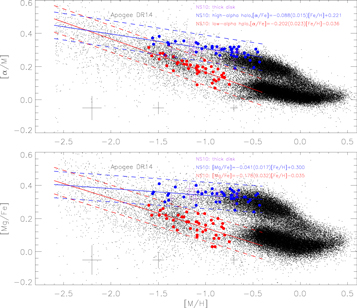}
    \caption{[$\alpha$/M] vs [M/H] (upper panel) and [Mg/Fe] vs. [Fe/H] (lower) diagrams from \citet{Chen2019ApJ...871..216C}. Low-$\alpha$ halo, high-$\alpha$ halo, and thick disc stars from \citet{nissen2010} are shown with red dots, blue dots, and purple pluses, respectively. }
    \label{fig:AlphaHalo}
\end{figure}

According to the standard cosmological scenario, galaxies grow their stellar mass on the one hand by internal mechanisms such as star formation and on the other hand by accretion of already formed stars in dwarf galaxies or stellar clusters.
The stars formed from molecular clouds inside the MW form the `in situ' population, while stars accreted from nearby satellite galaxies orbiting the Milky Way create the `ex situ' or accreted population. 
The halo retains memory of its merger history, and contains the remnants of both the first locally formed stars and the accreted ones. 
They can be separated, e.g.,  in the 
[$\alpha$/Fe]-[Fe/H] plane, where $\alpha$ indicates the average abundance of $\alpha$-elements such as Mg, Si, Ca and Ti, or in the [Mg/Fe] vs [Fe/H] plane.
An example is shown in Fig.~\ref{fig:AlphaHalo}, where 
we can distinguish the halo `in situ' population with high [$\alpha$/Fe] content, the accreted halo population with low  [$\alpha$/Fe]. The thick disc stars often overlap at low metallicity with halo stars formed `in situ', suggesting a common formation process \citep[see, e.g.][]{dimatteo2019A&A...632A...4D}.  


\subsubsection{The population of globular clusters}
\begin{figure}
    \centering  \includegraphics[width=0.75\linewidth]{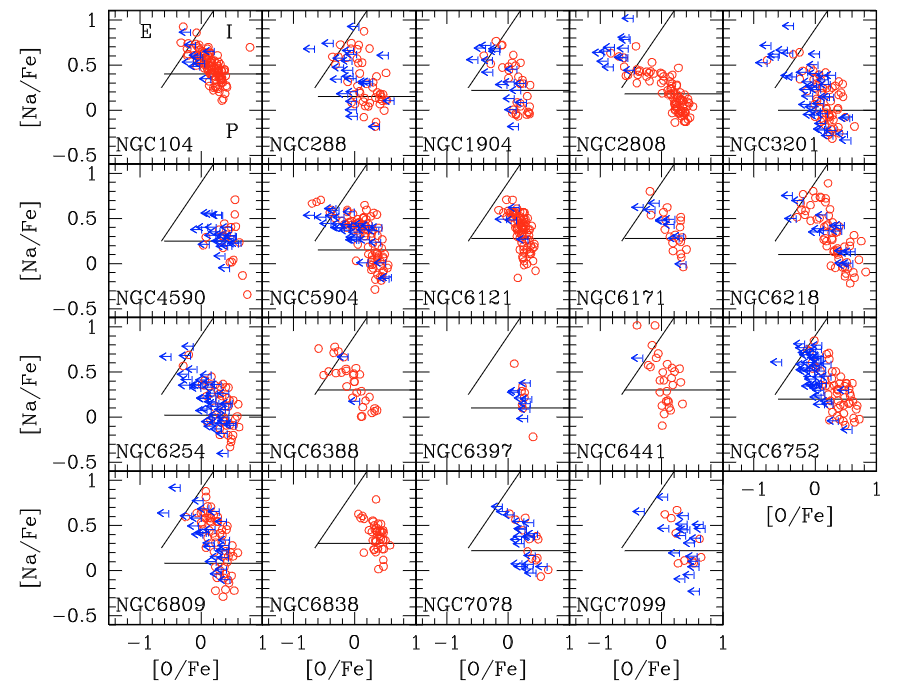}
    \caption{[Na/Fe] vs [O/Fe] anti-correlations in a sample of Galactic globular clusters. Figure adapted  from \citet{carretta2010properties}.}
    \label{fig:Na-Oanti}
\end{figure}

\begin{figure}
    \centering    \includegraphics[width=0.9\linewidth]{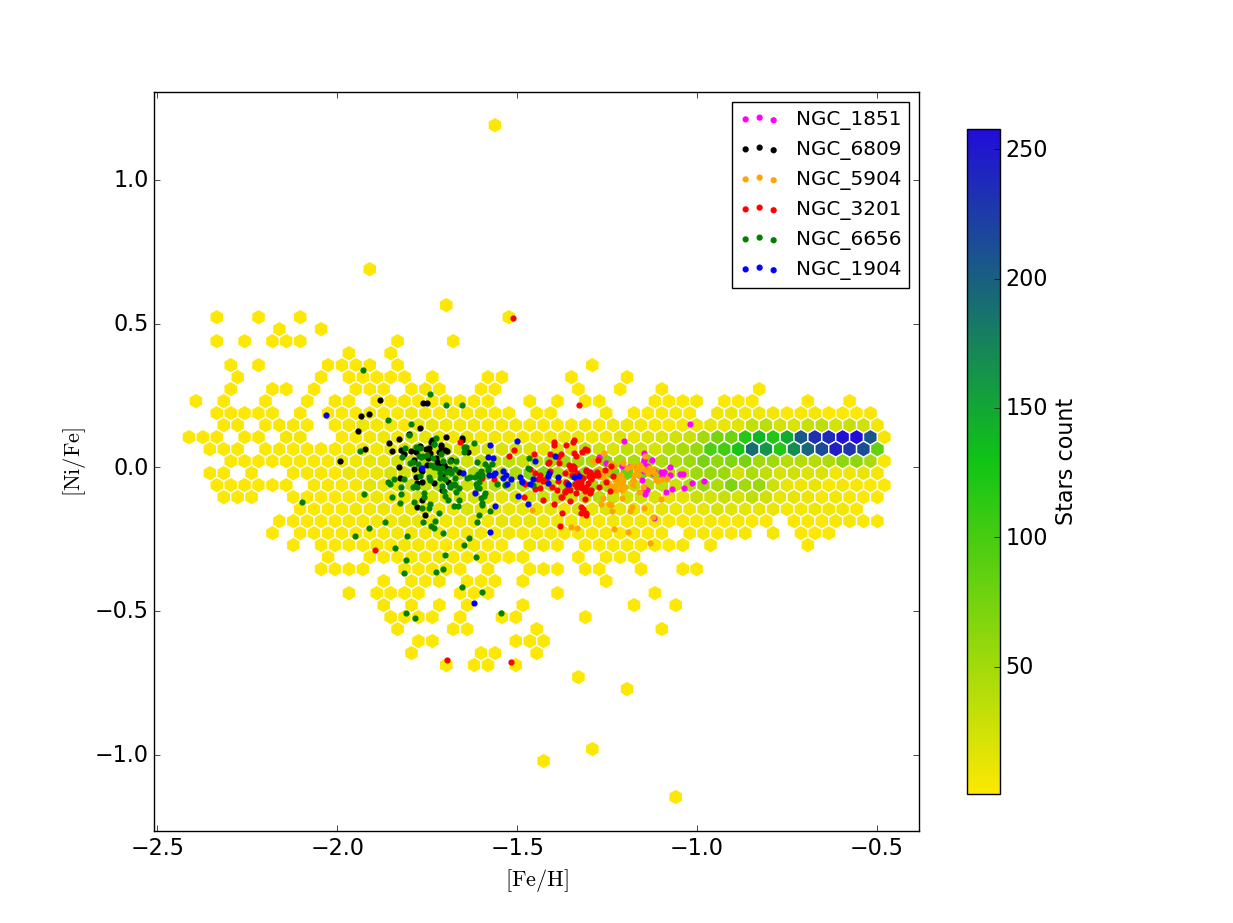}
    \caption{[Ni/Fe] vs [Fe/H] (iron-peak element) for the member stars of  GCs and for field stars, selected in this work from APOGEE database.}
    \label{fig:NiFe}
\end{figure}

\begin{figure}
    \centering
    \includegraphics[width=0.9\linewidth]{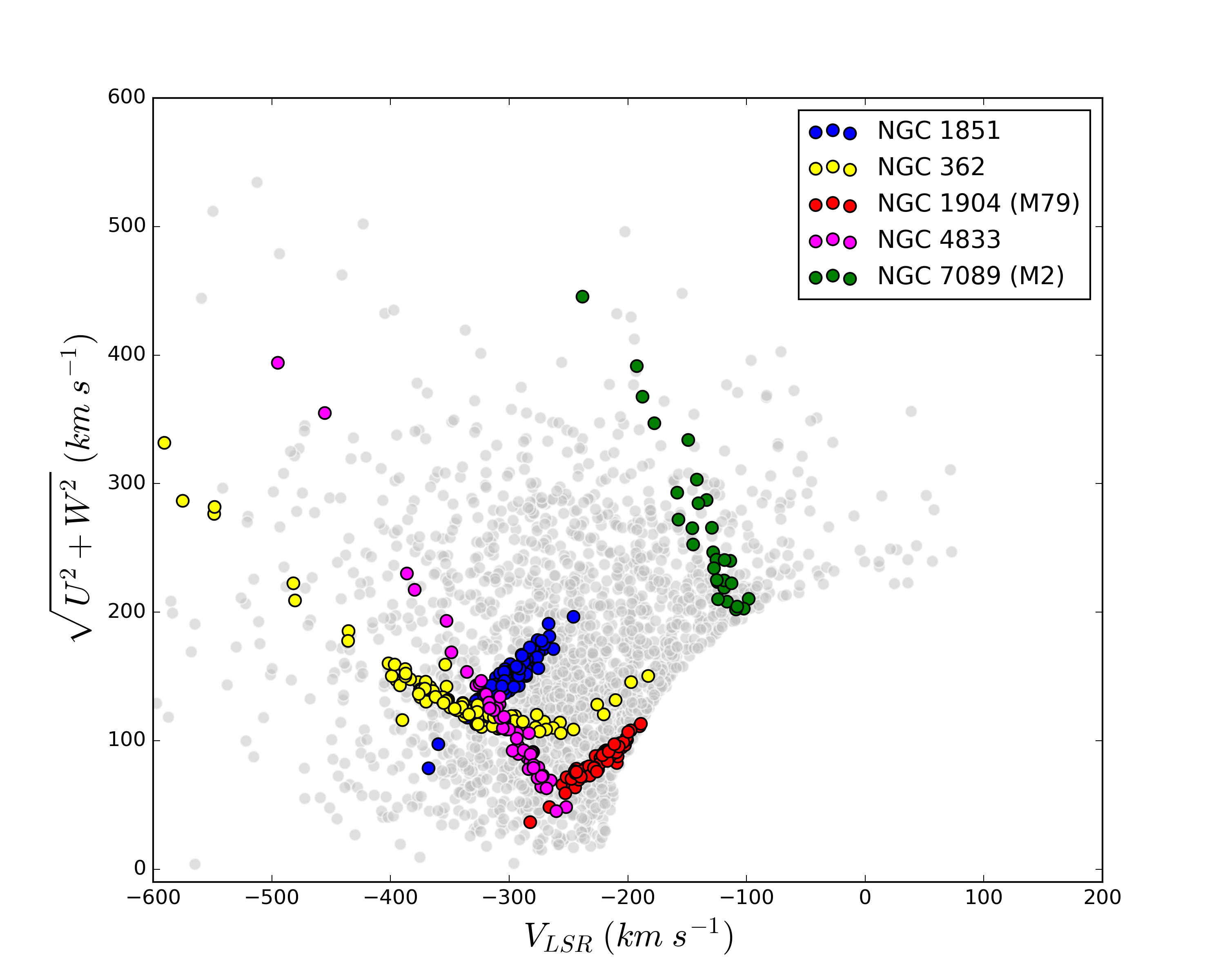}
    \caption{Toomre diagram for the halo field stars (in grey) and for members of GCs (colour-coded as in the legend). Selection done in  this work from the APOGEE database.}
    \label{fig:ToomreCluster}
\end{figure}

Globular Clusters are groups of approximately $10^4-10^6$ gravitationally bound stars, typically spherically symmetric, highly concentrated to the center and spread over a volume ranging from a few dozen up to more than 100 pc in diameter. Their age usually ranges from 10 to 12 Gyr. The Milky Way contains nearly 200 globular clusters orbiting within its galactic halo, with orbital patterns that are not confined to the plane of the galactic disc but rather spherically distributed. \newline
As far as iron-peak elements are considered, GCs are generally very homogeneous, as shown for Ni in Fig.~\ref{fig:NiFe}. They show, instead, a quite strong anti-correlation between Na and O that could be explained by their dense environment and the presence of multiple populations.
As explained by \citet{carretta2010properties}, the anti-correlation between Na and O (example in Fig.~\ref{fig:Na-Oanti}) and the one between Mg and Al can be justified with proton-capture reactions of the CNO cycle. Variations of these abundances, though, can be found also in unevolved stars currently on the main sequence. This means that this abundance pattern has been imprinted in the gas by a previous generation of stars as low mass stars in the main sequence are not able to reach the high temperature necessary for the nucleosynthesis of these elements.
Globular clusters are then considered to harbour at least two generations of stars \citep{Pancino2017A&A...601A.112P}.
The chemical composition of Globular Clusters, as for the MW Halo, is typically metal-poor. Galactic GCs, in fact, are population II objects with metal abundances  in the range $-$2.5 dex$<$[Fe/H]$<$-0.7 dex.\newline
The chemical homogeneity of iron-peak elements in GCs is a characteristic that distinguishes them from galaxies, that exhibit high variations in metallicity. However, not all GCs conform to the mono-metallic pattern. A notable example is Omega Centauri (NGC 5139), identified as the most massive GC in the Milky Way with a mass of approximately $M \simeq 4 \times 10^6 M_{\odot}$. Spectroscopic investigations have revealed a spread in metallicity of up to $\sim$ 1.0 dex and a complex metallicity distribution function. Current understanding suggests that Omega Centauri is not a typical GC but rather the nucleus of a dwarf galaxy that merged into the Galactic halo at some point in the past \citep{freeman2002new}. Additional instances of GCs with metallicity spreads include M22 and M54 \citep{Milone2022Univ....8..359M}. Globular clusters are also characterized by a little spread of velocities within the cluster. They can indeed be easily identified in the Toomre diagram, as shown in Fig.~\ref{fig:ToomreCluster} in which we highlight the location of some of the clusters studied in the present work. 

\begin{figure}
    \centering    \includegraphics[width=0.5\linewidth]{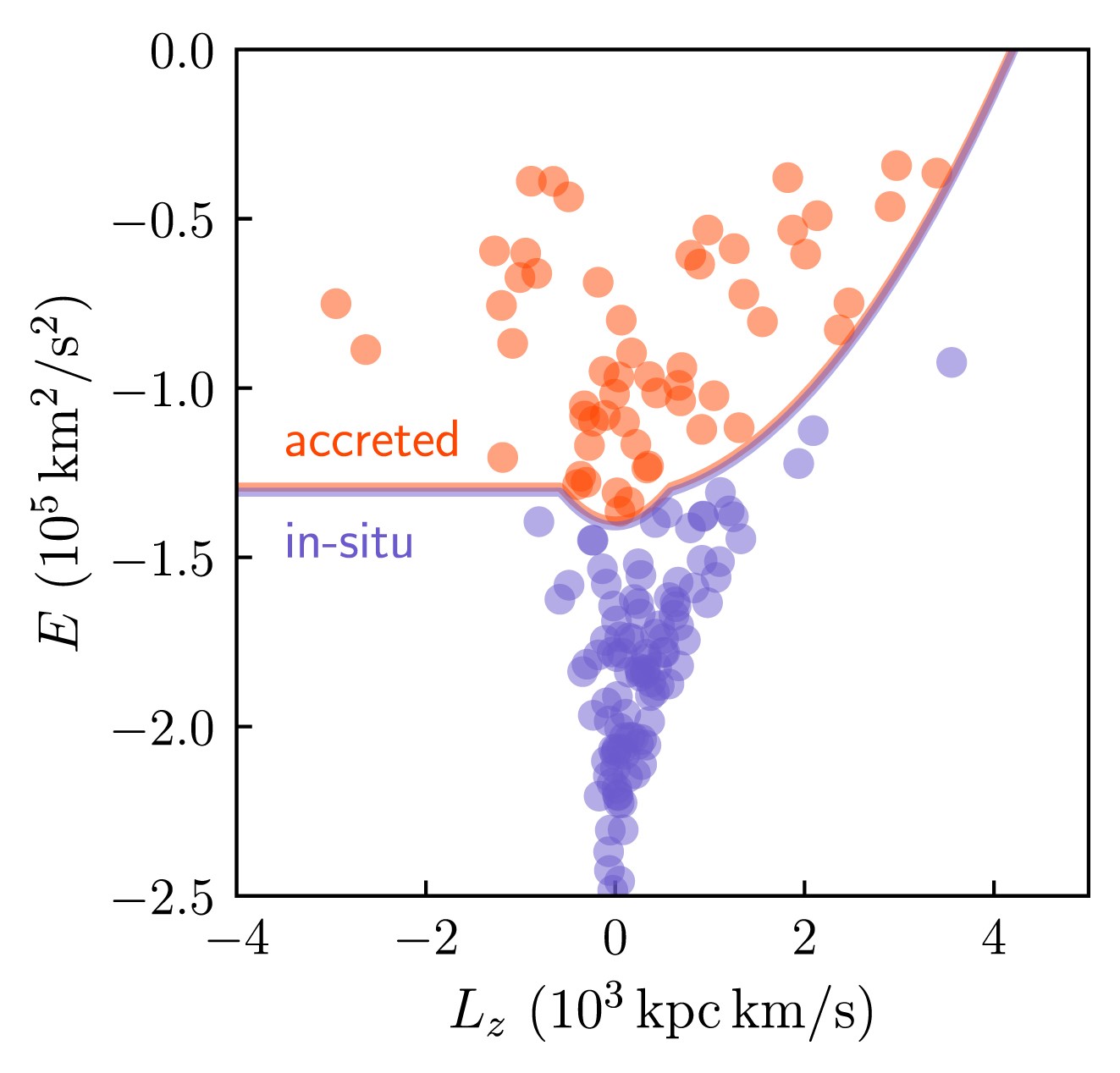}
    \caption{Total energy E and z-component of the orbital angular momentum of the in situ and accreted globular clusters. Figure adapted from \citet{Belokurov2024MNRAS.528.3198B}.}
    \label{fig:GCs}
\end{figure}

As for the origin of globular clusters, they can belong to the `in situ' or accreted population as well \citep[see.,e.g.][]{Belokurov2024MNRAS.528.3198B}. 
In Fig.~\ref{fig:GCs}, we show the distribution of the GCs in the total energy vs z-component of the angular momentum plane.  `In situ' GCs are distributed within the central 10 kpc of the Galaxy in a flattened configuration aligned with the Galactic disc, while the accreted GCs have a wide distribution of distances and a spatial distribution close to spherical. The two populations have also different metallicity distribution and are well-separated in the  [Al/Fe]-[Mg/Fe] and Lindblad planes.

\subsubsection{The halo streams}

\begin{figure}
    \centering
    \includegraphics[width=0.75\linewidth]{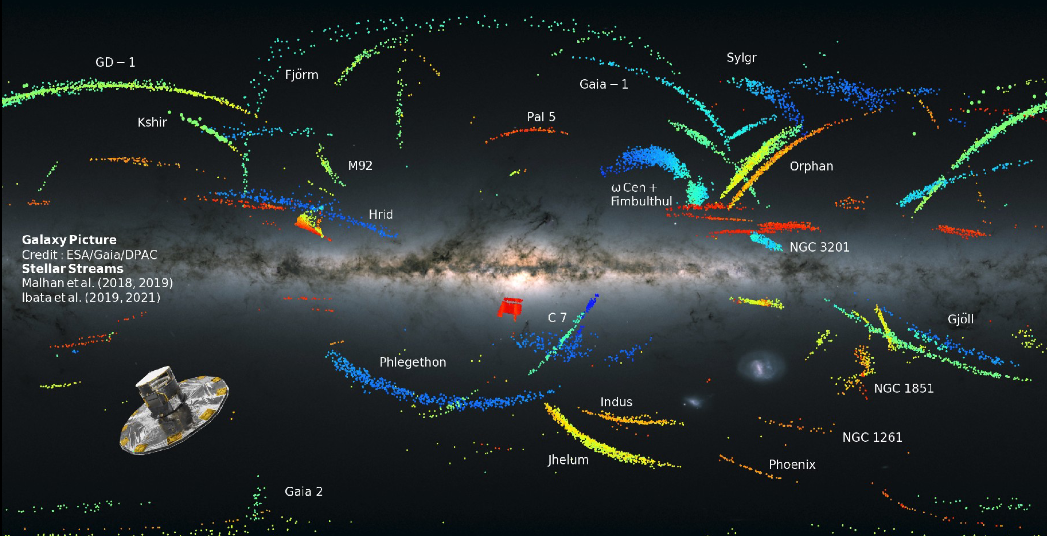}
    \caption{Stellar streams in the Milky Way found with Gaia DR2 and EDR3 data from \citet{malhan2018stream}, \citet{malhan2018} and  \citet{ibata2019}). The stellar streams are colour-coded by their heliocentric distance, where red (blue) means farther away (closer). A part of the streams are labelled by their name. Credit: ESA/GaiaDPAC}
    \label{fig:StreamPic}
\end{figure}

As globular clusters or dwarf galaxies orbit  the Galactic Centre, they undergo various perturbations, including tidal forces from the Milky Way, passages through the galactic plane, and potential star escape, thus creating stellar streams that are still identifiable and visible thanks to the recent surveys, in particular the {\em Gaia} mission (see Fig.~\ref{fig:StreamPic}). The present time population of GCs can be considered as the survivors of a much wider population, partially disrupted and spread out throughout the Galactic Halo and beyond.\newline
When a galaxy - or a cluster - is disrupted by tidal forces, its stars continue to follow closely the trajectory of the system they used to belong to. This means that their integrals of motion, such as energy and one component of angular momentum $L_z$ (in an axisymmetric system) are conserved. Since a galaxy is a collisionless system, the tidally stripped stars will follow very similar orbits to their progenitor. This explains why streams are long
and narrow if they originated from a small system or formed recently.
Numerous studies have however suggested that streams cannot be observed in the configuration space \citep[e.g][]{helmi2000mapping}. With time, streams can be disrupted, leading to the incorporation of their stars in the halo. Observations indeed suggest that a considerable fraction of Milky Way halo stars may originate from disrupted GCs, potentially ranging from 10\% to 50\%  \citep[see, e.g.][]{beasley2020}.\newline
In addition, debris of streams can mix spatially with time causing one single system to be responsible for multiple streams in a given location of the MW. Along these lines, predictions from numerical simulations suggest that a single accretion event can lead us to multiple substructures in phase space \citep[e.g.][]{koppelman2020}.
Some streams have been connected to their progenitors like a surviving globular cluster, but most of them have an unknown origin.

The spread in metallicity and in abundances in a stream depends on its origin. If the stream originates from a globular cluster, the abundances of its members are expected to be generally quite similar, reflecting the characteristics of GCs described above. Some streams, instead, show a large metallicity spread. This implies that its stars did not form in a single burst, but rather with a complex star formation history, like in a dwarf galaxy.


\subsection{Aim of this Thesis}

The unprecedented quality of {\em Gaia} data have initiated a revolution in Galactic astronomy. 
Studies of star clusters have been transformed by the precise proper motions and parallaxes measured by {\em Gaia}. 
A large number of new clusters have been discovered in the disc of the Galaxy \citep{cantat2022, Hunt2021A&A...646A.104H, Hunt2023A&A...673A.114H, Hunt2024arXiv240305143H}, as well as many new structures in the halo, such as clusters, streams, remnants of mergers \citep[see, e.g.,][]{Malhan2022ApJ...926..107M}. In addition to that, the large amount of data provided by {\em Gaia} is changing the way astronomers approach datasets as their dimensionality and volume increase.
Thus, the search for new star clusters and chemo-dynamical structures in general requires new techniques based on clustering algorithms that are more and more common. The possibility to use many different approaches to this search, allowed the astronomers to find us new objects even in regions where they have already been found.
Up until now, the search for globular clusters, streams and merger debris  has been implemented mostly in the kinematic space and, partly, in the chemical space through the application of clustering algorithms. \\
The aim of this thesis is to develop a method of clustering to find accreted structures in the MW halo by properly using both the kinematic and chemical information. 
The thesis is structured as follows: after introducing the context, in Chapter~\ref{chap2} we describe the machine learning techniques we have adopted. In Chapter~\ref{chap3}, we describe the preparation of the observational samples, extracted from the large spectroscopic survey APOGEE and {\em Gaia}-ESO, including the method used to calculate the stellar orbits. We provide a chemical characterisation, for the first time with {\em Gaia}-ESO, of several  halo streams.   In Chapter~\ref{chap4} we describe our algorithm, the {\sc creek}. 
In Chapter~\ref{chap5}, we show the results of our analysis applied to the two surveys, highlighting strengths and weaknesses of the method.
In Chapter~\ref{chap6}, we present future perspectives of our work in the landscape of the forthcoming large spectroscopic surveys and in light of the next generation of instruments. 
Finally, in Chapter~\ref{chap7} we give our summary and conclusions. 

%% file: 2-Metodi.tex
\section{Deep clustering with machine learning techniques}
\label{chap2}
\subsection{Machine Learning Algorithms: general concepts}

Human beings have always looked for ways to accomplish various tasks in a simpler way, and this is what has led to the invention of many machines. Machine learning (ML) techniques reside among them.
A simple definition given nowadays is that {\em a machine learning algorithm is an algorithm that is able to learn without being explicitly programmed}. A computer program is defined to be learning if its performance in a task improves with experience.
The present-time data volume in astrophysics is impressively large compared to the data to which astronomers had access to only 20 years ago and it will increase exponentially in the coming years thanks to the numerous new surveys and telescopes, as described in Chapter~\ref{chap1_landscape}. Methods that rely only on human work are no longer applicable to these huge datasets, both in terms of the amount of data to be analysed and the time required. 
We risk to miss or delay interesting and important discoveries due to our inability to accurately and consistently interrogate astronomical data.
ML has emerged as a powerful tool to improve analysis procedures, leveraging on the principle that an increased amount of data typically enhances the efficacy of machine learning algorithms. As it automates information extraction from data, ML is used in many fields.\\
In this chapter, we explore a selection of machine learning techniques within the vast emerging panorama of new methods. In particular, we focus on those that will later be used during the thesis work. \newline

\subsubsection{General classification}
The ML algorithms can be classified in two main groups:  supervised and unsupervised algorithms.

\paragraph{Supervised Learning}
Supervised Learning is a  machine  learning paradigm for  acquiring  the  input-output relationship information of a  system based  on  a  given  set of  paired input-output training  samples. The output can be considered as the labels of the input data. Therefore,  an input-output training sample is  called labelled training data or supervised data.
From a mathematical point of view, the ML algorithm looks for the best matching of the kind $Y = f(X)$, where $Y$ are the outputs and $X$ the inputs. As the answer is already known, the algorithm supervises the learning so that the predicted answers are similar to the actual answers.\newline
Supervised learning can be categorized into:
\begin{itemize}
	\item Classification: it allows  the separation of the data according to some patterns. The learning process provides data and labels and lets the ML create a model that assigns the labels to data based on some examples.
	\item Regression: it is an algorithm used to predict continuous numerical values based on input features. The goal is to learn a mapping function that maps input features to a continuous variable
\end{itemize}

\paragraph{Unsupervised Learning}
Unlike supervised learning, in unsupervised learning there is no target answer. Algorithms are only fed the input data and no corresponding output is provided. For instance, many ML algorithms group data instances according to their similarity. These are mainly used for clustering and feature reduction \citep[e.g.][]{mahesh2020}.\newline
The main difference between supervised and unsupervised learning is the type of input data, as unsupervised machine learning does not require a labelled dataset. Moreover, the goals of supervised ML are predetermined, meaning that the type of output of a model is already known before the algorithm is applied.\newline
An extremely popular family of Unsupervised Learning are the clustering algorithms. They can find similarities or hidden patterns in the dataset in order to partition it, resulting particularly useful for its easy visualization and interpretability.

The principal algorithms adopted in the present work  are Neural Networks (supervised) and {\sc OPTICS} (unsupervised), which are described below.

\subsection{Neural Networks}

\begin{figure}[H]
    \centering
    \includegraphics[width=0.5\textwidth]{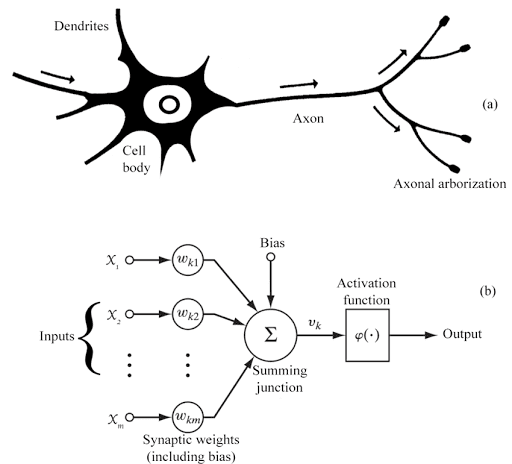}
    \caption{Representation of a human and an artificial neuron. Figure adapted from \citet{akgun2018modeling}.}
    \label{fig:NeurConfr}
\end{figure}

Artificial Neural Networks (ANNs) are computational models that reproduce the behavior of neurons in a human brain. They  allow computers to solve specific tasks.\newline
In a human brain, neurons receive a signal from the dendrites and bring it towards the center of the cell; axons, instead, move the signal to other neurons.
Exactly as a neuron in the brain, an artificial neuron receives an input signal, processes it and then sends the processed signal to the other neurons it is connected to, as shown in Fig.~\ref{fig:NeurConfr}.\newline

\begin{figure}
    \centering
    \includegraphics[width=0.99\textwidth]{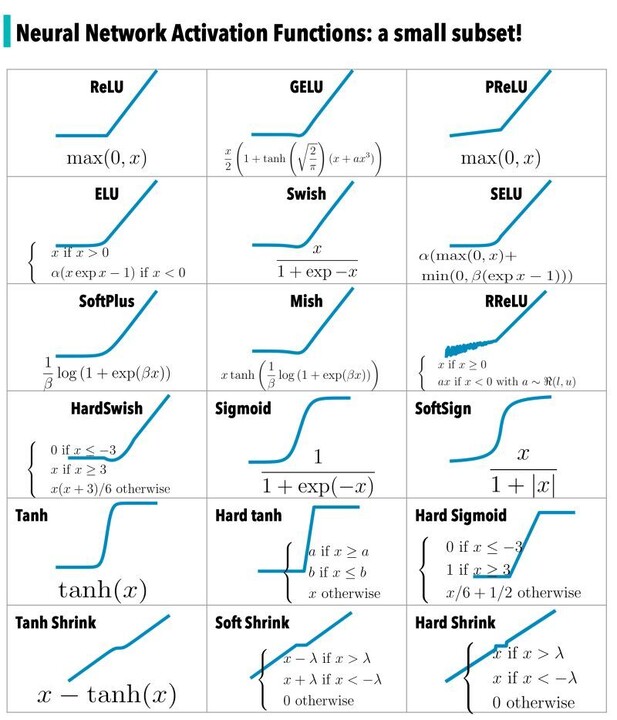}
    \caption{A few examples of activation functions (Figure adapted from \href{https://www.linkedin.com/pulse/activation-functions-heba-al-haddad/}{https://www.linkedin.com/pulse/activation-functions-heba-al-haddad/}).}
    \label{fig:ActivationFunction}
\end{figure}

An artificial neuron, called perceptron, takes a number of numeric inputs, $x_i$ and multiplies each of them for a corresponding weight $w_i$ that describes the importance that the ANN assigns to the input. This list of products is then summed and it gets added a bias value $b$, as described in Equation \ref{eq:weight}.
\begin{equation}\label{eq:weight}
    x_1w_1+x_2w_2+...+x_nw_n+b
\end{equation}
The result of this weighted sum is then passed through an activation function, that can have many shapes. The shape of the activation function usually depends on the purpose of the ANN. In our networks we mostly used a Rectified Linear Unit (ReLU) activation function, i.e. a function that has value zero for negative numbers and y=x for positive numbers, and the {\sc sigmoid} activation function, which is particularly useful to extrapolate probabilities from the network. Examples of other activation functions are depicted in Fig.~\ref{fig:ActivationFunction}.\newline

\paragraph{The structure of a neutral network}
A neural network (NN) is a network structured in layers of perceptrons. Perceptrons of a given layer are typically connected to those of the previous and subsequent layers. A NN receives in input the features of a given dataset and during training it optimizes the complex structure of weights and biases in order to perform a given task, such as to predict one or more parameters. Thus, a NN is formed by an input layer which - by definition - has as many neurons as the input features and by an output layer which is composed by a number of neurons equal to the number of values the algorithm has to predict. These latter are also known as ''target variables''. In between the input and the output layers there are other layers of neurons called hidden layers. The number of hidden layers and the number of neurons within each hidden layer are parameters that have to be specified by the user, depending on the type of task the algorithms has to accomplish and the available dataset. Typically, a larger number of layers and neurons increases the complexity of the algorithm, but also its ability to learn from smaller patterns of the  dataset.\newline
NNS are typically ``fully connected'', meaning that every neuron from a layer is connected to every neuron of the following layer. This is the simplest type of architectures NNs can have, which is applicable to a broad range of problems. However, there are special cases where NNs could be designed in more complex structures in order to capture specific patterns and relationships within the input data. An example of a non-fully connected NN is the Siamese Neural Network that is used in this work and it is described in Section \ref{SNN}. For simplicity's sake, the next few paragraphs focus on the functioning of fully connected NNs, however the general principles can be applied to all the other architectures of NNs.

\begin{figure}
    \centering
    \includegraphics[width=1\linewidth]{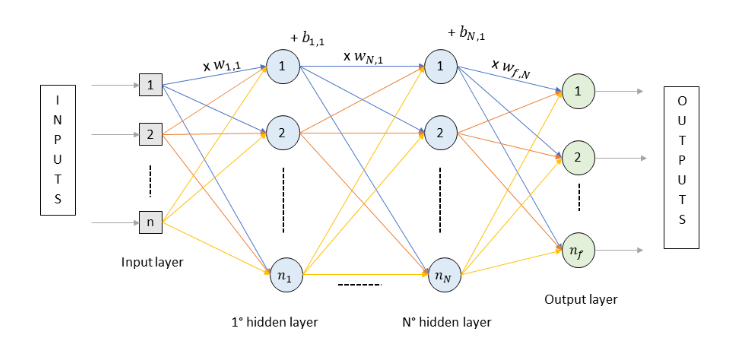}
    \caption{Operation diagram of a fully connected neural network. Inputs are represented as numbered squares and perceptrons of the hidden layers as numbered rounds in lightblue. The output layer is in the green circles. Arrows show the connections between the perceptrons, every perceptron is connected to the following in a different colour.}
    \label{fig:NN}
\end{figure}

The operation diagram of a fully connected neural network is depicted in Fig.~\ref{fig:NN}.
The information flows across a NN from the input layer, through the hidden layers, down to the output layer which produces the prediction.  If we assume that the input layer has $n$ inputs - the features we want to analyze - which is expressed by a vector X=[x$_1$, x$_2$, ..., x$_n$] and if the first hidden layer is composed by m neurons, then the layer will compute the following
\begin{equation}
    h_1 = x_{1}w_{11}+x_{2}w_{12}+...+x_{n}w_{1n}+b_{1}
\end{equation}
\begin{equation}
    h_2 = x_{1}w_{21}+x_{2}w_{22}+...+x_{n}w_{2n}+b_{2}
\end{equation}
\begin{equation}
    \vdots \newline
\end{equation}
\begin{equation}
    h_m = x_{1}w_{m1}+x_{2}w_{m2}+...+x_{n}w_{mn}+b_m\newline
\end{equation}
where w$_{\rm ij}$ is the weight of the i-th neuron associated to the j-th input feature, while b$_i$ is the bias term of the i-th neuron. This information can be expressed in a more compact form using the matrix notation. Namely, if W is the $m\times n$ matrix of all weights and B the bias array, the operation carried out by the first hidden layer can be written as 
\begin{equation}
    H = W^T X+B
\end{equation}

These values are then passed to the activation function $\Phi$, producing the output Z of the layer: 

\begin{equation}
    Z = \Phi(XW^T+B)
\end{equation}
The Z array of m elements is then passed to the next hidden layer. This process is repeated for as many hidden layers as the NN has. The final results are then passed to the output layer, which produces the network predictions. 
During the training phase, the neural network learns from data the optimal set of weights and biases to perform the given task. The dataset is generally composed by two vectors: the input features, that are usually called X, and the target features, which are the values that the model has to predict and that are usually called Y.
The data used to train the NN are called the training set. The successful analysis of the training set provides us with a model to be applied to the complete dataset.
In order to establish how good a NN is to generalize once applied to data, datasets are divided into two main subsets: the {\bf training set} and  the {\bf test set}. Unlike the training set, the algorithm does not learn from the test set, thus it provides an unbiased evaluation of the model performance on new, unseen data. The test set  is crucial for assessing how well the model generalizes to data it has never encountered before.

\begin{figure}[H]
    \centering
    \includegraphics[width=0.5\linewidth]{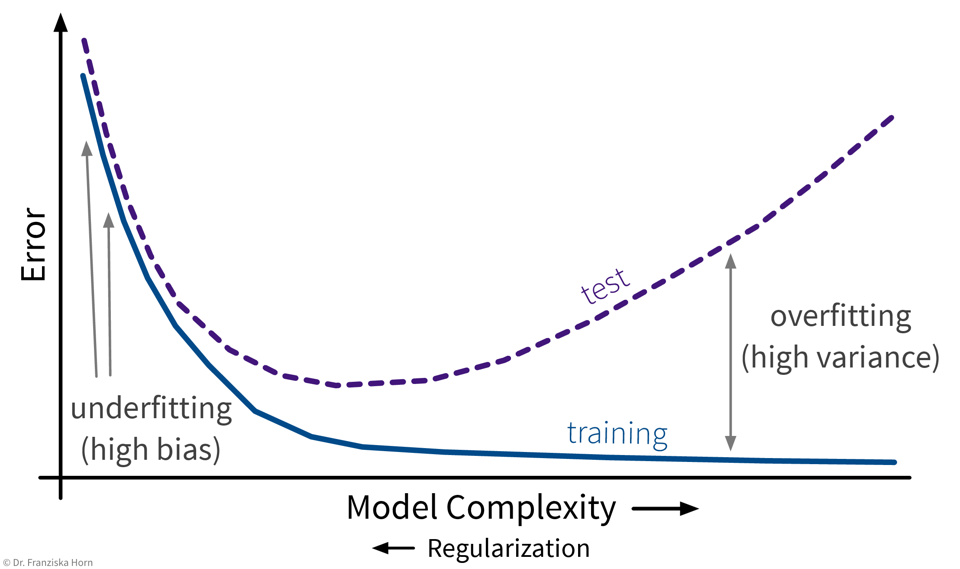}
    \caption{Error as a function of the model complexity. In purple is depicted the curve of error for the test set and in blue the curve for the training set. Figure adapted from \href{https://franziskahorn.de/mlbook/_model_does_not_generalize.html}{\texttt{https://franziskahorn.de/mlbook/\_model\_does\_not\_generalize.html}}.}
    \label{fig:test-train}
\end{figure}

In practice, once the NN is trained over the training set, one can apply the resulting model to the test set and derive a generalization error, which measures its ability in predicting the correct target features on unseen data. During the training phase the NN also measures the performance on the training set itself. The prediction error on the training set always diminishes with the complexity of the algorithm, i.e., the number of layers, the number of neurons, the number of the training iterations. When the level of complexity is low, the error on the test set decreases too, following that on the training set. However, when the model becomes too complex, error on the test set reaches its minimum and starts to climb up again. This regime is called {\bf overfitting} and it occurs when the model trained on the training set is so complex that is unable to generalize well on the test set. Ideally, we would like to stop the training at the minimum of the prediction error derived on the test set. This behaviour is depicted in Fig.~\ref{fig:test-train}.  \newline
Training a neural network consists in minimizing the prediction error by tuning the set parameters $\Omega$ = { $b_1$, $w_1$... $b_n$, $w_n$} which are the weights and biases of the NN. The prediction error is typically measured by a loss function, a derivable function that measures the difference between the prediction computed by the NN and the label Y. The loss $L$ itself is function of the set of parameters $\Omega$. The technique of minimizing $L$ is called {\bf gradient descent}, which is carried out by iteratively adjusting the parameters $\Omega$ until the absolute minimum of $L$ is reached. Namely, at each iteration the following steps are performed:  

\begin{enumerate}
\item the value of the loss $L$ function is calculated on the training set given the current set of parameters $\Omega$;
\item the derivative of $L$ relative to all parameters $\Omega$ is computed. This gradient $\nabla L$ is the direction of the figurative step in which  $\Omega$ needs to be adjusted in order to minimize $L$;
\item the parameters $\Omega$ are adjusted as
\begin{equation} \label{eq:grad}
\Omega^{(i+1)}=\Omega^{(i)}-\eta \nabla L
\end{equation} 
where $\Omega^{(i)}$ represents the set of weights and biases at the i-th iteration and $\eta$ is the quantity of the correction we want to apply to the parameters at each epoch (a.k.a., learning rate).
\end{enumerate}

One of the challenges lies in choosing an appropriate learning rate, as steps that are too small slow down convergence with the risk of stopping at a local minimum, while steps that are too large can overshoot the minimum. Batching data is a common practice to avoid being trapped by local minima, allowing the algorithm to analyze subsets of data and converge more efficiently towards the absolute minimum. In fact, while the local minima will significantly change for every batch, the absolute minimum will roughly be in the same position. Another advantage is that batches allow to use less memory storage, as we do not need to memorize weights and biases for the complete dataset. Instead, in order to avoid overshooting, the algorithm can be integrated with a learning rate scheduler, that diminishes the size of the step after a certain number of iteration where the loss function has not gotten better. Other strategies can be adopted as well in order to facilitate convergence. For instance, standardizing the dataset ensures consistent gradients in all direction, thus also a fastest descent towards the global minimum.\newline

\subsubsection{Graph Neural Networks}
\label{GNNs}
\begin{figure}
    \centering
    \includegraphics[width=1\linewidth]{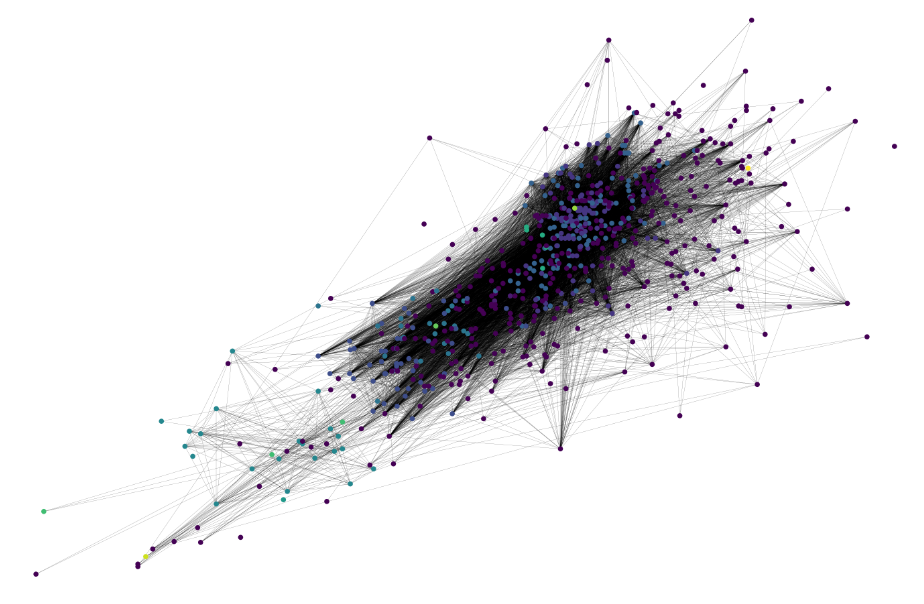}
    \caption{Example of a graph created in the present work. Stars are represented as dots in the abundance space, coloured based on the cluster they belong to. Links, or edges, are represented as grey lines.}
    \label{fig:Graph}
\end{figure}

In the realm of data science and machine learning, traditional datasets often represent information in a tabular format, where each row corresponds to an individual data point, and each column represents a feature or attribute of that data point. However, not all real-world data fits neatly into this structure. Many phenomena can be more accurately represented as graphs. \newline
Graphs consist of nodes and edges. Nodes typically represent entities or objects, while edges encode relationships or interactions between these entities. Links can be both directed and undirected. Also, nodes can have self-loops, meaning that they have edges pointing to themselves. Unlike traditional tabular datasets, where each data point is independent and identically distributed, graphs inherently capture relational information. In a graph, the relationships between entities are as crucial as the entities themselves, such as in the case of social networks or biological networks. Here we will pioneer the application of graphs to Galactic astronomy. An example of a graph produced in the present work is depicted in Fig.~\ref{fig:Graph}. The graphs used in our study are undirected and have self-loops. \newline
The simple statistical methods that only process the instances as independent entities may struggle to effectively leverage the inherent structure and dependencies present in social networks. Instead, recognizing and leveraging the underlying structure of graph data can lead to significant improvements in various machine learning tasks. Graph Neural Networks (GNNs) are a class of neural networks specifically designed to operate on graph-structured data. They extend traditional neural network architectures to handle the complexities of graphs, enabling the integration of both node and edge information into the learning process.\newline
Just like classical NNs, GNNs operate by propagating information between connected nodes in a graph. However, at each layer of the GNN, information from each node is aggregated and combined with that of all  nodes connected to it, including itself if it has a self-loop.  This process iterates over multiple layers, allowing nodes to gather and incorporate information from their connections as well as from the broader graph structure. There are  different strategies to aggregate the information of one node to all its connected nodes. In the present work, for our GNN, aggregation is realized through a simple weighted sum. In practice, a neuron aggregates information by summing up all their values associated to the node that is being processed and all its connections. However, the sum is not simply the addition of these values, but it is adjusted by the number of edges connected to the node. That is done to allow a balance between the case of a node with many edges and a node with few of no edges, so that nodes with more edges do not have a higher overall influence on the neuron's output.

To enhance the efficiency of a NN while optimizing the computer memory usage, it is a common practice to divide the dataset into mini-batches. The traditional approach that uses batches of constant size faces challenges when applied to GNNs as it presents high variability in the number of nodes and edges within graphs. The primary strategy for batching with graphs involves creating sub-graphs that retain essential properties of the larger graph. However, the application of this method introduces uncertainties \citep[see, e.g.][]{rozemberczki2020}. Finding an optimal sampling method for graphs is crucial, especially considering the anticipated growth of datasets in the future.

\subsubsection{Siamese Neural Networks}
\label{SNN}
\begin{figure}
    \centering
    \includegraphics[width=0.75\linewidth]{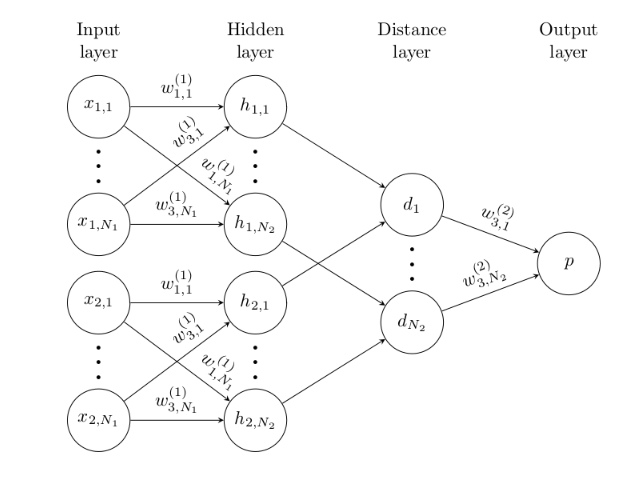}
    \caption{Operational  scheme of a siamese neural network. Figure adapted from \citet{koch2015}.}
    \label{fig:SNN}
\end{figure}
Siamese Neural Networks (SNN) are a class of neural network architectures that contain two or more identical sub-networks, where identical means that they have the same structure, the same biases and the same weights. Parameter updating through gradient descent is mirrored across both sub-networks and they are often used to compute distances or find similarities between the inputs.
SNN are usually used to find images similar to each other.
The structure of a typical SNN network is illustrated in Fig.~\ref{fig:SNN}.
The SNN takes pairs of input instances that need to be compared for similarity or dissimilarity.
Both inputs in a pair are passed through the same neural network, the siamese twins. 
Each input is encoded into a fixed-size feature vector by the shared neural network. This process extracts relevant features from the input data.
We can compare the results through the application of a metric, for instance the euclidean distance. The result is then passed to a neuron, that processes it and then passes it through a Sigmoid activation function. Thus, the output of the SNN is a value comprised between 0 and 1, which can be interpreted as a probability value that tells us how much the two input instances are identical. This value is then compared with the true value from the training set for the parameters' optimization, as in a normal ANN. Once the SNN is fully trained, for each pair of input instances it will return a probability that the two inputs are identical.\newline

\subsubsection{Auto-encoders}

\begin{figure}
    \centering
    \includegraphics[width=0.95\linewidth]{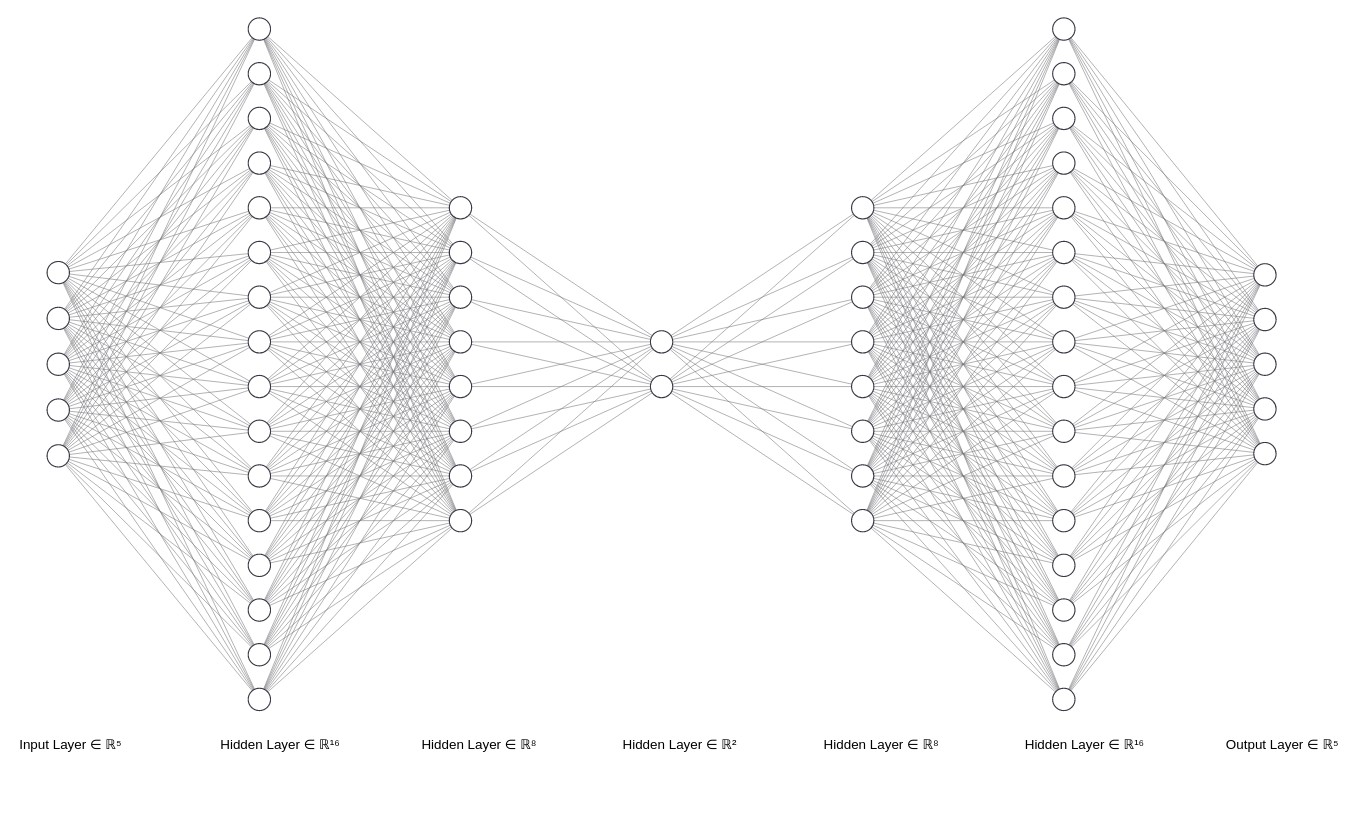}
    \caption{Schematic representation of an Auto-Encoder. The first layer and the last layer are respectively the input and output layers and the central layer is called latent space. Input and output layers are supposed to be as similar as possible.}
    \label{fig:auto-encoder}
\end{figure}

Auto-encoders are artificial neural networks particularly used in the domain of feature learning and data compression.
The fundamental idea behind an auto-encoder is to provide a compact, lower-dimension representation of the input data by encoding it into a latent space and then reconstructing it as realistically as possible. The network consists of an encoder and a decoder, working together to get the best possible representation in the latent space, i.e. the bottle-neck of the NN, as showed in Fig.~\ref{fig:auto-encoder}.
The encoder is composed by hidden layers that gradually reduce the number of neurons. The purpose of the encoder is to enclose the information in a clearer or more compact way. 
The inner layer, with the smallest number of neurons should then contain all the fundamental information about the input layer, reducing them to their essence. This is called the latent space.
This information has subsequently to be decoded through a mirrored set of hidden layers, so this time with an increasing number of neuron for each layer. The NN is trained such that the resulting target output is as similar as possible to the input layer. By doing this, the encoder is able to map the information into a  latent space that is informative enough for the decoder to recover the original dataset.\newline

\subsection{Clustering algorithms: {\sc DBSCAN} and {\sc OPTICS}}

Clustering analysis consists in grouping a set of data points into clusters based on the similarity of some of their properties.  All fields that deal with large amounts of data can benefit from the application of clustering techniques. These techniques manage to automatize the search for structures, substructures or patterns in data. They are also useful for anomaly detection \citep{syarif2012}, i.e. the process of finding outliers, customer segmentation \citep{kansal2018}, i.e. the process of clustering customers depending on their purchases and activity and so on.\newline
{\sc DBSCAN} (Density-Based Spatial Clustering of Applications with Noise) is a clustering algorithm that groups points based on their closeness to dense regions. Given the input set of data, it groups them in a way that reflects the underlying data density. It can then find arbitrary shaped clusters, handling different amounts of noise and not requiring prior information about the number of clusters to find.\newline

Based on two fundamental hyperparameters called minPoints and Epsilon ($\varepsilon$), {\sc DBSCAN} classifies the dataset points into:
\begin{itemize}
    \item Core points: a point p is a core point if at least minPoints other points are within distance $\varepsilon$ of it.
    \item Directly reachable points: a point q is directly reachable from point p if point q is within distance $\varepsilon$ from core point p.
    \item Reachable points: a point q is reachable from point p if there is a set of points that form a path from point p to point q and are directly reachable from point p. This means that all points that form a path, along with a point p, must be core points.
    \item Noise points (or outliers): if a point is not reachable from any other point, it is considered to be an outlier or noise point.
\end{itemize}

\begin{figure}[H]
\centering
\includegraphics[width=0.8\textwidth]{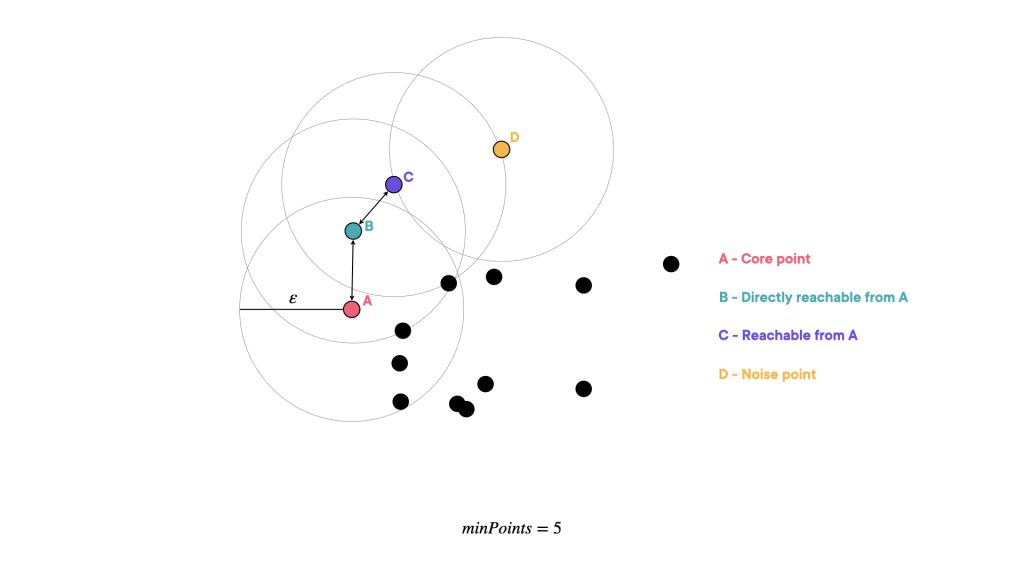}
\caption{\label{fig:DBSCANpoints} Visual representations of Core point, Directly reachable point, Reachable point and noise point. \\
Figure adapted from \href{https://www.atlantbh.com/clustering-algorithms-dbscan-vs-optics/}{\texttt{https://www.atlantbh.com/clustering-algorithms-dbscan-vs-optics/}}}
\end{figure}

Therefore, the two hyperparameters of {\sc DBSCAN} can be described as it follows:
\begin{itemize}
    \item Epsilon ($\varepsilon$): the maximum distance between two points for one to be considered in the neighborhood of the other.
    \item minPoints: the minimum number of points in a neighborhood for a point to be considered as a core point. The variation of minPoints results in a variation of the local density level. For instance,  if we consider too few points, we end up with seeing randomic density fluctuations. On the other hand,  if we consider too many points, we cannot see the clusters anymore.
\end{itemize}
A visual representation of this is shown in Fig.~\ref{fig:DBSCANpoints}.\newline
In practice, {\sc DBSCAN} starts with the first point of the dataset and counts how many instances are located within a distance $\varepsilon$ from it. This region is called the instance's $\varepsilon$-neighborhood. If there are at least minPoints points in its $\varepsilon$-neighborhood, including itself, it is classified as a core point. It then moves on to the nearest point and repeats the analysis.\newline
All instances in the  $\varepsilon$-neighborhood of a core point belong to the same group. Any point that is not a core point and does not have one in its  $\varepsilon$-neighborhood is considered as noise. 

\begin{figure}
    \centering
    \includegraphics[width=0.99\linewidth]{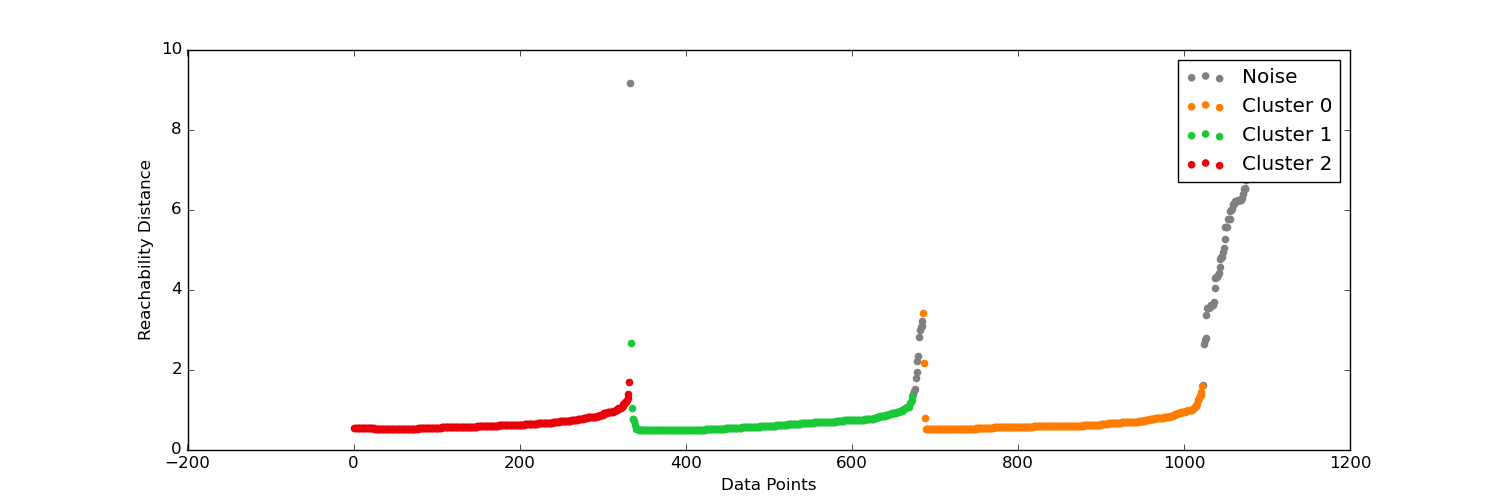}
    \caption{Example of a reachability plot for three synthetic clusters (in red, green, orange). The noise is  plotted in grey. Drops in the reachability plot are what {\sc OPTICS} interprets as cluster.}
    \label{fig:ExampleReachPlot}
\end{figure}

\begin{figure}[H]
    \centering
    \includegraphics[width=0.99\textwidth]{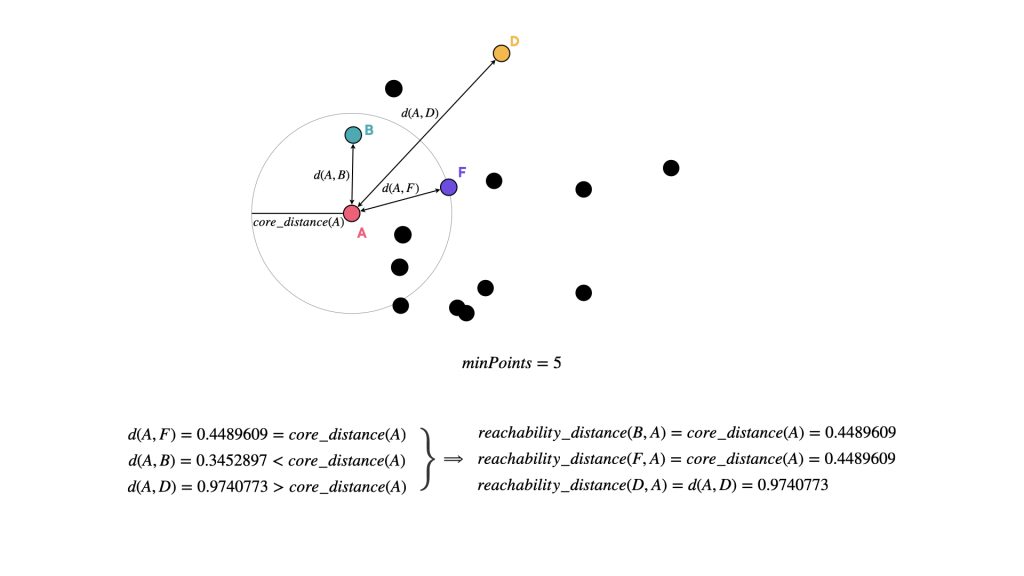}
    \caption{\label{fig:OPTICSpoints} Visual representations of core distance and reachability distance. Figure adapted from \href{https://www.atlantbh.com/clustering-algorithms-dbscan-vs-optics/}{\texttt{https://www.atlantbh.com/clustering-algorithms-dbscan-vs-optics/}}}
\end{figure}

{\sc DBSCAN} works well if clusters are dense enough and they are separated by low density regions, but it is unable to detect clusters of varying density. This is due to the fact that it uses a constant $\varepsilon$ value together with one density threshold minPoints to determine whether a point is in the neighborhood.
To solve this problem, in 1999 some of the {\sc DBSCAN} authors developed {\sc OPTICS} (Ordering Points To Identify Clustering Structure). The main difference is that {\sc OPTICS} generates a hierarchical clustering result for a variable neighborhood radius.\newline
{\sc OPTICS} is useful because instead of assigning cluster memberships, it stores the order in which the points are processed. It does not need the $\varepsilon$ parameter.\newline
In addition to the concepts mentioned above of the {\sc DBSCAN} algorithm, {\sc OPTICS} introduces two more terms shown in Fig.~\ref{fig:OPTICSpoints}, namely:
\begin{itemize}
    \item Core distance: the minimum distance required for a data point p to be considered as a core point. If the p is not a core point, then its core distance is undefined.
    \item Reachability distance: the reachability distance of point q with respect to another point p is the smallest distance such that q is directly reachable from p if p is a core point. This distance cannot be smaller than the core distance of point p, since for smaller distances there are no points that are directly reachable from point p. If p is not a core point, then the reachability distance of point q with respect to point p is undefined. The reachability distance is inversely proportional to the density of the data points
\end{itemize}
We should also notice that what was called minPoints is in {\sc OPTICS} the parameter min\_samples.\newline
The OPTICS algorithm does not immediately divide a dataset in clusters. Instead, it generates a plot from which we can extract information of the various high-density aggregates within the dataset. This plot is called ''reachability plot``, which consists in the sequence of points sorted by their procession order and plotted as a function of their reachability distances. An example is in Fig.~\ref{fig:ExampleReachPlot}. In order to build a reachability plot, {\sc OPTICS} begins with an empty seed-list and picks the first point p of the dataset, finds its $\varepsilon$-neighborhood, and determines its core distance. The reachability distance of this first point is set to undefined, and current point p is written to the output list. If point p is not a core point, then the algorithm simply picks another random point from the input dataset. If point p is a core point, then the reachability distance of each neighboring point q with respect to point p is calculated. All neighboring points are then inserted into the seed-list, and the list is sorted in ascending order by the reachability distance value.\newline
In the next iteration, {\sc OPTICS} takes the point that is at the top of the seed-list and, in case it is a core point, finds its $\varepsilon$-neighborhood, determines its core distance, and calculates the reachability distance for each of its neighboring points. If the newly calculated reachability distance of some unprocessed point is smaller than the one present in the seed-list, the value is updated to a smaller value, and the seed-list is sorted. If the current point is not a core point, then {\sc OPTICS} moves to the next point from the seed-list. The process continues until each point is processed and the seed-list is empty.\newline
The ordering output from the {\sc OPTICS} algorithm can be visualized using a reachability plot, a 2-D plot with points in the ordering returned by OPTICS on the x-axis and their reachability distances on the y-axis.
We can extract clusters from the reachability plot: generally, clusters appear as valleys so that deeper valleys represent dense clusters, while shallow valleys represent sparse clusters. The parameter Xi can extract clusters by detecting valleys by steepness between adjacent points, using it as threshold. This allows to extract clusters at different levels of granularity.\newline
Once the reachability plot is ready, the parameters that can be set in order to extract clusters are the minimum cluster size - i.e. we can decide to get the {\sc OPTICS} groups with more than a certain number of members - and xi - i.e. the steepness discriminating whether a drop in the reachability plot is a cluster.

\subsection{The choice of the metric}

In {\sc OPTICS}, as in many other clustering algorithms, the choice of a distance metric plays a significant role in the effectiveness of the algorithm and the quality of the clusters it produces.
Different metrics measure distances in different ways, leading to varying interpretations of proximity between data points. For example, Euclidean distance is often used as a default metric because it is one of the most familiar and intuitive metrics in geometry and linear algebra. However, the choice of metric heavily depends on the data context. For instance, some metrics are more sensitive to noise and outliers than others.  
The most widely used metrics are the Euclidean distance, the Manhattan Distance, the Chebyshev Distance, Cosine Similarity,  the Minkowski Distance, and the Mahalanobis Distance.
Here we describe, one of the possible choice, the Manhattan metric, which has been proven to work very well in the detection of disrupted clusters in the Galactic disc \citep[see, e.g.][]{Spina2022A&A...668A..16S}.

\subsubsection{The Manhattan metric}

\begin{figure}[htbp]
    \centering
    \includegraphics[width=0.5\linewidth]{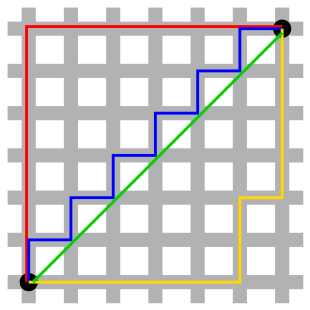}
    \caption{Red and blue lines are in Manhattan geometry and they are the same length, the green line represents the Euclidean geometry and it is shorter. Figure adapted from \href{https://en.wikipedia.org/wiki/Taxicab_geometry}{\texttt{https://en.wikipedia.org/wiki/Taxicab\_geometry}}.}
    \label{fig:Taxi}
\end{figure}

The Manhattan geometry, or taxicab geometry, is a geometry whose usual distance function is replaced by a new metric in which the distance between two points is the sum of the absolute differences of their Cartesian coordinates.
This is the metric that a taxi would use to compute distances in a city like Manhattan, were streets are all orthogonal. For instance, in Fig.~\ref{fig:Taxi}, the red and blue line that a taxi would follow are the distances computed in the Manhattan metric and they are the same length. The green line instead, is the Euclidean distance and it is shorter. It is also the distance used in chess for the rook.\newline
The taxicab distance between $p = (p_1, p_2,..., p_n)$ and $q = (q_1, q_2,..., q_n)$ in an n-dimensional real vector space, is the sum of the lengths of the projections of the line segment between the points onto the coordinate axes:
\[
d_T(p,q) = \abs{\abs{p-q}}_T = \sum_{i=1}^n{\abs{p_i-q_i}}
\]
and in $\mathbb{R}^2$, the taxicab distance between $p=(p_1, p_2)$ and $q = (q_1, q_2)$ is $d_T=\abs{p_1-q_1}+\abs{p_2-q_2}$.

\begin{figure}[H]
\centering
\includegraphics[width=0.99\textwidth]{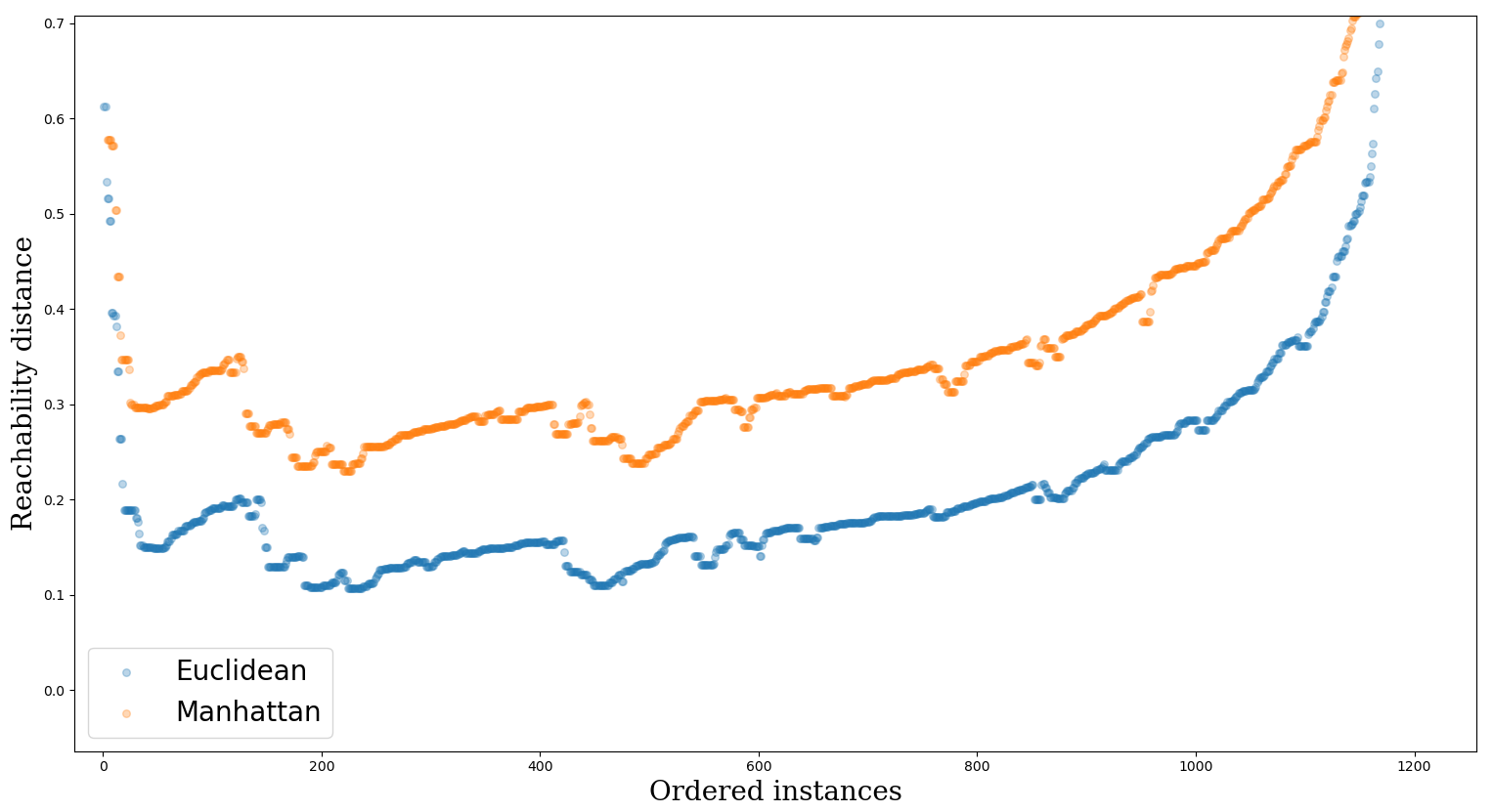}
\caption{\label{fig:ManhvsEucl} Example of the difference of the {\sc OPTICS} algorithm applied with Euclid and Manhattan metrics: in orange the results obtained with the Manhattan distance and in blue with the Euclidean one. The orange dots present a 0.1 vertical offset for better readability. }
\end{figure}

In our case, this geometry is particularly useful because it is more sensitive to clusters in noisy datasets, such as that of chemical abundances generated by automatic pipelines of large spectroscopic surveys.\newline
It is possible to demonstrate that, independently of the dimension of the space we are considering (in our case the abundances space), the following relation is satisfied:
\[
\frac{\Delta d_M}{d_M}>\frac{\Delta d_E}{d_E}
\]
Where $d_E$ is the Euclidean distance. This means that for a small distance variation, the distance is 'amplified'. A cluster then, will stand up more in a Manhattan geometry than in a Euclidean geometry.
In our clustering algorithm, {\sc OPTICS}, we have the possibility of changing the metric used to compute distances and the difference when applying the Euclid or Manhattan geometry to clusters data is quite evident, as we can see from the reachability plot in Fig.~\ref{fig:ManhvsEucl}, in which it is possible to appreciate that the depressions that allow clusters to be identified are deeper using the Manhattan metrics. 

%% file: 3-Dataset.tex
\section{The observational datasets}
\label{chap3}
\subsection{The era of large spectroscopic surveys}

The recent observational advancements are propelling astronomy into big data era. Large stellar spectroscopic surveys provide detailed information about the structure, composition, dynamics, and evolution of the MW. They, indeed, supply essential data for constraining theoretical models of Galaxy formation and evolution and to improve our knowledge on stellar physics and evolution. They allow us to perform an efficient cataloguing and characterisation of stars in the different components of the Galaxy. They also have the great advantage of being public, so that many people can simultaneously act on the same catalogue, designing a large number of  different experiments and testing theories.\newline
Here, we describe the two public spectroscopic surveys we used for this work. We outline the selection of our samples of stars and the computation of their orbits by combining spectroscopic data with {\em Gaia} astrometric data.  

\subsection{The APOGEE survey}

APOGEE (Apache Point Observatory Galactic Evolution Experiment) is a large-scale, medium-resolution stellar spectroscopic survey \citep{majewski2017apache}. It employed high signal-to-noise (SNR > 100) infrared spectroscopy in order to avoid dust effects in the bulge and disc of the Galaxy. It observed over 600\,000 red giant stars.
It consists of two 300-fiber cryogenic spectrographs that operate onboard the 2.5 meter Sloan Foundation Telescope and the 1 meter NMSU Telescope at Apache Point Observatory (APO) in New Mexico, United States, and the 2.5 meter Irénée du Pont Telescope of Las Campanas Observatory (LCO) in Atacama de Chile. APOGEE is unique among large spectroscopic surveys because it is able to acquire medium resolution spectra ($R\simeq 22\;500$) across the entire Milky Way.\newline
APOGEE data products are spectra, radial velocities, stellar parameters - including effective temperature, surface gravity, metallicity - and chemical abundances. These abundances are produced by the APOGEE Stellar Parameters and Chemical Abundance Pipeline (ASPCAP)\citep{perez2016aspcap}. Chemical abundances precision varies based on stellar parameters.\newline
ASPCAP works in two steps. First, it determines atmospheric parameters through a global fitting over the whole spectral range. Once atmospheric parameters are derived, it determines elemental abundances with a local fit.
The data release used in this work, {\sc DR 17}, includes data for over 400\,000 stars, and it is comprehensive of chemical abundances for up to 20 ions: C, CI, N, O, Na, Mg, Al, Si, S, K, Ca, Ti, Ti II, V, Cr, Mn, Fe, Co, Ni and Ce. We note that C is measured in two different ways: from molecular bands (C), and from atomic lines  of neutral (C~I).  Ti is measured both from neutral titanium lines, and Ti II is measured from singly ionized titanium lines.
We note that abundances of cool stars, especially below T$_{\rm eff}<$3500 K are particularly challenging, perhaps due to significant presence of  molecular absorption and challenges in the determination of the stellar continuum. 
Warmer stars face some challenges too: at warmer temperatures, lines from many elements become weak or disappear, and in addition, the hottest stars usually rotate, making the use of high resolution ineffective with a consequent blending the lines. So it is difficult to determine abundances over 7000 K.

Abundances are classified according to their quality and reliability: 
\begin{itemize}
    \item Most Reliable: Species that are precisely measured, and available over a wide range of stellar parameters. They follow very well trends expected from the literature; 
    \item Reliable: species that are less precisely measured, available over a narrower range of stellar parameters, and follow trends consistent with literature with some scatter; 
    \item Less Reliable: less precisely measured, available over a narrow range of stellar parameters, and have apparent chemical trends consistent with literature with large scatter around the expected trend; 
    \item  Deviant: measured, but the results deviate from literature expectations.
\end{itemize}
According to these criteria, the most reliable abundances are C (from molecular bands), N, O, Mg, Al, Si, Mn, Fe, Ni;  reliable abundances are C I, Na, K, Ca, Co, Ce;  less reliable are S, V, Cr, while deviant abundances are Ti, Ti II.\newline
In this work, we restrict our sample following the recommendations in  \citet{Horta2023}. Our adopted  criteria are shown in Table \ref{tab:Criteria}. While our selection closely aligns with \citet{Horta2023}, we adopted  a  wider range of values for surface gravity (0<log~g<5)   with respect to their selection.  

\begin{table}[H]
    \centering
    \small
\begin{tabular}{|l|l|}
 \hline
SNR &   > 70\\
\hline
TEFF &   3500-6500 K\\
\hline
STARFLAG & = 0\\\hline
ASPCAPFLAG&< 256\\\hline
\end{tabular}
\caption{Criteria applied in the present work for the selection of stars in the APOGEE catalog.}
    \label{tab:Criteria}
\end{table}
We verified that our selection did not introduce trends or offsets in the abundances as a function of stellar parameters. To do that, we plotted the abundances of member stars in clusters versus temperature and gravity.
In principle, since clusters are homogeneous in many elements (not all, in the case of globular clusters, as shown in Fig.~\ref{fig:Na-Oanti} in which there are some well-known anti-correlations), we expect no trends between [El/Fe] and stellar parameters. 
We then checked that there were no correlations between the abundances and parameters in the selected T$_{\rm eff}$ and log~g intervals. To do this, we performed linear fits between stellar parameters and [El/Fe], choosing the elements for which the slope was null, within the errors.
In general, we find only small correlations between the abundances of elements and either temperature or log g with the exception of $\Omega$~Cen, that is a peculiar cluster as it has a complex star formation history.
Contrarily to most abundance ratios, like [Mg/Fe], we notice a quite strong correlation between [C/Fe] and log~g. This effect is expected, because in the sample we have mainly red giant stars, in different stages from red giant branch (RGB) to red clump (RC). In the red giant phase, the carbon undergoes various mixing episodes, called dredge-ups, which can change its photospheric abundance.

\subsubsection{Kinematic selection of our APOGEE sample}

After applying the quality flag in the APOGEE sample, we computed the orbits of the selected sample. 
To calculate the orbits, we need precise positions, all the velocity components (not just the radial one, which is measured from the APOGEE spectra) and distances. To obtain the proper motion, parallax and  distance, we cross-matched our sample with {\it Gaia} {\sc DR3}. 
The orbits were computed assuming the McMillan potential \citep{mcmillan2016mass} already described in Chapter~\ref{chap1_orbits}, by means of the code AGAMA \citep{vasiliev2019agama} using the prescriptions described in \citet{massari2019origin}. AGAMA requires as inputs the right ascension (RA), declination (Dec), distance, radial velocity (with its error), proper motion components in right ascension and declination (with errors) and the correlation coefficient for errors in the two proper motion components. The input data are provided by the cross match of our APOGEE sample with {\it Gaia} {\sc DR3}, where distances are those computed in \citet{bailer2021estimating}. AGAMA provides as outputs the maximum and minimum Galactocentric radii, the total energy, the angular momentum components, the actions, the circularity and the velocities of stars in the Galactic coordinate system (see Chapters~\ref{chap1_reference} and  \ref{chap1_motions}). 

The code also creates multiple Monte Carlo realizations of the positions and velocities of each star by sampling from their measured uncertainties to provide uncertainties for the orbital quantities.\newline

\subsubsection{Our APOGEE sample of halo stars}

\begin{figure}
\centering
\includegraphics[width=1.0\textwidth]{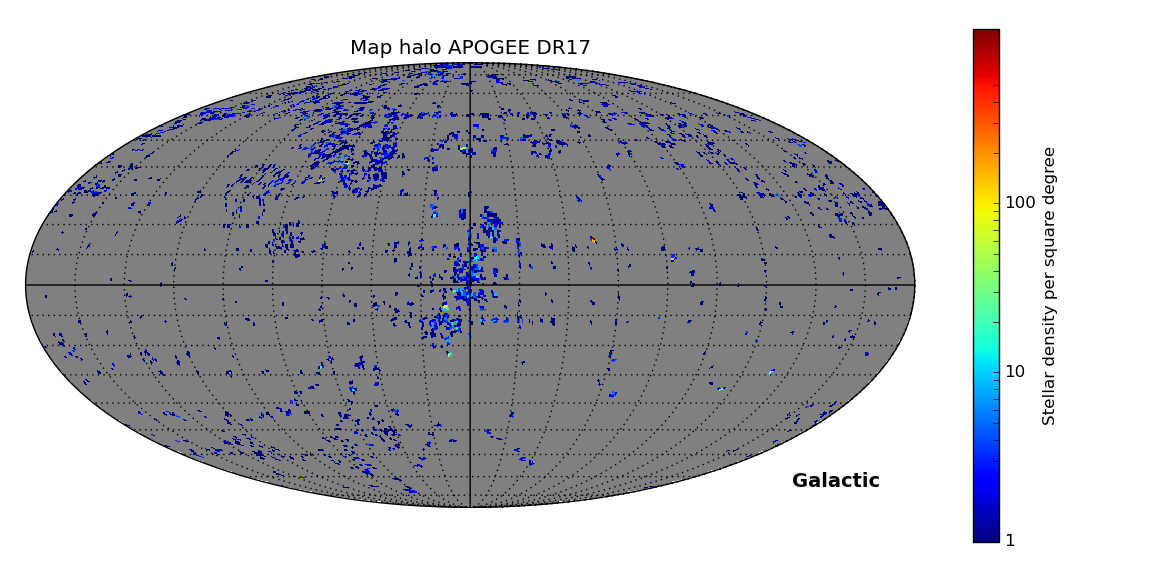}
\caption{\label{fig:MollweideApogee} Mollweide projection of the selected halo stars colour-coded by their number density.}
\end{figure}

\begin{figure*}
\centering
\includegraphics[width=1.\textwidth]{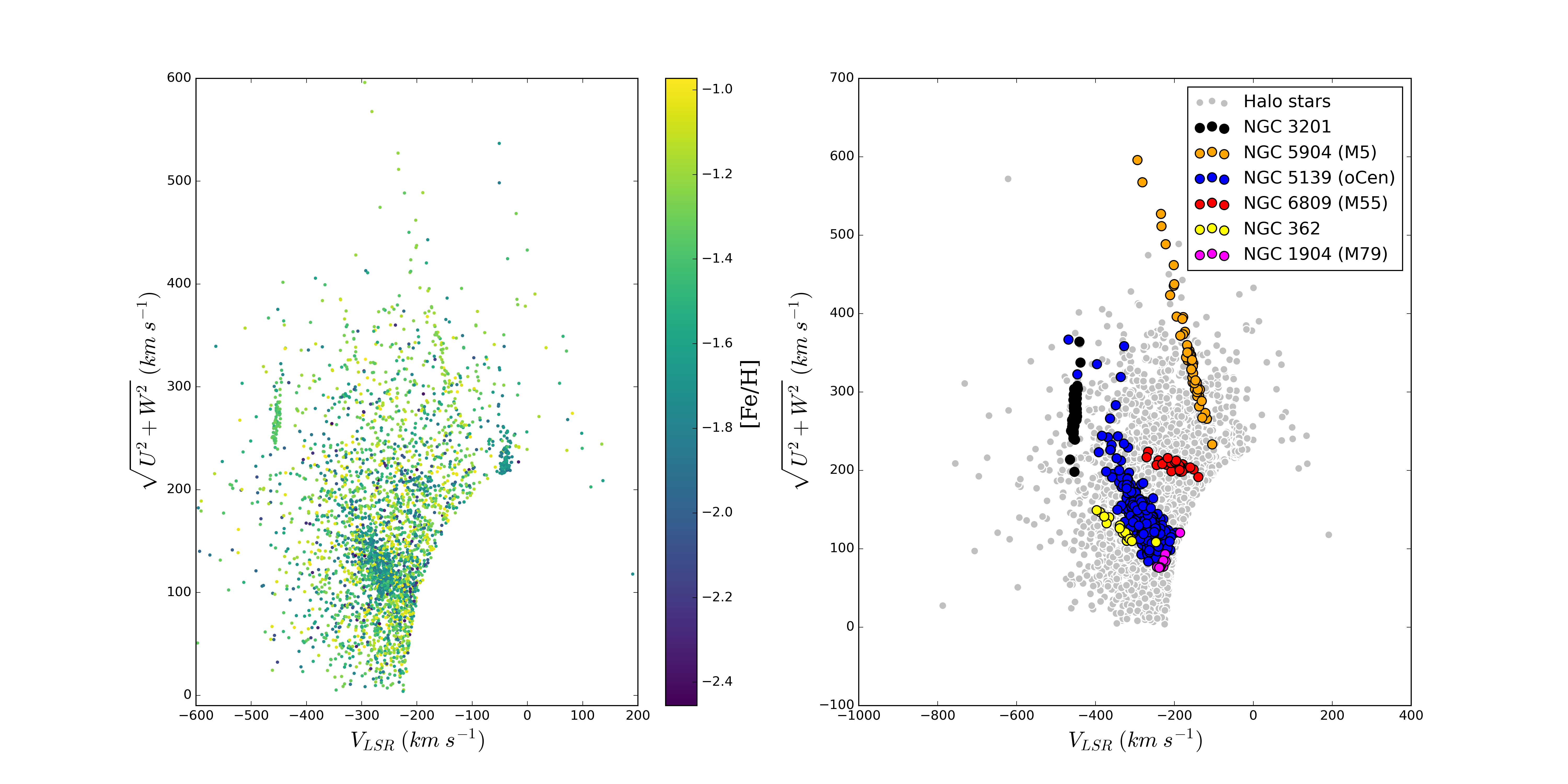}
\caption{\label{fig:ApogeeToomre} Left Panel: Toomre diagram for the selected halo stars in APOGEE. Stars are colour-coded by  [Fe/H]. Right panel: Toomre diagram in which the selected globular clusters are highlighted. The clusters are colour-coded as explained in the legend. Field stars are in grey.}
\end{figure*}

Our aim is to select a subsample of halo stars, which contains both stars in globular clusters and in the field. We used both chemical and kinematic information to select them. The first step in selecting halo stars is based on their total velocity. 
We select, indeed, stars with total velocities over 220 km~s$^{-1}$. This kinematic selection allows us to focus only on stars that are compatible with the typical velocities of the halo of the Milky Way. The second step is to consider the typical metallicity of the halo:  we make a cut in metallicity, selecting stars with [Fe/H]$<-1$. We restrict the sample, though, by excluding stars belonging to the Magellanic Clouds with a cut in declination and right ascension.
The metallicity cut reduces substantially the size of the sample of stars, getting to a total of 3548 stars. The spatial distribution of our sample stars is represented in the Mollweide view\footnote{The Mollweide projection  represent the distribution of stars across the sky minimizing the distortion and maintaining the integrity of their spatial distribution} in Fig.~\ref{fig:MollweideApogee}, which is colour-coded according to the stellar density and shows a slightly higher concentration of stars towards the bulge.
The selected stars are shown in the Toomre diagram of Fig.~\ref{fig:ApogeeToomre}.  In this plot,  we can recognise the kinematic cut executed to select halo stars. Stars are colour-coded according to their [Fe/H].
The selection criteria we have chosen are rather tight, because we preferred to avoid the thick disc contamination that would occur, for example, by including total velocities between 150 and 220 km~s$^{-1}$, or by extending metallicity -1$<$[Fe/H]$<$-0.5. In this way, we certainly lose some halo stars, but avoid including a large number of disc stars. 

Of the selected 3548 halo stars, 916 stars belong to 28 different globular clusters, according to the memberships from \citet{vasiliev2021gaia}. Globular clusters with less than ten members are not considered  in our analysis. In practice, their members are considered as field stars, and we do not attempt to recover them in our analysis. Instead, globular clusters with more than ten stars are shown in Table \ref{tab:tabClusters}. 
Their properties in the kinematic space are highlighted in the right panel of  in Fig.~\ref{fig:ApogeeToomre}. In the figure, some GCs are shown with coloured symbols, while field halo stars are in grey. From this plot, we can see that stars in the same cluster generally have coherent velocities and they can be easily identified in the Toomre diagram. In addition, GCs correspond to clumps at low metallicity (see left panel of Fig.~\ref{fig:ApogeeToomre}), highlighting the fact that GCs are both chemically homogeneous -- at least in [Fe/H]-- and kinematically coherent.\newline

\begin{table}[H]
    \centering
    \small
\begin{tabular}{| l|l |}
 \hline
cluster name & Number of members\\
\hline
NGC5139 $\Omega$Cen &    558\\
\hline
NGC6656 M22&         77\\
\hline
NGC3201&     61\\
\hline
NGC5904 M5&           49\\
\hline
NGC6809 M55&          35\\ \hline 
 NGC 362&20\\ \hline 
 NGC 6544&15\\ \hline
NGC6205 M13&     14\\
\hline
NGC1904 M79&     10\\
\hline
NGC6121 M4&           10\\
\hline
\end{tabular}
\caption{Globular clusters with more than 10 members and the respective number of member stars in our APOGEE sample.}
    \label{tab:tabClusters}
\end{table}

\subsubsection{Properties of the halo field stars} 

\begin{figure}
\centering
\includegraphics[width=0.75\textwidth]{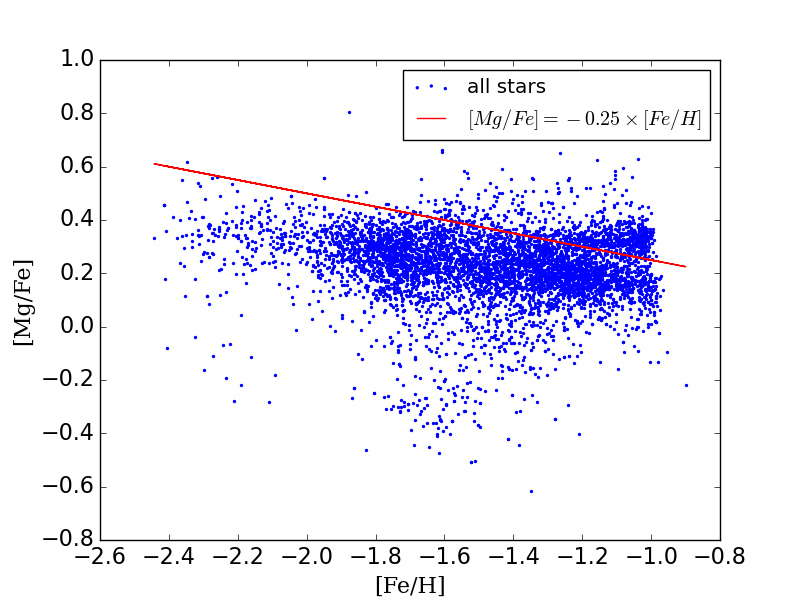}
\caption{\label{fig:MgRetta} [Mg/Fe] vs [Fe/H] for our APOGEE sample halo stars. The red line marks the separation  between LMg and HMg stars.}
\end{figure}

We can analyse the chemical properties of our halo sample using their distribution in the [El/Fe]  abundance ratios vs. metallicity [Fe/H] plane.  
In some of these planes, in particular those involving $\alpha$ elements,  halo stars are separated in two branches. For instance, at a given metallicity, there are two sequences corresponding to stars with a higher [Mg/Fe] abundance (high-$\alpha$ sequence) than others (low-$\alpha$ sequence). 
The separation in [Mg/Fe] has been used by \citet{Hayes2018} to split stars with  different origins: in situ stars and accreted stars. Stars in the Low-Mg sequence (LMg) generally show a halo-like kinematics with little rotation and large velocity dispersion. They are likely accreted stars from other galaxies. On the other hand, stars in the high-Mg sequence (HMg) that show a small velocity dispersion are likely born in situ, i.e. originated within the Milky Way. There can be some contamination between the two though, as the HMg can contain stars with halo-like orbits. This overlap can be due both to a chemical overlap between the two populations.
In Fig.~\ref{fig:MgRetta}, we plot our sample halo stars in the  [Mg/Fe] vs. [Fe/H] plane. The red line is the separation between the LMg sequence and the HMg sequence as set in \citet{Hayes2018}. 
There is an offset of 0.05 dex between the dividing line used by \citet{Hayes2018} to separate the two sequences and the line that would be obtained using our abundances.
It is likely due to an offset between the different APOGEE data releases used by us ({\sc DR~17}) and by \citet{Hayes2018} ({\sc DR~13}) and to the different criteria used to select the samples.

\begin{figure}
    \centering
    \includegraphics[width=0.85\linewidth]{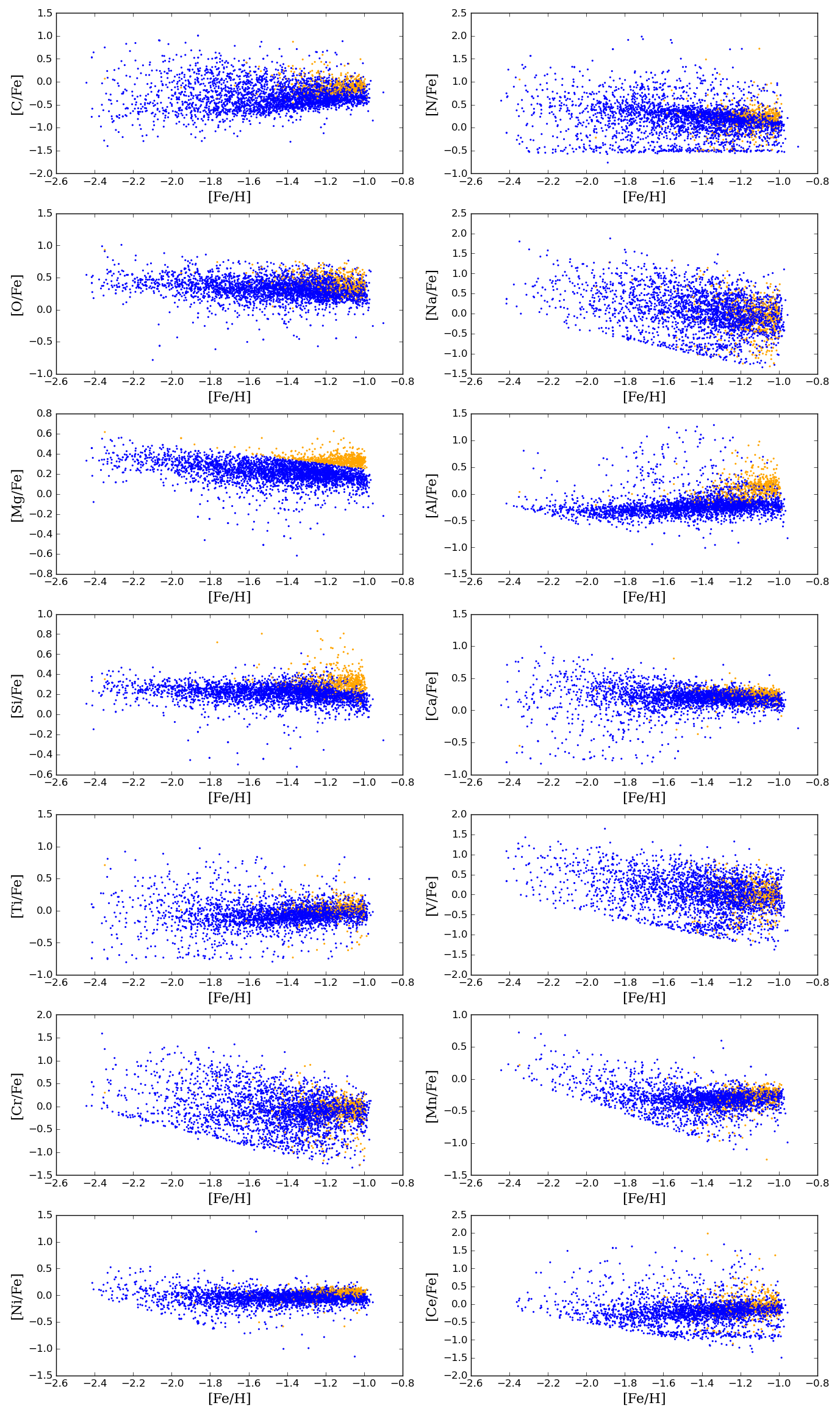}
   \caption{[El/Fe] vs [Fe/H] for our sample of  APOGEE halo stars. Stars belonging to the HMg are plotted in orange, while stars belonging to the LMg are in blue.}\label{fig:plotHMgLMgElements}
\end{figure}

The separation between HMg and LMg is visible also in the other abundances planes, as shown in Fig.~\ref{fig:plotHMgLMgElements}. In this plot, the separation is done using only [Mg/Fe] abundances, and stars are coloured according to their belonging to the HMg (orange) or LMg (blue). The stars in the two sequences are then displayed in other abundances planes. In particular, the planes that contain the $\alpha$-elements (O, Mg, Si, Ca and Ti),  the separation is evident. In the planes containing the iron peak elements, the separation is less clear because they are less prone to variations due to different star formation histories and they follow closely the evolution of iron, as, for instance,  Cr and V.\newline

\subsubsection{Properties of the Globular Cluster population
}
\begin{figure}
     \centering
     \includegraphics[width=0.75\linewidth]{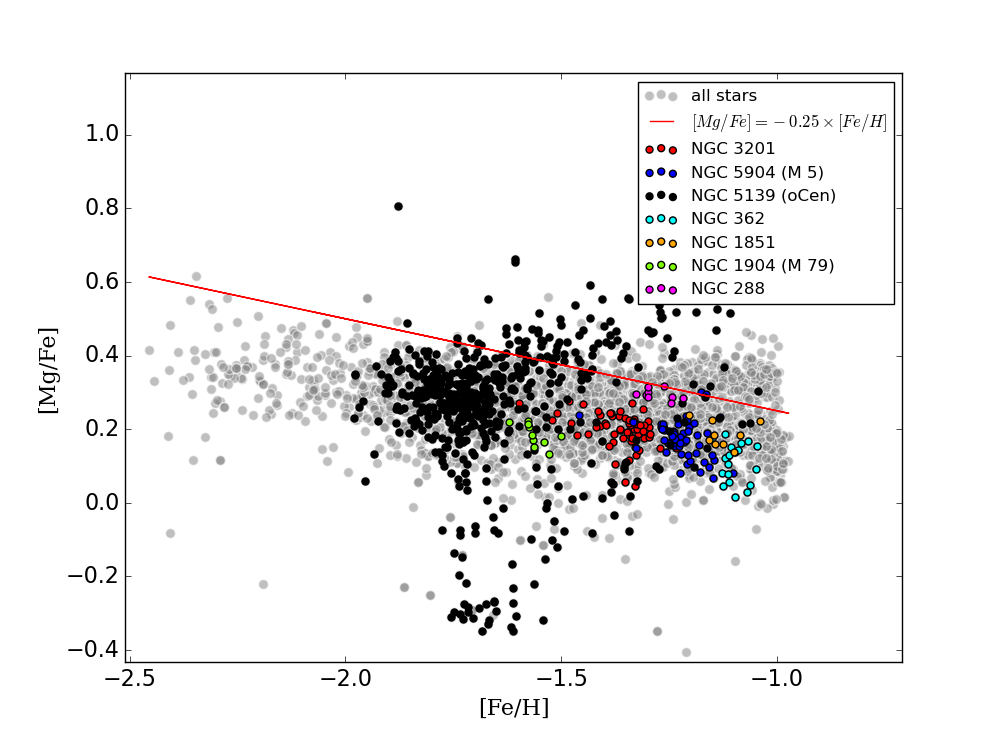}
     \caption{[Mg/Fe] vs [Fe/H] for the member stars of the most populated GCs of our  APOGEE sample (colour-coded as in the legend) and for the field stars.}
     \label{fig:clusterMgFe}
 \end{figure}

We can see in Fig.~\ref{fig:clusterMgFe} that most GCs in our sample are located  in the LMg. The external origin of metal-poor branch clusters is indeed reinforced (or at least not excluded) by their lower  [$\alpha$/M] abundances  \citep[see. e.g.][]{Recio2018A&A...620A.194R, Belokurov2024MNRAS.528.3198B}.
As we have already mentioned, $\Omega$ Cen is an anomalous cluster, more similar to a galaxy than a globular cluster. This also appears in the wide range in metallicity and abundance ratios that its stars cover \citep[see., e.g.][]{Villanova2014ApJ...791..107V}. 

\begin{figure}
\centering
\includegraphics[width=0.75\textwidth]{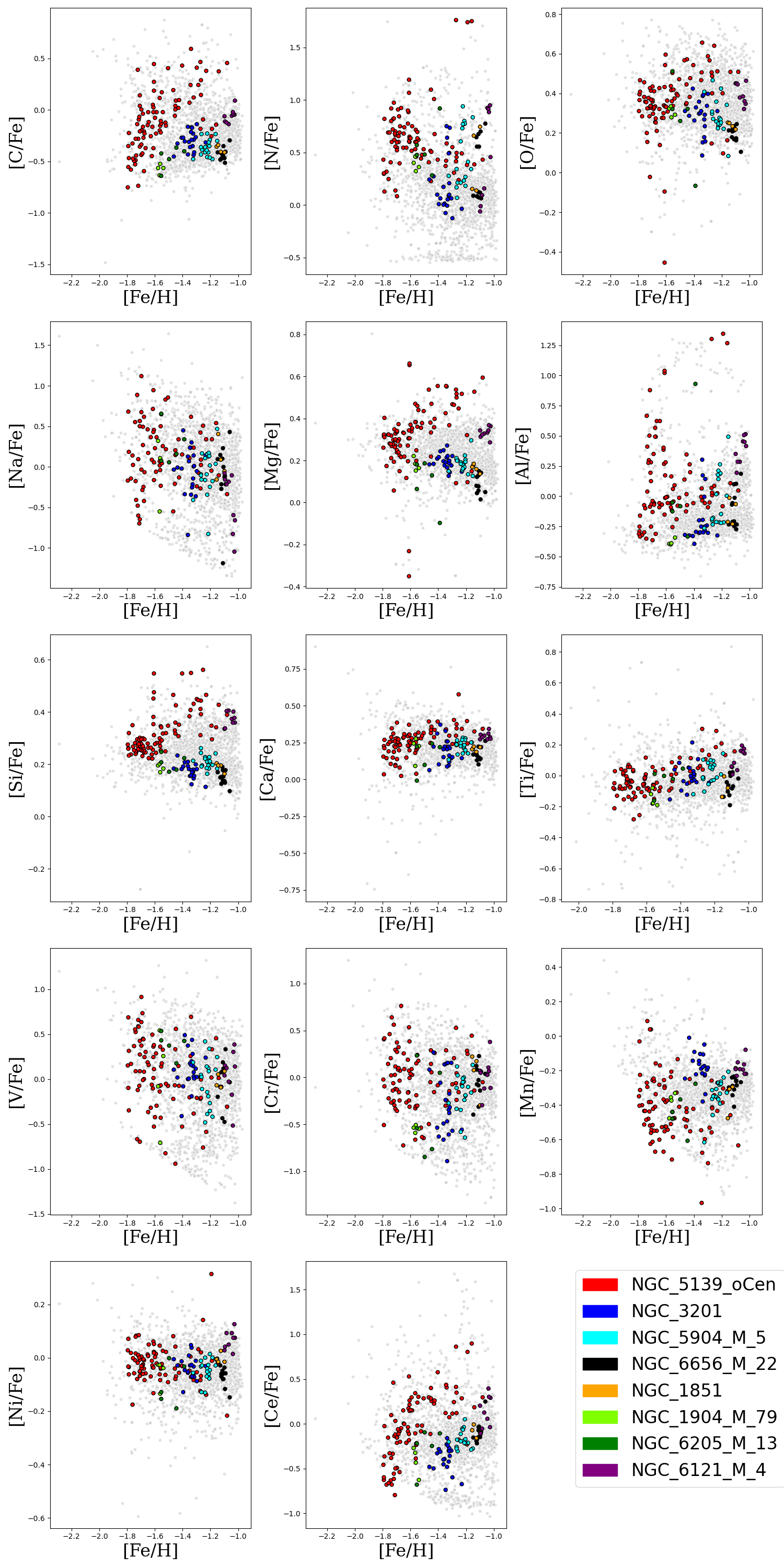}
\caption{\label{fig:elementscluster} Abundance ratios vs [Fe/H] of member stars of some of the most populated clusters in APOGEE used in  the present work, colour-coded as in the legend.}
\end{figure}

As  seen in the Toomre diagram, clusters can be identified in the velocity space: in Fig.~\ref{fig:ApogeeToomre} we see that they are well grouped in this kinematic space.  In the left panel of Fig.~\ref{fig:ApogeeToomre}, we can see how kinematic structures also have a metallicity distinguishable from that of field stars. 
We can also question the possibility of distinguishing clusters by using individual abundances instead of metallicity, as shown in in Fig.~\ref{fig:elementscluster} for some representative clusters of our sample. 
The behaviour is not the same for all elements. In fact, we should remember that for some elements, in particular O, Na, Mg, stars in GCs  show a large dispersion in abundance. These abundance ratios show anticorrelations, due to the presence of multiple populations in globular clusters \citep{Gratton2012A&ARv..20...50G}. 
We find that for elements not affected by observational spreads due to the quality of the data and anticorrelations due to the presence of multiple populations (e.g., Ca, Fe, Ti, Ni) member stars are quite homogeneous and well-separated from the  other clusters.  
By using kinematic and some specific chemical properties, it might be possible, then, to distinguish clusters among them and from field stars. 

\subsubsection{Properties of the Galactic halo streams}
\begin{figure}
    \centering
    \includegraphics[width=0.85\linewidth]{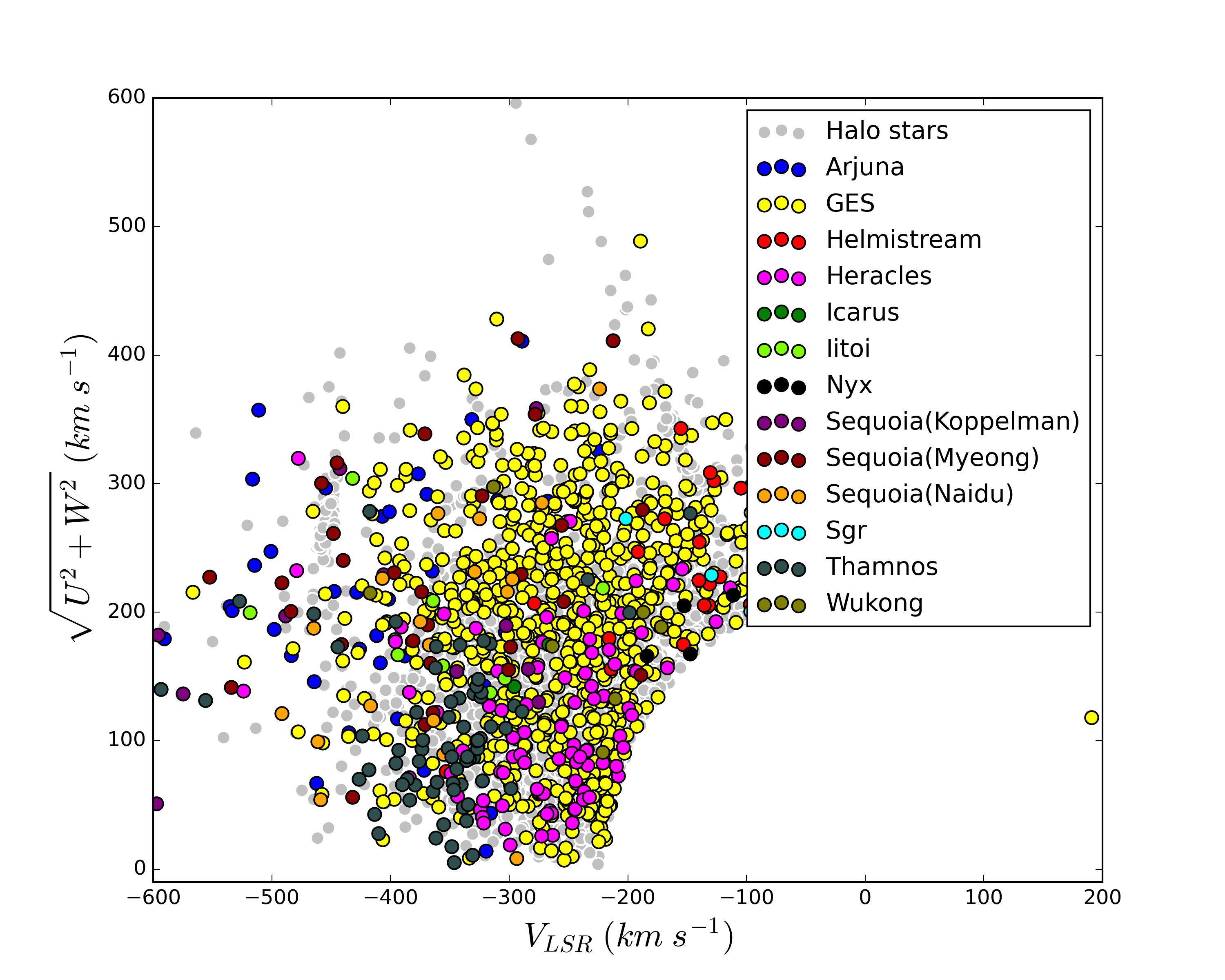}
    \caption{Toomre diagram with highlighted streams from \citet{Horta2023}. Stars in gray represent the field halo stars.}
    \label{fig:StreamToomre}
\end{figure}

\citet{Horta2023} discussed the properties of some Galactic halo streams in the APOGEE catalog. Using the members of streams  selected by \citet{Horta2023} and reference therein, we end up with a list of members present in our sample of halo stars. In particular, we have stars belonging to: Gaia Enceladus (GES), Thamnos, the Helmi Stream, Heracles, Arjuna, Wukong, I'itoi, Nyx, Sagittarius, Icarus and three different selections of Sequoia.
We plotted in Fig.~\ref{fig:StreamToomre} the velocities of the members of the streams in the Toomre diagram. The result is quite different with respect to that of GCs. 
Globular clusters are clearly identifiable and grouped.  Streams, instead, are sparse and spread over the plane, even when excluding stars from the largest merger, Gaia Eneceladus/Sausage (GES). GES is one of the most important known streams in our Galaxy. It represents the dynamical record in the velocity space of a major collision that the MW experienced more than 10 Gyr ago with a quite massive dwarf galaxy \citep{Belokurov2018MNRAS.478..611B} . Its high variability in velocities, but also - as we will see later - in chemistry is due to its past as a galaxy, with an extended star formation history, with an elevated star formation rate between $1$ and $3.5 M_{\odot}/yr$ \citep{Vincenzo2019}. 
We then expect GES to be more sparse and spread compared to other streams, but also the other streams found in APOGEE are difficult to be identified in the Toomre diagram.
This is because with the time passing, the stars belonging to mergers have lost kinematic coherence. However, they may have retained some quantities that are orbit invariant, such as the total Energy or the vertical component of the angular momentum $L_Z$ or their chemical properties.  

\begin{figure}[ht]
    \centering
    \includegraphics[width=1.00\linewidth]{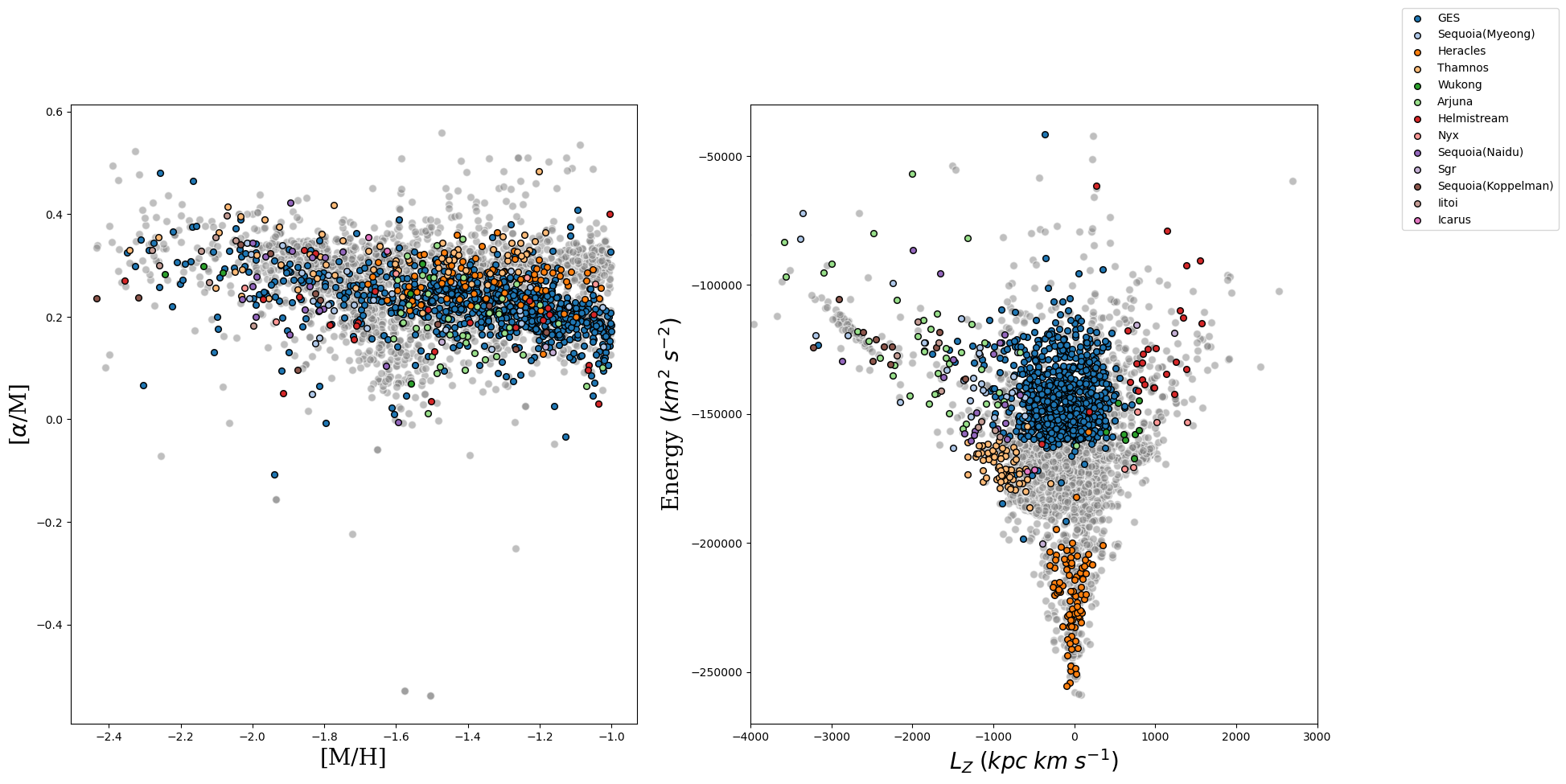}
    \caption{Left panel: [$\alpha$/M]  vs. [M/H] for the selected streams in the APOGEE sample. Right panel: streams are represented in the Energy-Lz plot. Prograde streams are positioned on positive values of Lz and retrograde streams are on negative values.}
    \label{fig:StreamsAlphaEnergy}
\end{figure}

On the left panel in Fig.~\ref{fig:StreamsAlphaEnergy}, we show the abundance ratio [$\alpha$/M] vs metallicity [M/H]\footnote{[M/H] is the total metallicity, and in FGK stars is often approximated with [Fe/H]. In this plot we use [M/H] instead of [Fe/H] because it is directly related to the global [$\alpha$/M].}.
Streams generally show a low [$\alpha$/M] abundance since they are usually an accreted population. For instance  the Sagittarius stream,  the Helmi stream,  Sequoia and in Gaia Enceladus are low in [$\alpha$/M], but we find a few notable exceptions in Thamnos and Heracles that seem to be $\alpha$-enhanced or located at the limit between the two sequences.\newline
In Fig.~\ref{fig:StreamsAlphaEnergy}, right panel,  we plot the members of the streams in the Energy vs $L_Z$ plane.  
Due to the coordinate system considered to compute orbits in this work, prograde stars are situated in the positive angular momentum side. Consequently, disc stars are positioned on the right branch.
Most of the streams, such as Sequoia, Thamnos and Arjuna are located in the retrograde zone and around $L_Z = 0$, where the broad Gaia Enceladus remnant is particularly noticeable. Still, there are some exceptions. Nyx and the Helmi stream are on the prograde side of the plot.

\begin{figure}[ht]
    \centering
    \includegraphics[width=1.00\linewidth]{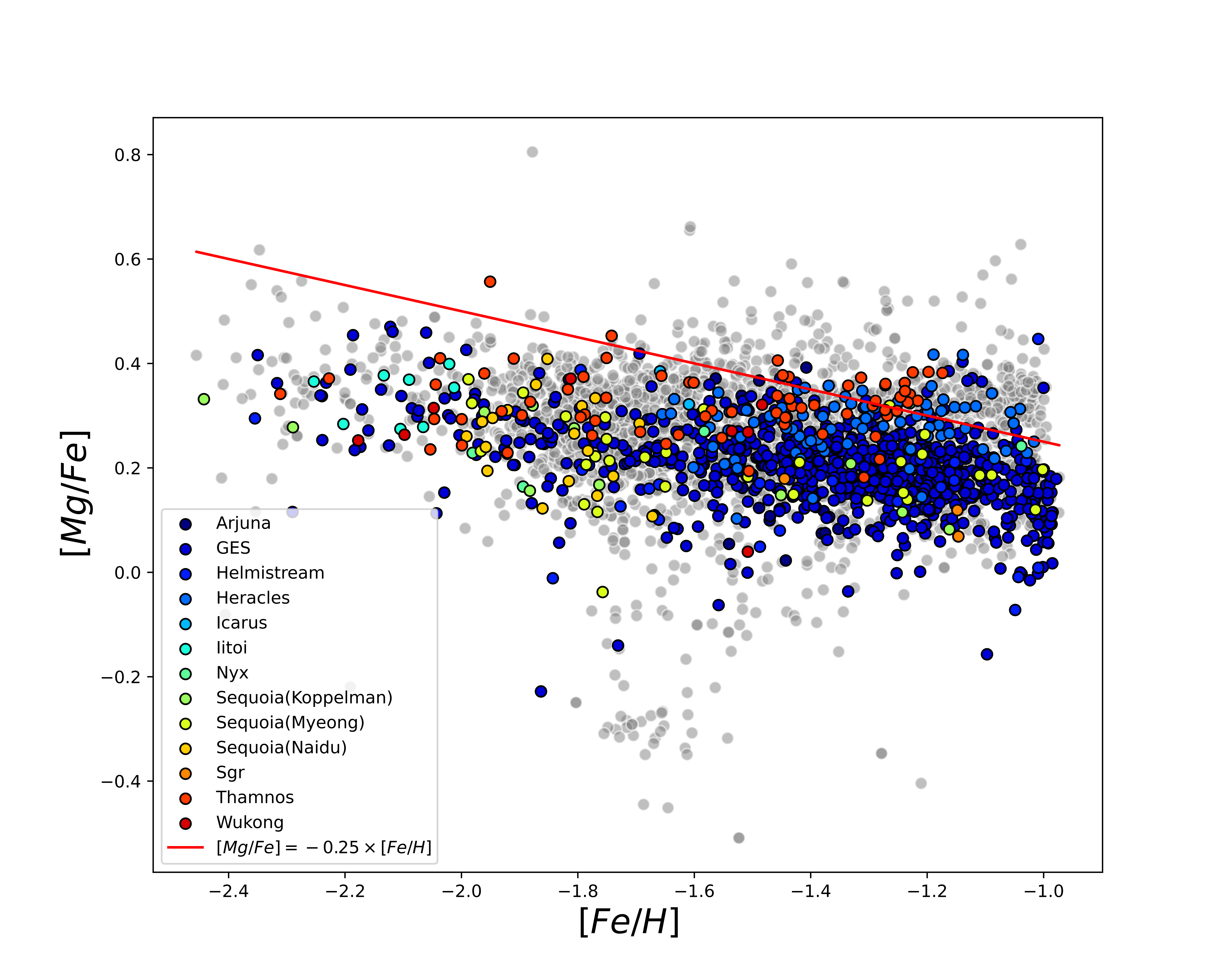}
    \caption{[Mg/Fe] vs [Fe/H] in our APOGEE sample.  The line that separates the LMg and HMg sequence is in red and the streams are colour-coded as indicated in the legend over the grey of field halo stars.}
    \label{fig:StreamMg}
\end{figure}

Another interesting plane in which we can look for differences between streams and field population is the [Mg/Fe] vs [Fe/H] one. As we already said, [Mg/Fe] allows us to separate between accreted and in situ objects. Since streams are thought to be formed from the disruption of Globular Clusters and dwarf galaxies accreted by the Milky Way, we expect them to be found in the LMg portion of the plot.\newline
The plot is presented in Fig.~\ref{fig:StreamMg}: most streams belong  predominantly to the LMg sequence. The Nyx stream has a high [Mg/Fe], but with low [Fe/H] and still below the dividing line between in sito and accreted populations. 
Hercules  is instead a bit above the separation line, and it might be  part of  the category of stellar substructures created by past interactions in the disc or secular processes, as discussed in  \citet{zucker2021}.

\subsection{The {\em Gaia}-ESO survey}

\begin{figure}
    \centering
    \includegraphics[width=0.75\linewidth]{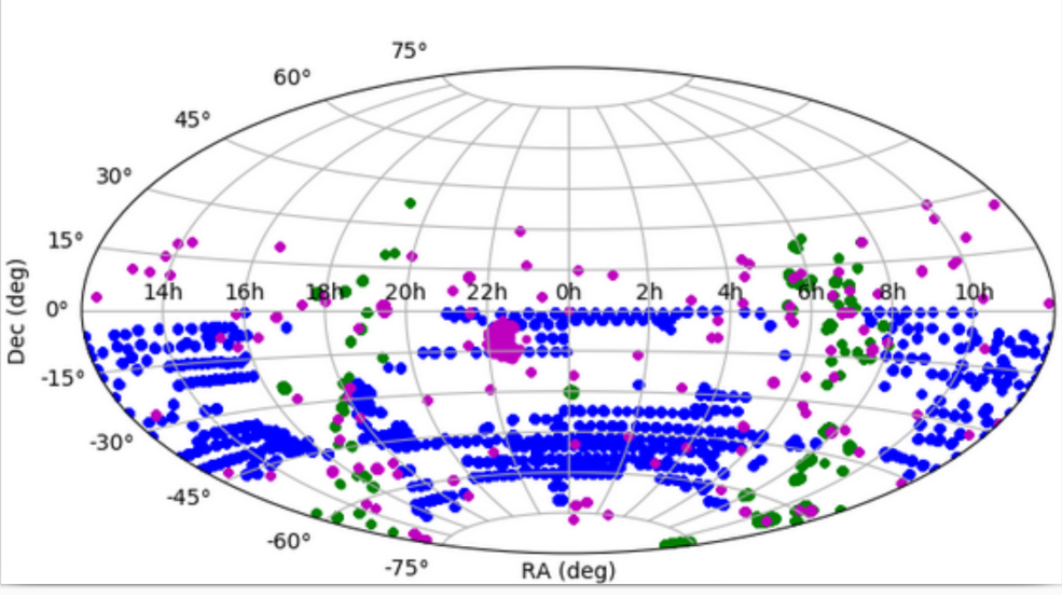}
    \caption{Map of observed targets in the {\em Gaia}-ESO survey. Blue, green and magenta dots indicate the Milky Way field stars, Clusters and Calibration fields. Figure adapted from \citep{randich2022gaia}}
    \label{fig:GaiaEsoCov}
\end{figure}

The {\em Gaia}-ESO survey is a public spectroscopic survey carried out with FLAMES, at medium resolution with  GIRAFFE and at high resolution with UVES on the VLT (Very Large Telescope) in Chile between 2012 and 2018. It observed around 115000 stars across all major components of the Milky Way. The data products of {\em Gaia}-ESO include radial and projected rotational velocities, stellar parameters such as effective temperature, surface gravity and metallicity, abundances of several elements, specific parameters for tracing accretion and activity in young stars and cluster probability and membership thanks to the combination with astrometry from {\it Gaia} EDR3.
GIRAFFE has a spectral resolution of $R \sim 20000$ and allows the observation of up to 130 targets at the same time. UVES instead has a spectral resolution $R \sim 47000$ and provides observations of up to 7 stars simultaneously.\newline
{\em Gaia}-ESO is designed to observe three classes of targets in the Milky Way: field stars, candidate members of open clusters and calibration standard stars. The primary goal is to select a robust sample of field stars in all the main components of the Milky Way
Fig.~\ref{fig:GaiaEsoCov} shows the map of the observed targets in the {\em Gaia}-ESO survey. 

The final Gaia-ESO catalogue (DR5.1) is publicly available and provides stellar parameters, radial velocities and a large number of abundances: Fe, C, N, O, Ne, Na, Mg, Al, Si, Ca, Sc, Ti, V, Cr, Mn, Cu, Co, Zn, Y, Zr, Ba, La, Ce, Sm, Eu.

To select high quality abundances, we  use the homogeneity properties of star clusters. We verified that the abundances versus stellar parameters in cluster members do not show a strong correlation to check the quality of the data and of the selection, in a parallel to that done in APOGEE. 

\subsubsection{Kinematic selection of our {\em Gaia}-ESO sample}

For the computation of orbits we used the same method applied for the APOGEE sample. 
As input, we used the radial velocity from {\em Gaia}-ESO and the astrometry from {\em Gaia} {\sc dr3}, including RA, DEC, distance, and the two components of the proper motion. We assumed with the McMillan Galactic potential, and we used the AGAMA code to compute the orbital parameters.

\subsubsection{Our {\em Gaia}-ESO sample of halo stars}
\begin{figure}[H]
\centering
\includegraphics[width=0.8\textwidth]{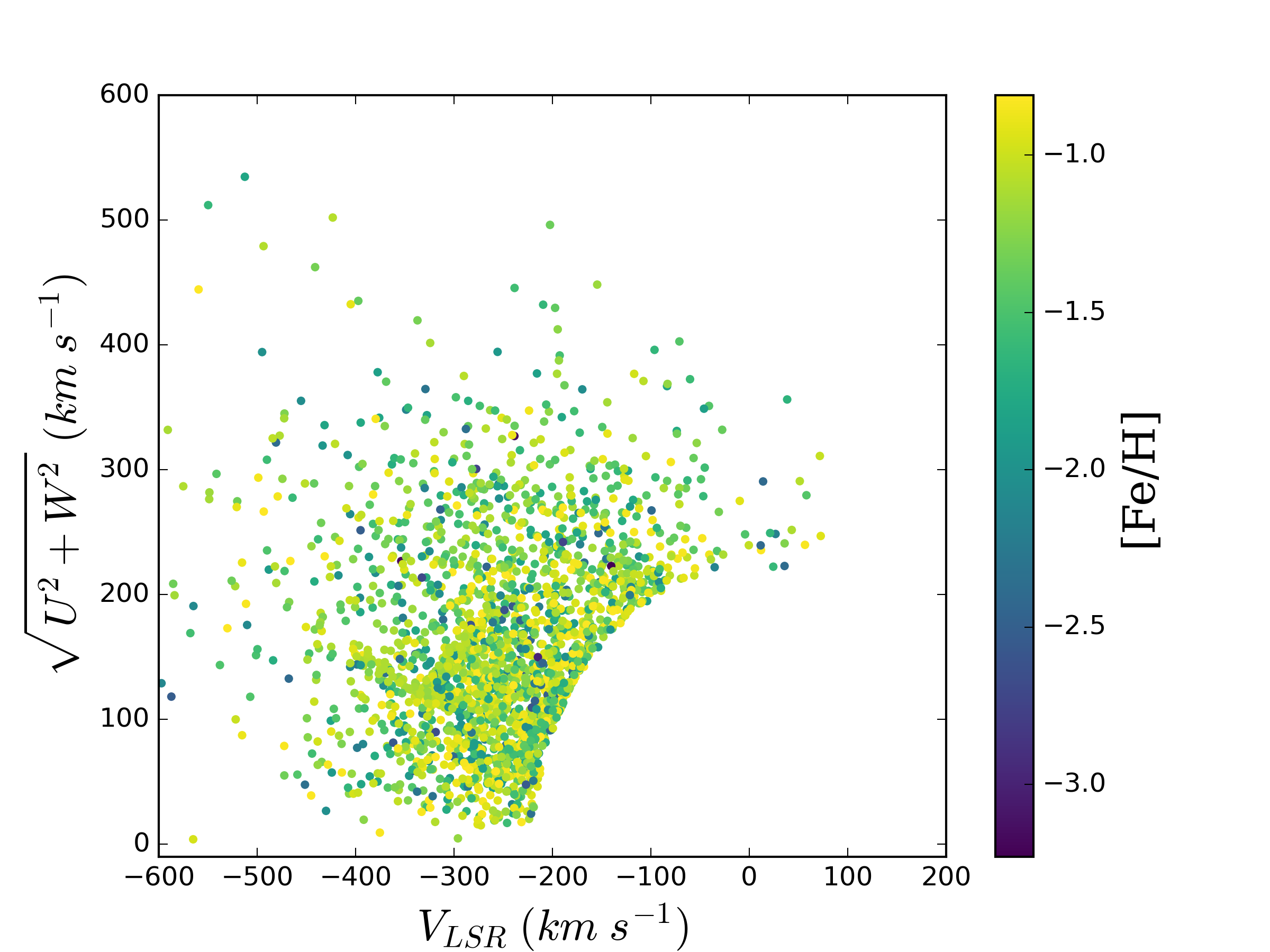}
\caption{\label{fig:GaiaEsoToomre} Toomre diagram of the selected {\em Gaia}-ESO  halo stars. Stars are colour-coded by their [Fe/H]. }
\end{figure}

We applied as for the APOGEE sample a two-step selection: a first cut based on metallicity and a second one the kinematic. 
To select halo stars, we included in our sample stars with [Fe/H]$<$-0.8 (the cut is slightly different to that used in APOGEE because with {\em Gaia}-ESO we started from a smaller sample, and the cut with [Fe/H]$<$-1.0 would have reduced it too much). With the same procedure used for the APOGEE stars, we computed their orbits (about 13\,000 stars). 
From these stars, we selected those with total velocities > 220 km~s$^{-1}$. 
The selection of stars with total velocities over 220 km~s$^{-1}$ is visible in the Toomre diagram in Fig.~\ref{fig:GaiaEsoToomre}.
Stars were further selected based on the error on the main stellar parameters (T$_{\rm eff}$,  log~g), [Fe/H] and of the principal $\alpha$ element available in our sample, Mg. The selection criteria employed are outlined in Table \ref{tab:CriteriaGaia} resulting in a total of 2307 stars that compose our sample of halo stars. 

\begin{table}
    \centering
    \begin{tabular}{|c|c|} \hline 
         E\_TEFF& < 90\\ \hline 
         E\_LOGG& < 0.18\\ \hline 
         E\_MG1& < 0.3\\ \hline
    \end{tabular}
    \caption{Criteria applied for the selection of {\em Gaia}-ESO stars.}
    \label{tab:CriteriaGaia}
\end{table}

\begin{figure}
    \centering
    \includegraphics[width=0.75\linewidth]{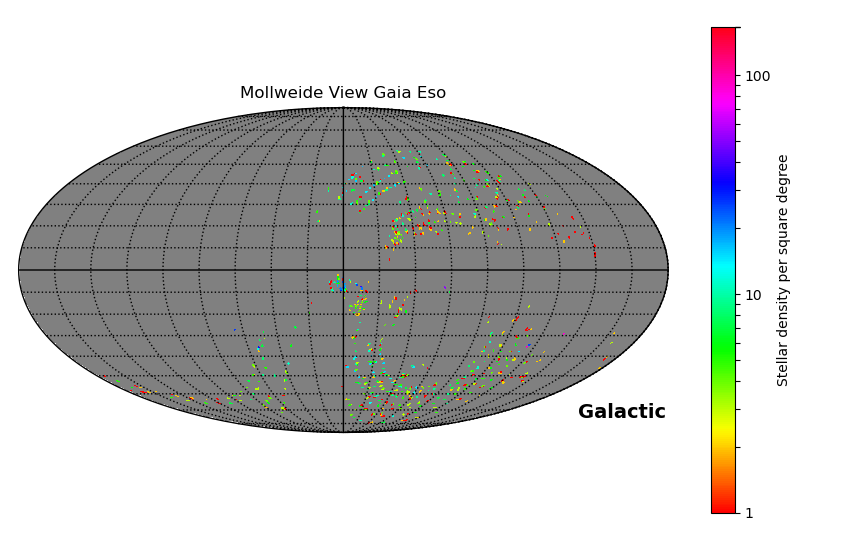}
    \caption{Mollweide view of the selected halo stars in the {\em Gaia}- ESO survey. Stars are colour-coded according to stellar density.}
    \label{fig:MollviewGaia}
\end{figure}

The location in the Galaxy  of our sample stars  is represented in the Mollweide view plot in Fig.~\ref{fig:MollviewGaia}, where the stars are coloured according to their density.
Since {\em Gaia}-ESO has mainly focused its observations on the disc population, including mainly thin and thick disc field stars and open clusters, the halo population is very limited in number. Such a small sample does not allow us to study general properties of the halo field population. In the next sections, we outline the properties of the population of globular clusters, which are well represented in {\em Gaia}-ESO because they are also used as calibrators, and we characterise some of the streams that are present in our sample. 

\subsubsection{Properties of the Globular Cluster population}
As we did for the APOGEE data, we aim at characterizing both globular clusters and streams in the {\em Gaia}-ESO dataset. With our criteria,  we find a total of 397 stars belonging to six well-populated globular clusters. 
There are more GCs in {\em Gaia}-ESO, but they follow outside the range of our chemo-kinematic selection criteria or they contain too few members that fulfill all the quality requests. The numbers of member stars in each of the six clusters  are reported in Table \ref{tab:clusters}.

\begin{table}
    \centering
    \begin{tabular}{|l|l|}
         \hline
         Cluster name& Number of members\\
         \hline
         NGC 1851& 117\\
         \hline
         NGC 362& 107\\
         \hline
         NGC 1904 M79& 69\\
         \hline
         NGC4833& 49\\
         \hline
         NGC 7089 M2& 30\\
         \hline
         NGC 7078 M15& 25\\\hline
    \end{tabular}
    \caption{Globular clusters with more than 10 stars found in the {\em Gaia}-ESO metal-poor halo. }
    \label{tab:clusters}
\end{table}

\begin{figure}
    \centering
    \includegraphics[width=0.75\linewidth]{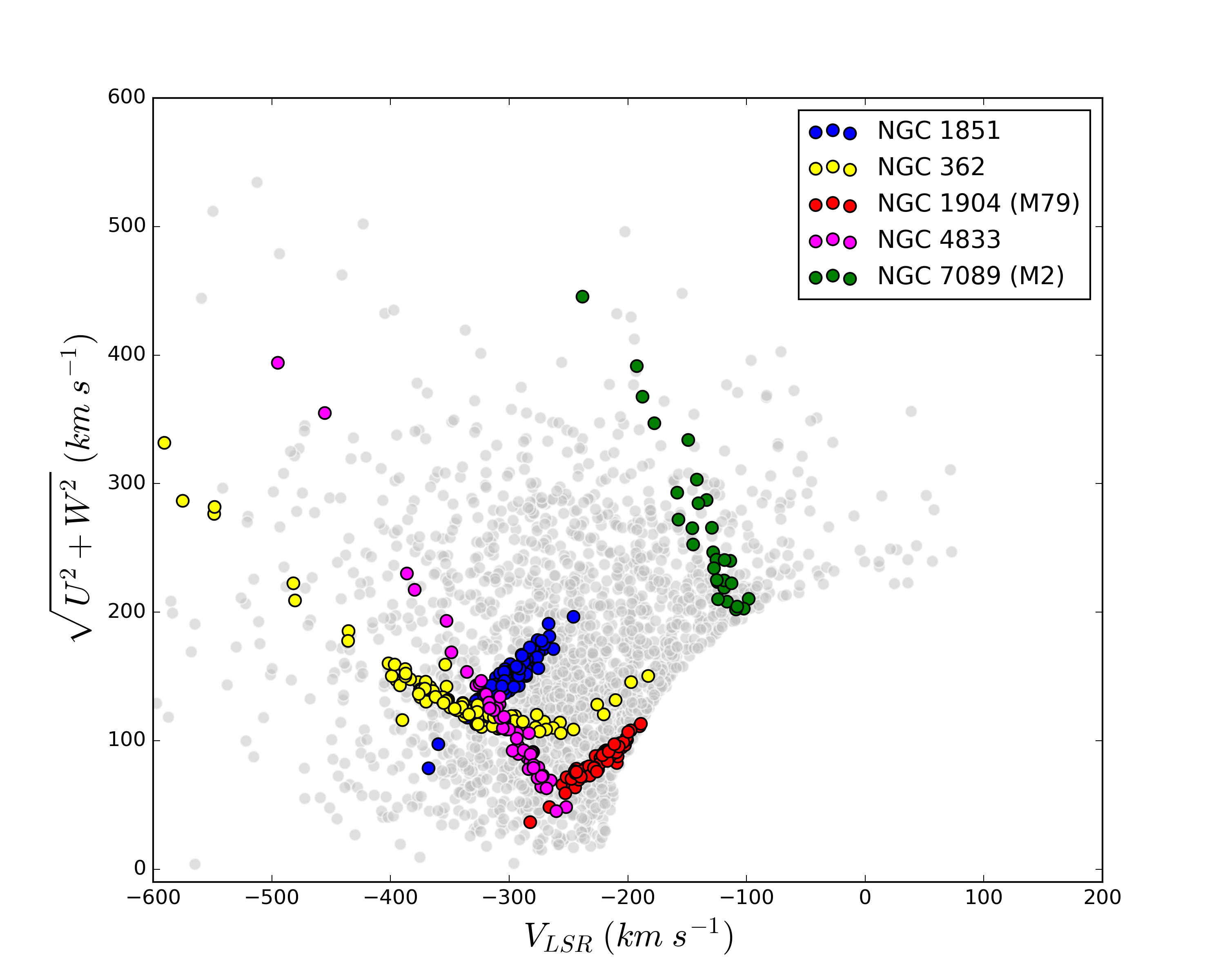}
    \caption{Toomre diagram of our sample of halo stars in the {\em Gaia}-ESO survey. They are colour-coded as in legend. Field stars are in grey.}
    \label{fig:GaiaToomreClusters}
\end{figure}

\begin{figure}
    \centering
    \includegraphics[width=0.95\linewidth]{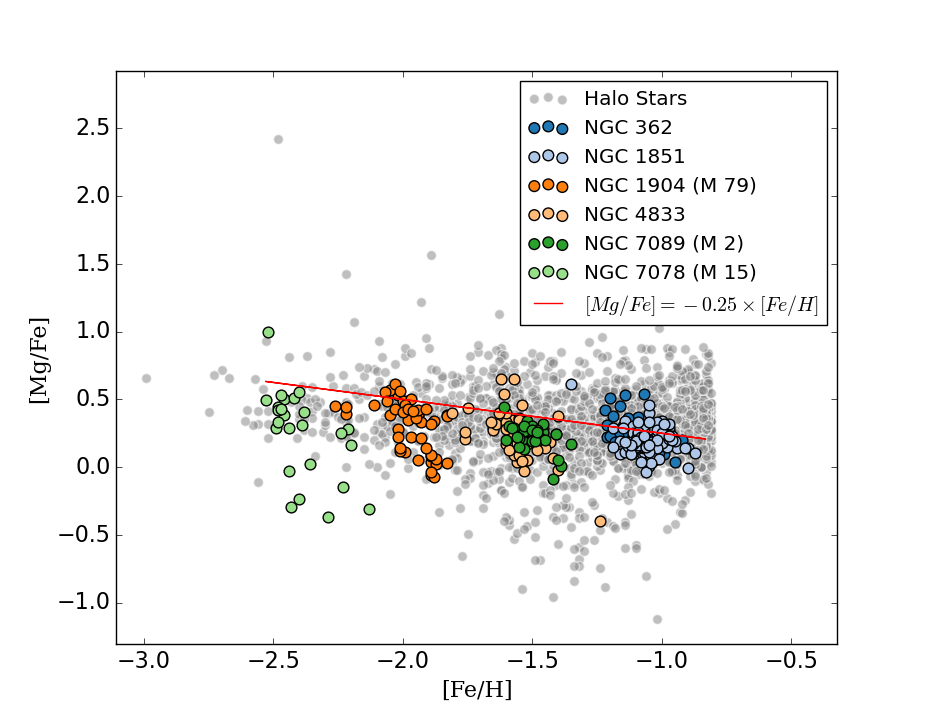}
    \caption{[Mg/Fe] vs [Fe/H] for the most populated clusters in {\em Gaia}-ESO. Stars are colour-coded as in legend, while field stars are in grey. The red line shows the separation between LMg and HMg.}
    \label{fig:MgFeGaiaClusters}
\end{figure}

Fig.~\ref{fig:GaiaToomreClusters} shows the most populated Globular Clusters present in our dataset coloured in the Toomre diagram. As we already saw with the APOGEE sample, GCs are kinematically grouped and easily identified in the Toomre diagram.\newline
Using the same line separator found in APOGEE, we colour the most populated clusters in {\em Gaia}-ESO and find that they mostly lie below the line, in the accreted region, as shown in Fig.~\ref{fig:MgFeGaiaClusters}.

\begin{figure}
    \centering
    \includegraphics[width=1\linewidth]{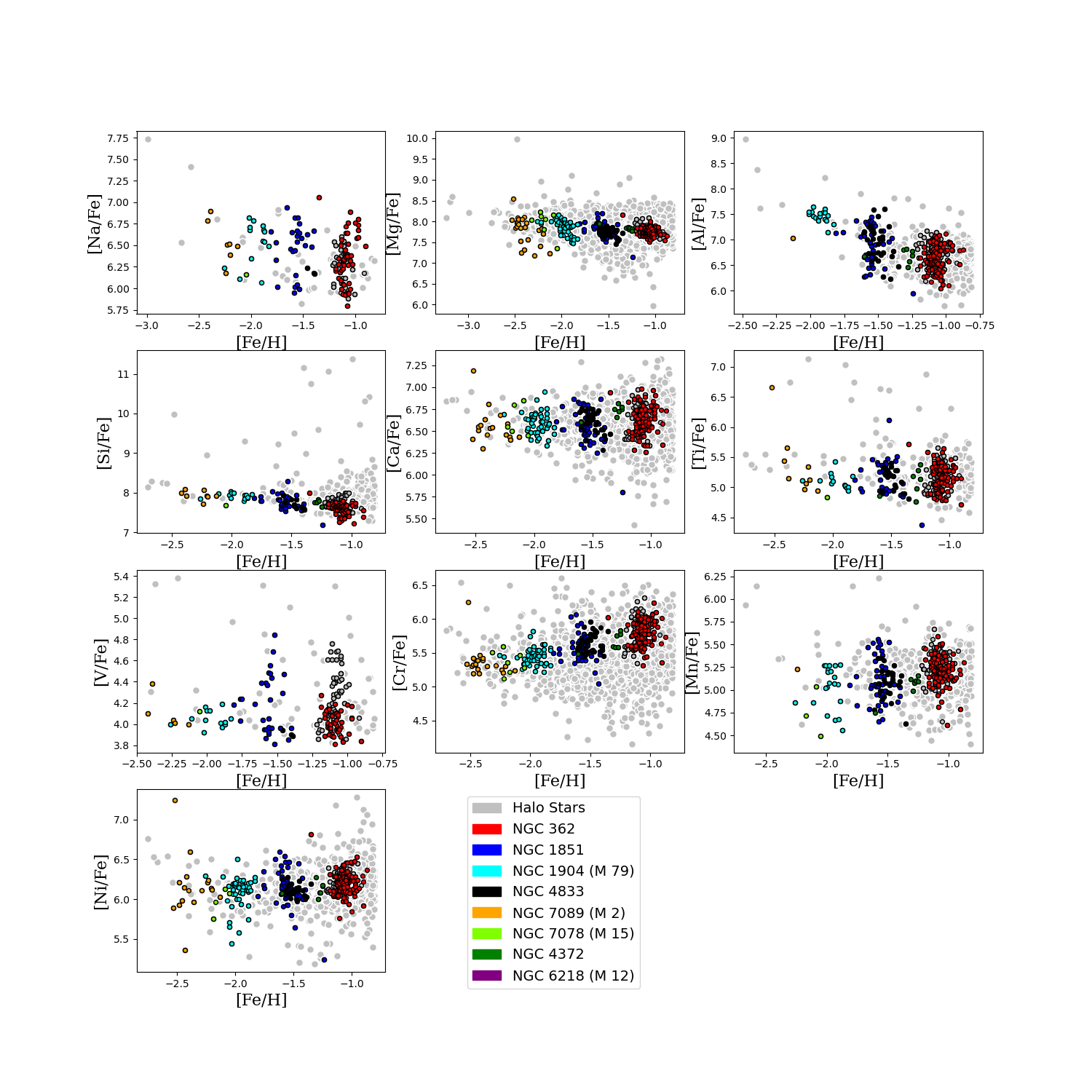}
    \caption{Abundance ratios vs [Fe/H] of member stars of some of the most populated clusters in {\em Gaia}-ESO. Stars in grey are halo stars, member stars of globular clusters are colour-coded as in legend.}
    \label{fig:ElementsClusterGaia}
\end{figure}

As shown for APOGEE, we see in Fig.~\ref{fig:ElementsClusterGaia} that clusters can be distinguished through individual abundances in addition to  [Fe/H]. Again, we see that some elements, such as Na and V show a great spread within the clusters, both for the presence of anticorrelations (Na) or for the quality of the abundances (V).  Elements such as Ni, Cr, Si and Ca separate quite well the clusters one from each other, instead.

\subsubsection{Properties of the Galactic halo streams}
So far, there has been no attempts to classify halo stars in Gaia-ESO as stream members. We therefore applied the criteria indicated in \citet{Horta2023} to associate stars in our catalog with known streams. The adopted criteria are displayed in Table \ref{tab:StreamGaiaCriteria}. The differences in signs compared to the original criteria depend on the different coordinate system. 

\begin{table}
    \centering
    \begin{tabular}{|l|l|}
    \hline
        {\em Gaia}-Enceladus / Sausage & $\abs{L_Z}<0.5(\times10^3 kpc \;km\; s^{-1})$,\\
         & $-1.6<E<-1.1(\times10^5 km^2 \;s^{-2})$\\
         \hline
        Helmi Stream & $0.75<L_Z<1.7(\times10^3 kpc\; km \;s^{-1})$,\\
        & $1.6<L_{\perp}<3.2(\times10^3 kpc \;km\; s^{-1})$\\
         \hline
        Icarus&  [Fe/H]<-1.45, $1540<L_Z<2210(kpc\; km\; s^{-1})$,\\
        &$L_{\perp}<450(kpc\; km\; s^{-1})$, $z_{max}<1.5$, [Mg/Fe]<0.2\\
         \hline
        Sequoia (N20) & $\eta<-0.15$, $E>-1.6(\times10^5 km^2\; s^{-2})$,\\
         & $L_Z<0.7(\times 10^3 kpc \;km \;s^{-1})$, -2<[Fe/H]<-1.6\\
         \hline
        Arjuna &$\eta<-0.15$, $E>-1.6(\times10^5 km^2\; s^{-2})$, \\
        & $L_Z<0.7(\times 10^3 kpc\; km\; s^{-1})$, [Fe/H]>-1.6\\
         \hline
        I'itoi & $\eta<-0.15$, $E>-1.6(\times10^5 km^2\; s^{-2})$\\
        & $L_Z<-0.7(\times 10^3 kpc\; km\; s^{-1})$, [Fe/H]<-2\\
         \hline
    \end{tabular}
    \caption{Criteria used to identify streams in {\em Gaia}-Eso.}
    \label{tab:StreamGaiaCriteria}
\end{table}

\begin{figure}
    \centering
    \includegraphics[width=0.95\linewidth]{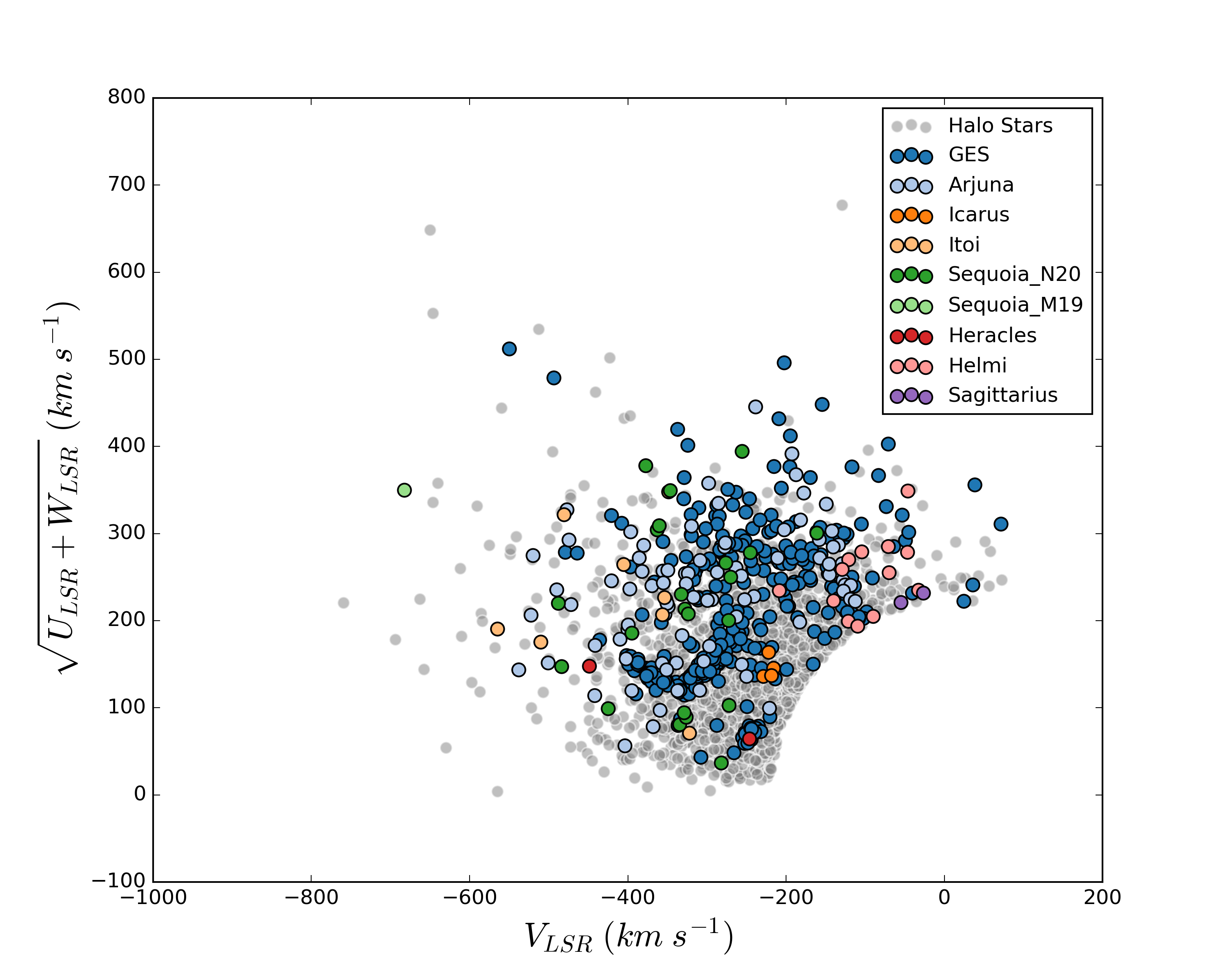}
    \caption{Toomre Diagram with streams from Gaia colour-coded as explained  in the legend. Halo stars are represented in grey.}
    \label{fig:ToomreStreamsGaia}
\end{figure}
\begin{figure}
    \centering
    \includegraphics[width=0.85\linewidth]{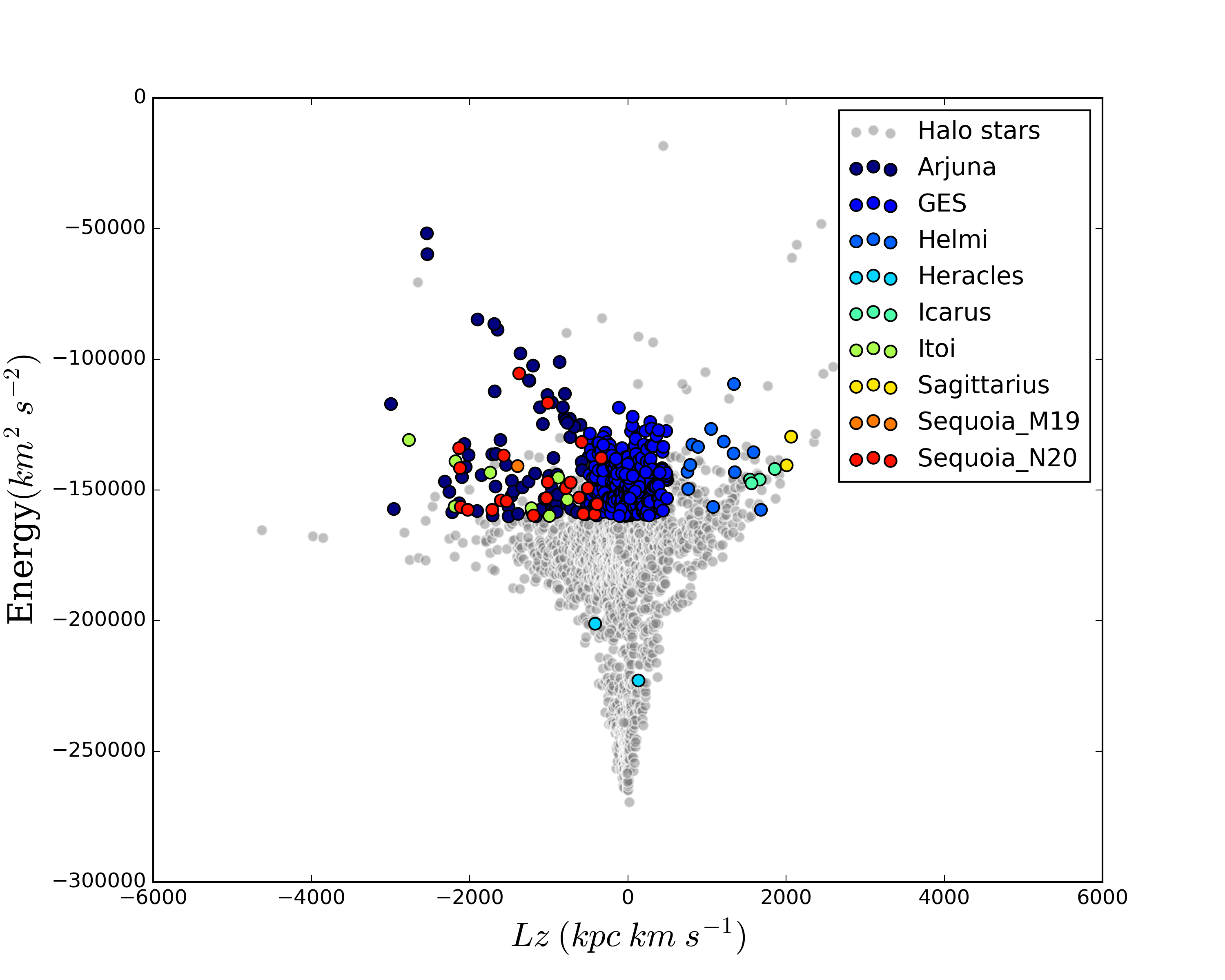}
    \caption{Lindblad diagram for {\em Gaia}-ESO streams. Streams are colour-coded as in the legend. Halo stars are in grey.}
    \label{fig:LindbladGaiaEso}
\end{figure}

As already shown for APOGEE stars, when we plot streams in the Toomre diagram, in fig.~\ref{fig:ToomreStreamsGaia}, we find that they are quite spread through the plane and sparse. 
As shown in Fig.~\ref{fig:LindbladGaiaEso}, the streams we select in {\em Gaia}-ESO have  usually  high total energies (over-170000). Some streams are  retrograde, such as Arjuna, I'itoi and Sequoia, whereas the Helmi stream and Icarus are prograde. We remember that prograde stars in this work are positioned on the right side of this plot. The two categories are separated by GES, situated around $L_Z = 0$.

\begin{figure}
    \centering
    \includegraphics[width=0.75\linewidth]{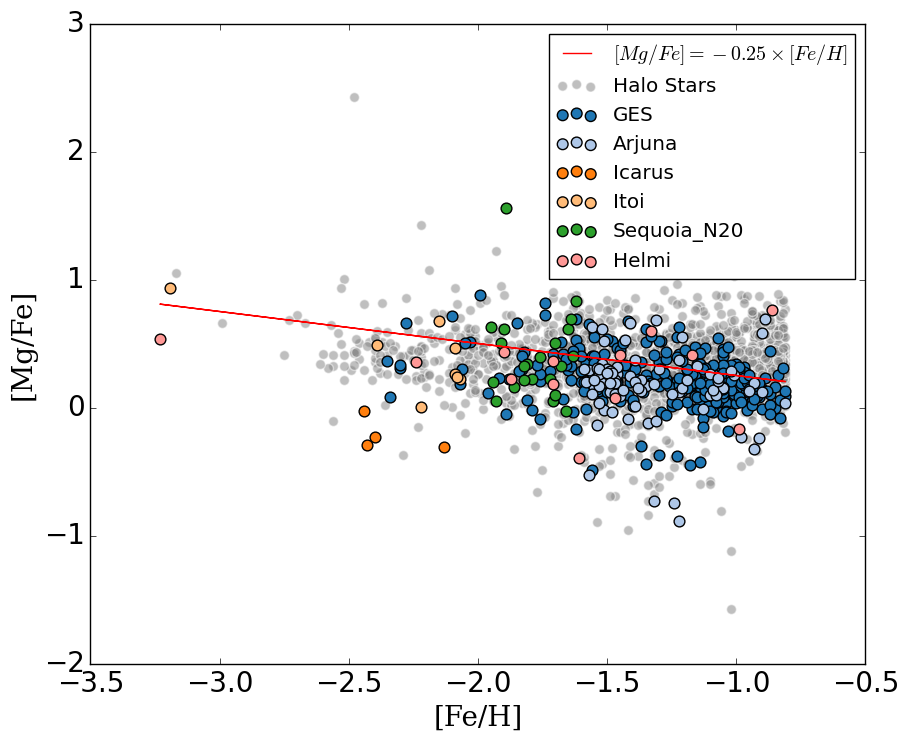}
    \caption{[Mg/Fe] vs [Fe/H] for the streams present in {\em Gaia}-ESO and for field halo stars. The line between accreted and in situ object is plotted in red and stars are colour-coded as in the legend.}
    \label{fig:MgFeStreamsGaia}
\end{figure}

In {\em Gaia}-ESO the stars observed are not enough to see the separation between accreted and in situ stars in the [Mg/Fe] vs [Fe/H] plane. We added the separation line between the two sequences as described above and  we noticed that that most of the streams are located, as expected,  in the LMg region, as shown in Fig.~\ref{fig:MgFeStreamsGaia}.

\subsubsection{The first chemical characterization of stellar streams with {\em Gaia}-ESO data}

 The number  of  stream members found in our sample are presented in Table~\ref{tab:StreamGaia}. As a stream with a limited  number of stars is not significant in our statistical study, we will not consider streams with less than 4 members.

\begin{table}[H]
    \centering
    \begin{tabular}{|c|c|} \hline 
         Stream& Members\\ \hline 
         GES& 383\\ \hline 
         Arjuna& 83\\ \hline 
         Sequoia(N20)& 22\\ \hline 
         Helmi& 13\\ \hline 
         I'itoi& 8\\ \hline 
         Icarus& 4\\ \hline 
         Heracles& 2\\ \hline 
         Sagittarius& 2\\ \hline 
         Sequoia(M19)& 1\\ \hline
    \end{tabular}
    \caption{Number of stars found in each stream in {\em Gaia}-ESO.}
    \label{tab:StreamGaia}
\end{table}

\begin{figure}
    \centering
    \includegraphics[width=1\linewidth]{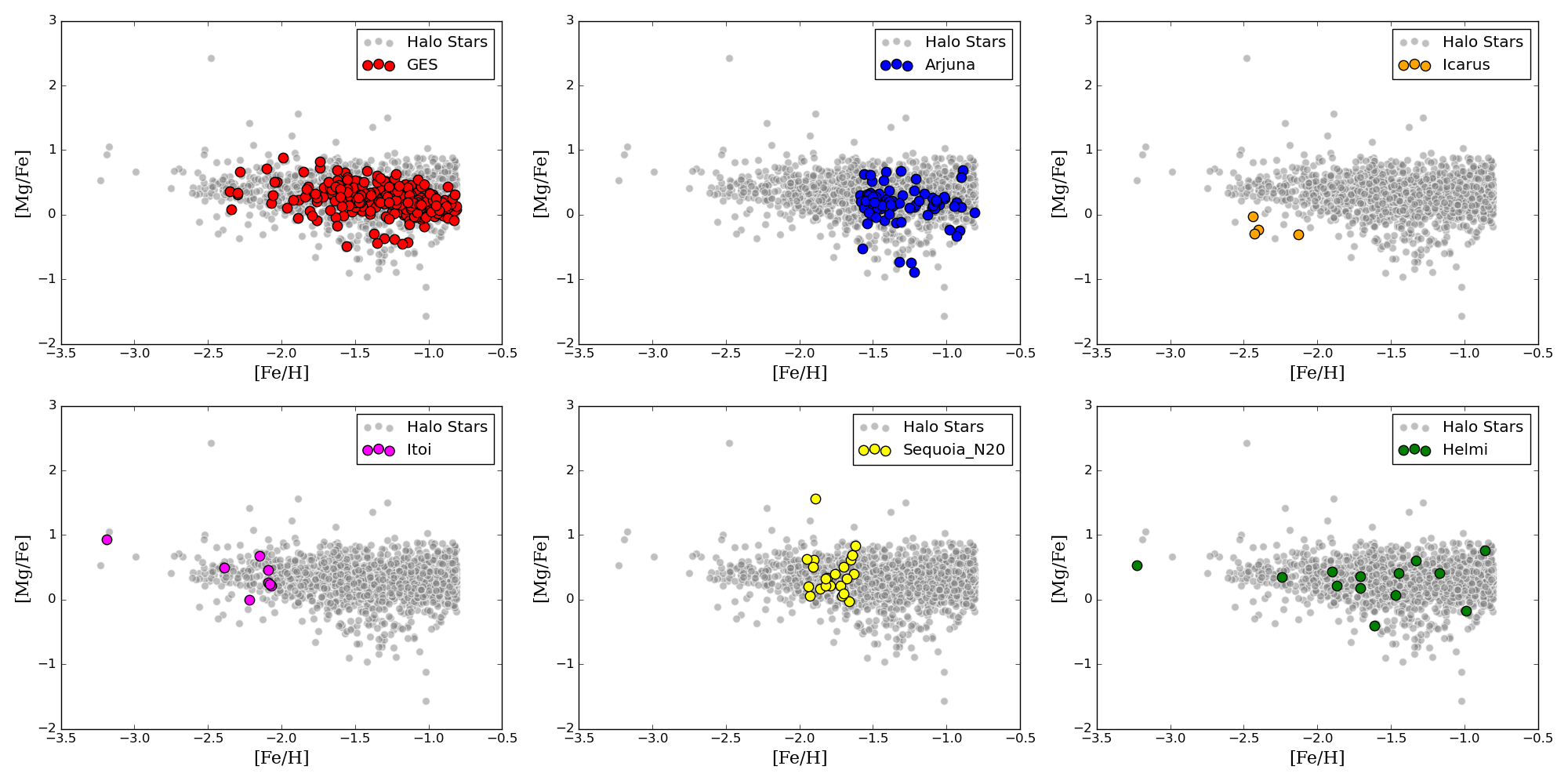}
    \caption{[Mg/Fe] vs [Fe/H] for streams in {\em Gaia}-ESO, grey stars represent halo stars of our dataset, coloured stars are members of the selected streams.}
    \label{StreamMggaiaeso}
\end{figure}

\begin{figure}
    \centering
    \includegraphics[width=1\linewidth]{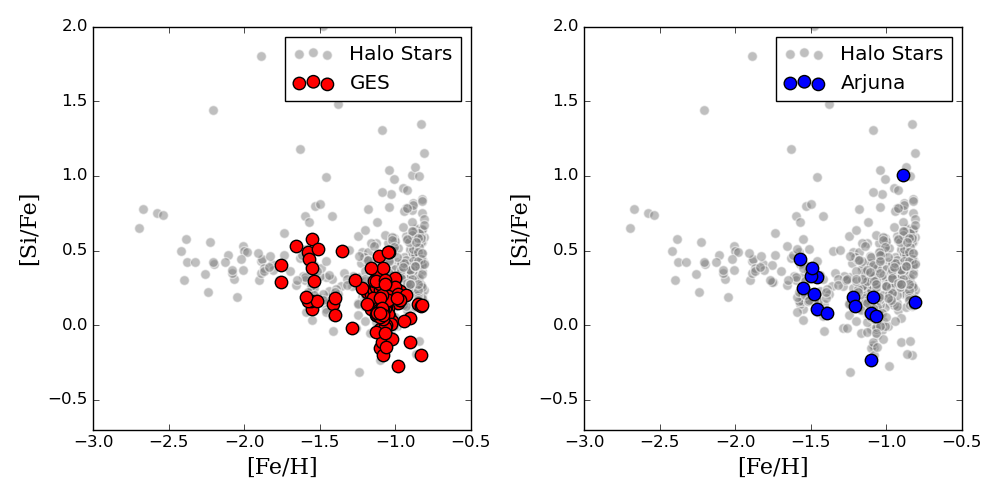}
    \caption{[Si/Fe] vs [Fe/H] for streams in {\em Gaia}-ESO, grey stars represent halo stars of our dataset, coloured stars are members of  streams. Other streams are not shown as they have a too low number of stars with measured Si.}
    \label{StreamSigaiaeso}
\end{figure}

\paragraph{$\alpha$ elements} 
We first investigate the behaviour of their $\alpha$-element abundances in the [$\alpha$/Fe] vs [Fe/H] plane. This plane is interesting as it provides insight of the star formation history and the chemical enrichment process of the substructures. As a collective $\alpha$ abundance is not present in the {\em Gaia}-ESO dataset, in Fig.~\ref{StreamMggaiaeso} we analyse the abundance of Mg, the $\alpha$-element measured in the largest number of  stars in our sample.
In Fig.~\ref{StreamSigaiaeso} we show [Si/Fe] vs [Fe/H] which is available for a significant amount of stars for GES and Arjuna. 
From Figs.~\ref{StreamMggaiaeso} and \ref{StreamSigaiaeso} we find that GES and Arjuna are particularly low in their $\alpha$-element abundances over iron. This is also valid for Icarus, though only Mg is measured in its stars. In GES we confirm the results presented in \citet{Horta2023}: GES reaches almost solar metallicities (not visible in our plots, where they reach the highest [Fe/H] of the sample) and presents an $\alpha$-knee below [Fe/H]<-1, i.e. lower compared to the disc stars. Even though they are characterized by a higher spread, at low metallicity it seems that GES and Arjuna are also characterized by a lower plateau in [$\alpha$/M] compared to the halo one. The similarity and partial overlapping between GES and Arjuna suggests a common origin or even a belonging to the same structure.\newline
The larger extension of GES in [Fe/H] suggests a longer star formation history, thus a more massive progenitor.
When examining Sequoia and I'itoi, we observe a similarity in Mg abundance. The difference in metallicity between the two structures depends on the selection criteria. The Helmi stream, instead, shows a great dispersion in both metallicity and Mg abundance precluding any definitive conclusion.
Other $\alpha$-elements - O, Ca, S, Ti - are  measured in fewer stars and often not in stars belonging to streams.

\begin{figure}
    \centering
    \includegraphics[width=1\linewidth]{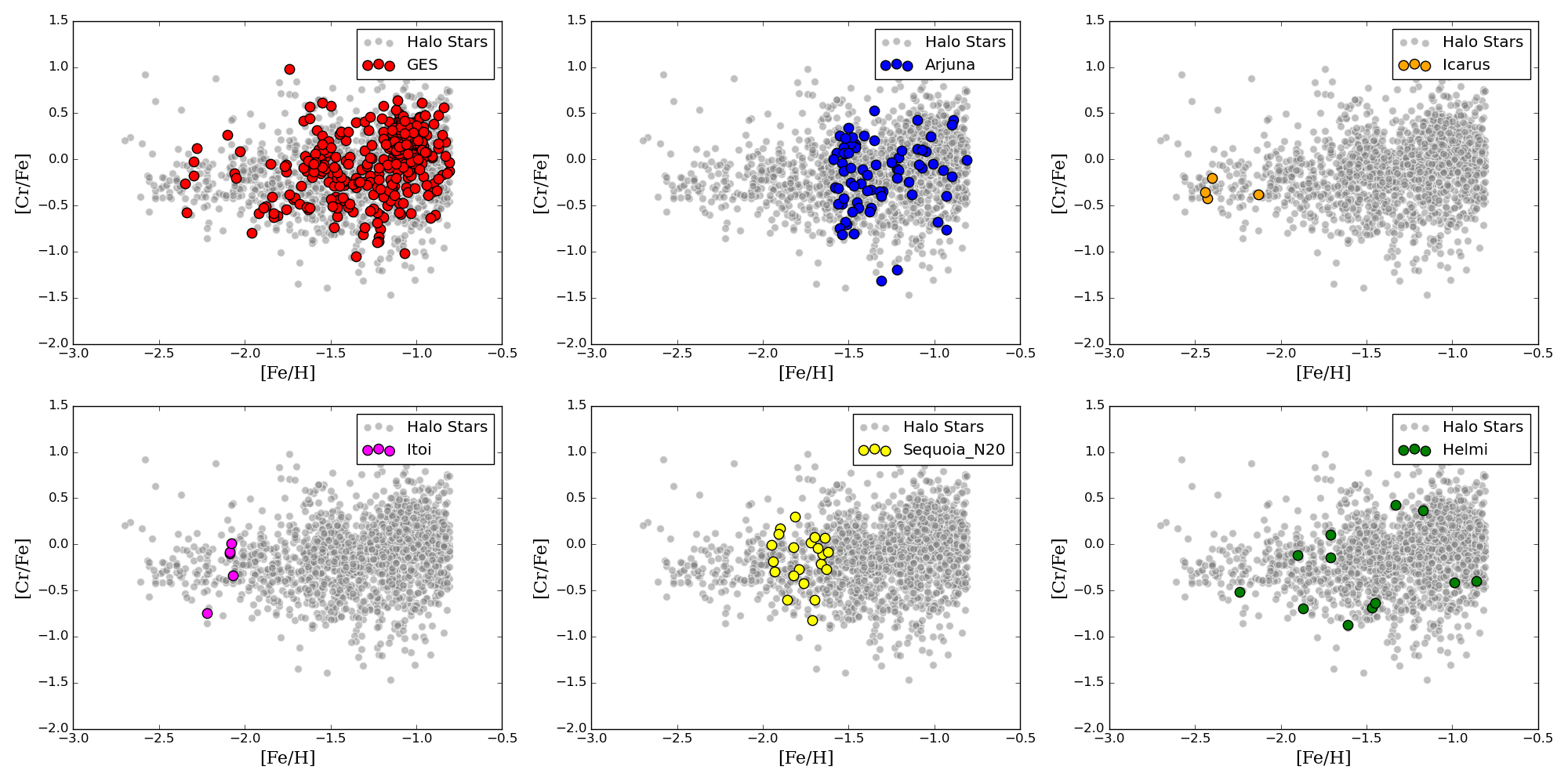}
    \caption{[Cr/Fe] vs [Fe/H] for streams in {\em Gaia}-ESO, grey stars represent halo stars of our dataset, coloured stars are members of the selected streams.}
    \label{fig:StreamCr}
\end{figure}

\begin{figure}
    \centering
    \includegraphics[width=1\linewidth]{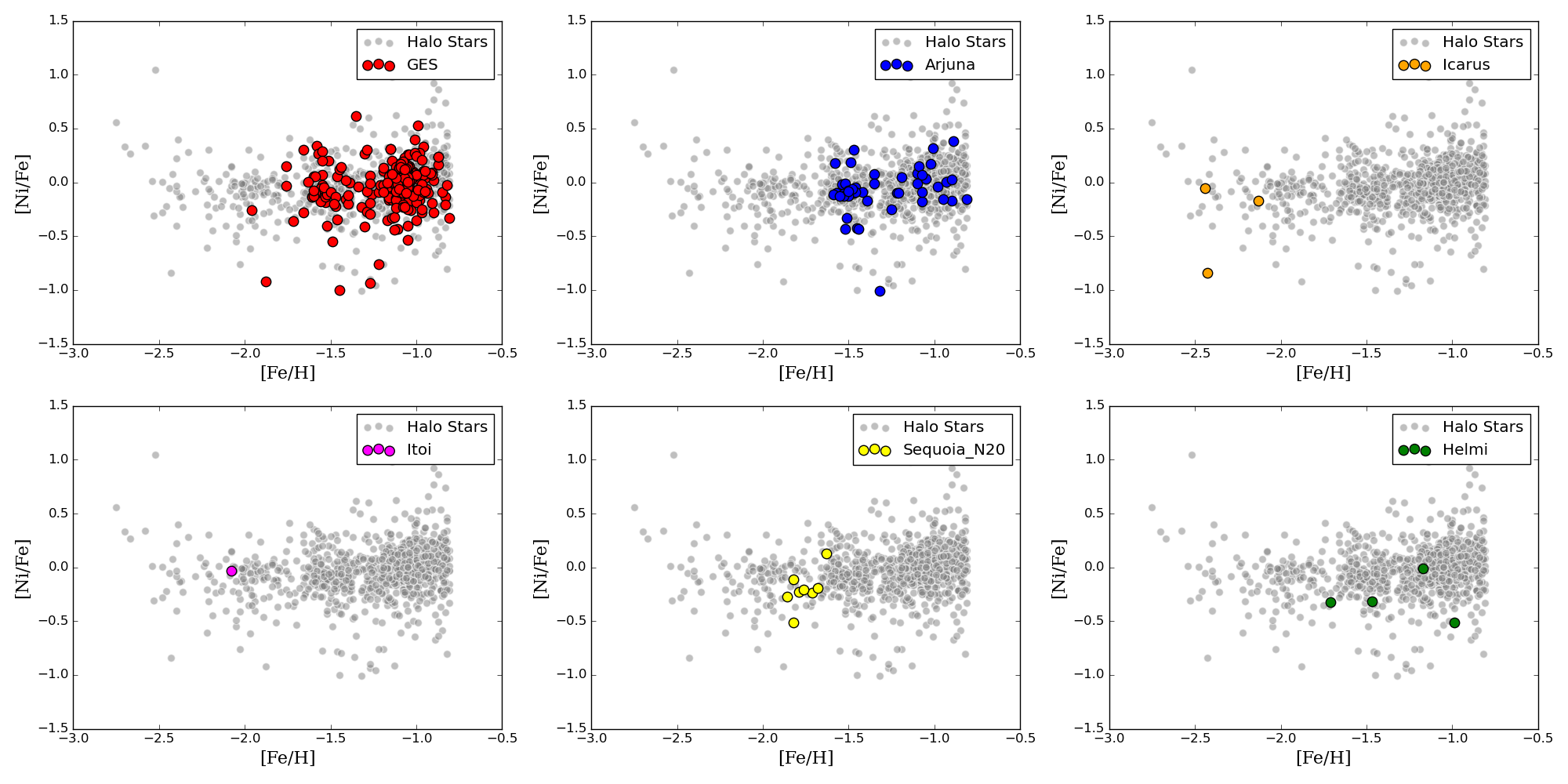}
    \caption{[Ni/Fe] vs [Fe/H] for streams in {\em Gaia}-ESO, grey stars represent halo stars of our dataset, coloured stars are members of the selected streams.}
    \label{fig:StreamNi}
\end{figure}

\paragraph{Iron-peak elements} 
We also analysed the behaviour of the iron-peak element abundances in the streams. As already explained, these elements probe the contributions of SNe type Ia. The best iron-peak elements are Cr and Ni as they have the highest number of measurements for stars in our sample. Cr and Ni are shown in Fig.~\ref{fig:StreamCr} and Fig.~\ref{fig:StreamNi} respectively. Streams have iron-peak abundances similar to those of the other halo stars. We can expect this behaviour because iron-peak elements and iron are generated by the same nuclosynthesis channels, and thus their ratio is independent of SFH.

\begin{figure}
    \centering
    \includegraphics[width=1\linewidth]{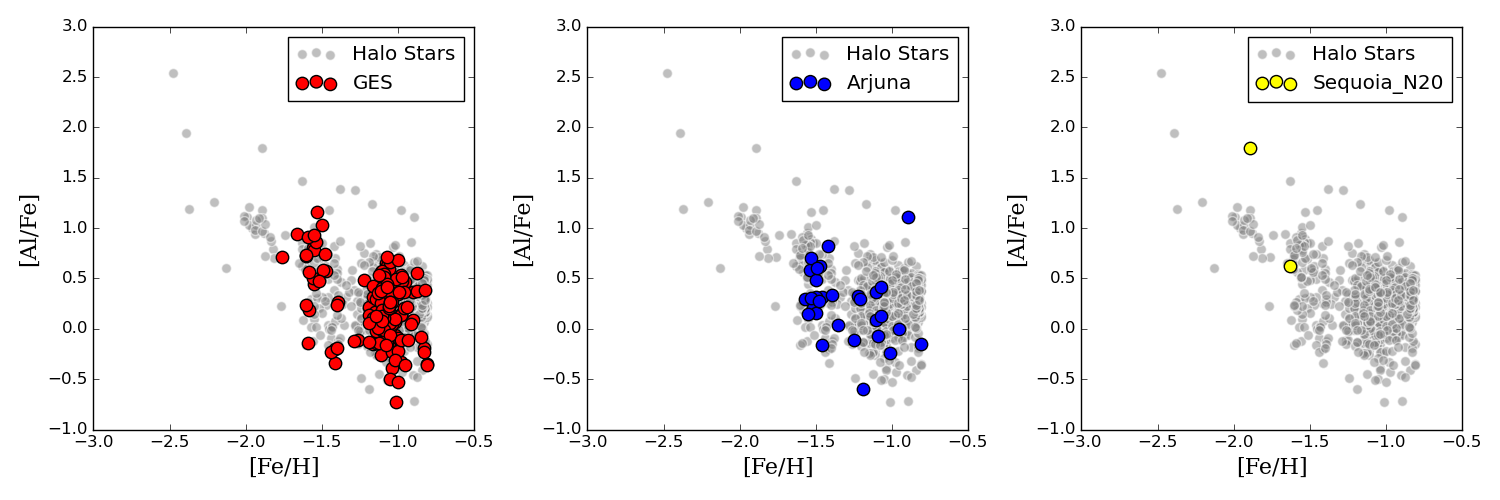}
    \caption{[Al/Fe] vs [Fe/H] for streams in {\em Gaia}-ESO, grey stars represent halo stars of our dataset, coloured stars are members of the selected streams.}
    \label{fig:alges}
\end{figure}

\paragraph{odd-Z elements} 
Another important element to discuss is aluminum. Aluminum is an odd-Z element. It is usually depleted in satellite galaxies of the Milky Way and in accreted systems compared to in situ populations \citep{Hawkins10.1093/mnras/stv1586}. 
From Fig.~\ref{fig:alges} we can see that accreted halo stars have a peculiar distribution in the [Al/Fe] vs [Fe/H] plane, usually they have a low [Al/Fe] with respect to the halo in situ field stars. Though the cause is unclear, Al is generally depleted in accreted populations, as we see in GES but especially in Arjuna in this figure. Stellar nucleosynthesis models posit that Al is produced mainly by SN type II with a metallicity dependence \citep{woosley2002evolution}. It has also been theorised that Al production increases with metallicity at a similar rate to the dilution of Type II supernovae ejecta by Type Ia supernovae or that we are missing a parameter in low-mass systems that causes Al to stay low compared to Mg\citep{feuillet2021selecting}.
For observational reasons, the abundance of Al is easier to observe at higher metallicity, so we do not have many abundance measurements in the low metallicity tail of our sample.  

\paragraph{Neutron-capture elements and future follow-ups}
Other elements would be interesting to analyse, but they are not present in this dataset, such as neutron capture elements of the s-process family, i.e. Ce, or the plot [Al/Fe] vs [Mg/Mn], that helps discriminate accreted populations from in situ populations.
To this purpose, we requested observation time with UVES@VLT to acquire new high-resolution spectra of stream members that had been observed with GIRAFFE at medium resolution in {\em Gaia}-ESO.

%% file: 4-Algorithm.tex
\section{The {\sc creek} - {\underline C}lustering with g{\underline R}aph n{\underline E}ural n{\underline E}tworK}
\label{chap4}
\subsection{Selecting the dataset: best abundances and best clusters}
One of the crucial aspects of this work lies in the selection of the dataset. The variation of the number and type of abundances used produces indeed a big variation in the results.
In the previous sections, we have seen that halo structures that are still spatially coherent, such as globular clusters, can be identified through kinematics alone. However,  remnants of past mergers and streams have lost velocity and spatial coherence thus requiring the use of orbit invariant quantities to be identified. These invariants may be some dynamical properties such as the total energy E and the vertical angular momentum $L_Z$, or their actions $J_R$, $J_{\phi}$ and $J_Z$ and the chemical properties.
Dynamics alone, indeed, is not enough, as ancient mergers can incur phase mixing and thus lose dynamical coherence \citep{Mori2024arXiv240113737M}. The chemical photospheric composition, on the other hand, is a property that is not lost. Therefore, we base our analysis on the use of the best chemical tracers, with the help of orbital properties. 
For this reason, in order to recover past mergers and disrupted clusters through their chemistry, we need to choose the best elements. In addition, to test the performances of the method we need to check it on the best clusters, i.e. the ones that are more compact in the chemical space.

The selection of the best elements takes into account two aspects: the ability of a given element to separate populations with different characteristics due to their different SFH, and the precision with which we can measure the abundances of the various elements. 
Indeed, stars will be more clustered in some abundances compared to other, as different elements come from different processes. This is a crucial point, as depending on the observed clusters we find that a different element separates them best.
On the other hand, some elements might be excellent for separating stellar populations from a theoretical point of view, such as neutron capture elements, but they are extremely difficult to measure with high precision. The determination of abundances for low metallicity stars is indeed remarkably challenging. In {\em Gaia}-ESO, for example, the low metallicity sample is complete only for Fe, Mg and Cr, while the other elements are measured only in a small fraction of the sample.

\begin{figure}[H]
\centering
\includegraphics[width=1.1\textwidth]{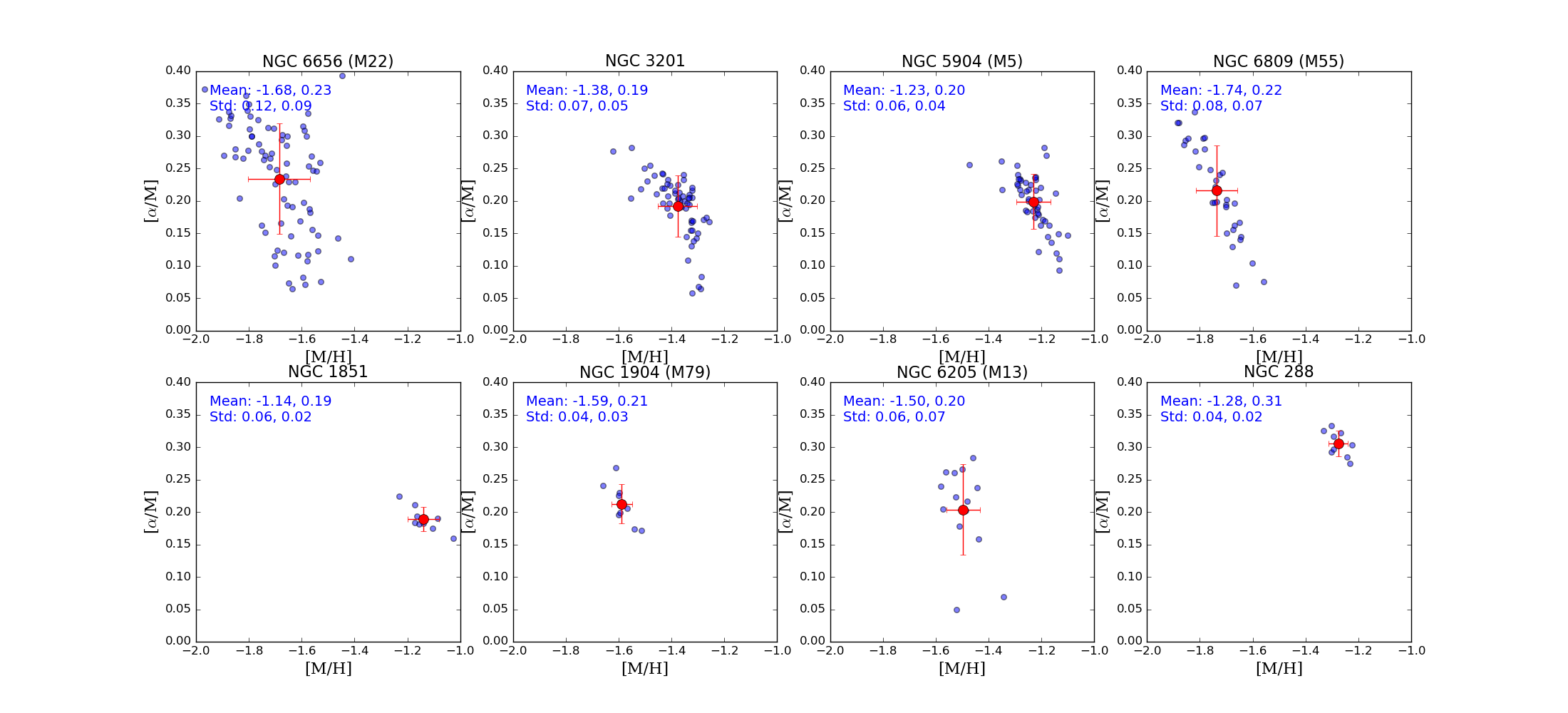}
\caption{\label{fig:abuclusterAl} [$\alpha$/M] vs [M/H] in our sample globular clusters. Member star of each cluster are shown with  blue circles. The red dot represents the mean value and the error bar is the standard deviation.}

\end{figure}

For both APOGEE and {\em Gaia}-ESO, to choose the best elements, we based our selection on the abundances of globular clusters, selecting those elements that do not have large variations due to the presence of multiple populations. 
An example of the kind of plot used to determine the best elements to separate clusters  from each other and from field stars is in Fig.~\ref{fig:abuclusterAl}.

This visual classification is complemented by a more rigorous method that consists in the computation of a metric inspired by \citet{Mitschang2013}.
We want to establish the best abundances to use, i.e. the abundances that show the biggest variations in distances in the chemical space between stars computed inside the cluster with respect to stars belonging to two different clusters.
In order to do this, we define the following metric:
\begin{equation} \label{Metric}
    \delta_c = \sum_{i,j}^{i\neq j}{\frac{\abs{A_C^i-A_C^j}}{N_{stars}}}
\end{equation}

Where $i$ and $j$ are indices of the stars, $A_C$ are the abundances of the element and $N_{stars}$ is the number of stars considered. We compute this $\delta_C$ for stars inside the clusters, obtaining an intra-cluster metric, and for stars in different clusters, obtaining an inter-cluster metric.
The ratio between intra-cluster and inter-cluster metric gives us a parameter to determine whether the element in question is good for the application of  {\sc OPTICS}. The closer to 0 this parameter is, the better.
\[
p = \frac{\delta_C^{intra-cluster}}{\delta_C^{inter-cluster}}
\]
This metric is extremely sensitive to the number of clusters and to the clusters themselves used to compute it. For instance, we get different values if we only consider the most massive clusters or if we include all of them.
An example of the values obtained for APOGEE for every element is in Table \ref{tab:metric}.

\begin{table}[H]
    \centering
    \begin{tabular}{|l|l|l|}\hline
        Abundance& $p$&Stars\\
        \hline
        \hline
        [C/Fe] & 0.68 &3539\\
        \hline
        [N/Fe] & 0.76 &3547\\
        \hline
        [O/Fe] & 0.93 &3541\\
        \hline
        [Na/Fe] &0.84 &2841\\
        \hline
        [Mg/Fe] & 0.55 &3548\\
        \hline
        [Al/Fe] & 0.72 &3509\\
        \hline
        [Si/Fe] & 0.49 &3548\\
        \hline
        [Ca/Fe] & 0.82 &3476\\
        \hline
        [Ti/Fe] & 0.73 &3319\\
        \hline
        [V/Fe] & 0.94 &3073\\
        \hline
        [Cr/Fe] & 0.96 &3354\\
        \hline
        [Mn/Fe] & 0.62 &3012\\
        \hline
        [Ni/Fe] & 0.83 &3482\\
        \hline
        [Ce/Fe] & 0.83 &3160\\
        \hline
        [Fe/H] & 0.20 &3548\\ 
        \hline 
        [M/H]&0.24 &3548\\
        \hline
        [$\alpha$/M]&0.68 &3548\\
        \hline
    \end{tabular}
    \caption{Metric values obtained for our elements in APOGEE, the closer the metric is to 0, the better. The last column indicates the number of stars in which abundances where measured. Note that the total number of stars present in the APOGEE dataset is 3548.}
    \label{tab:metric}
\end{table}
The table clearly illustrates that iron abundance [Fe/H] and global metallicity [M/H] are the most effective quantities in distinguishing clusters, as they are available for all stars in both datasets and their metric is the closest one to zero, indicating their ability to distinguish different populations. 
[$\alpha$/M] is also quite good to separate different structures. In addition it is available for all stars of the sample. 
Out of the abundances measured by APOGEE, those that have a smaller value than the metric are the abundances of individual $\alpha$ elements (particularly Mg and Si), and the average [$\alpha$/M]. In fact, we expect that compared to iron-peak elements, which behave like Fe independently of SFH, these elements allow clusters from different regions or galaxies to be distinguished. 
For APOGEE, we opted to use [M/H] and [$\alpha$/M] because they have good metric values, are available for the entire sample, and have lower uncertainties than individual abundances. 

In {\em Gaia}-ESO, as in APOGEE, the best metric is obtained for metallicity, expressed as [Fe/H]. In this case, the $\alpha$-element abundances are not provided as an average quantity. The $\alpha$-element abundance with lower uncertainties, available for most of the sample, and with a good metric value is Mg. We included in our analysis also [Cr/Fe], an iron-peak element present for a high number of stars.

In summary, the selected abundance ratios are [$\alpha$/M] and [M/H] for APOGEE and for {\em Gaia}-ESO [Mg/Fe], [Fe/H] and [Cr/Fe].
Once we choose the chemical elements to use in our analysis, we select only the stars that have no flag for those elements and that have a small error on them and this further reduces the dataset.
The choice to use a limited number of abundances is justified by the fact that we are approaching this study as a preliminary investigation into a new method of analysis. We want to show that the method is reliable and works but we do not aim at making an extensive use of it, since available datasets are still small. Future surveys, e.g. 4MOST and WEAVE, with more precise measurements and a higher number of stars will allow us to apply this method on a broader and more significant set of abundances.

\subsection{The Algorithm Structure}

\begin{figure}
    \centering
    \includegraphics[width=1\linewidth]{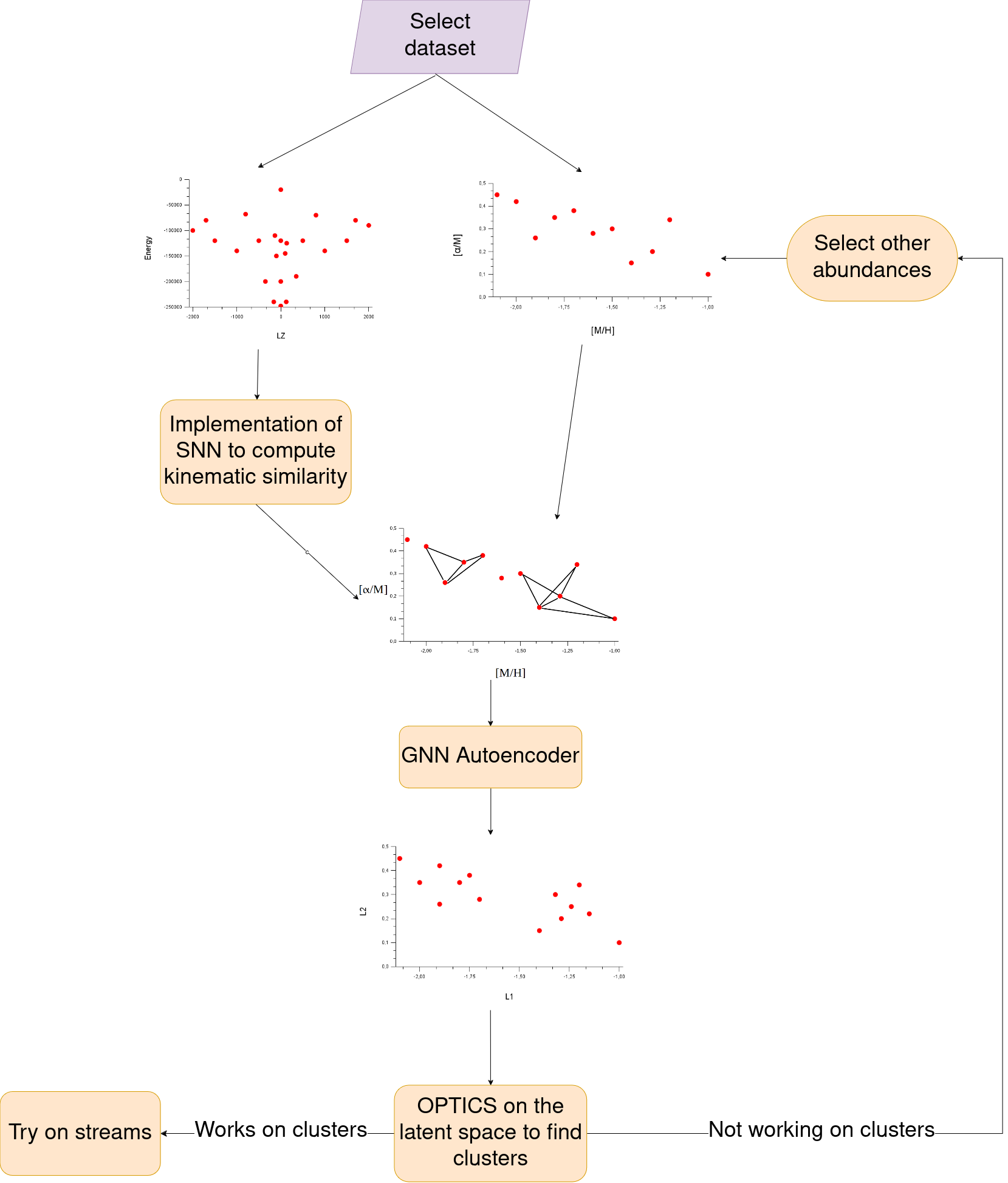}
    \caption{Flowchart of our algorithm used for deep clustering.}
    \label{fig:chart}
\end{figure}

In this section we describe the procedure followed for the data elaboration. 
A simple flowchart that represents the final structure of the algorithm used for deep clustering, named the creek ( {\underline C}lustering with g{\underline R}aph n{\underline E}ural n{\underline E}tworK),  is depicted in Fig.~\ref{fig:chart}.
The name `the {\sc creek}' also recalls the structure of the streams, which appear as rivers or creeks in the halo of the MW.
%

The standard analysis for a chemical clustering consists in applying a clustering algorithm (e.g. {\sc OPTICS}) directly to the abundances of the samples of stars. As explained in Chapter \ref{chap2}, OPTICS groups stars based on the distance between some properties, for instance,  abundances.
However,  whatever the element combinations may be, we find that OPTICS alone is not able to recover most of the clusters. In fact, although homogeneous, clusters are often superimposed one to  each other in the chemical space. Furthermore, the quality of the abundances introduces a significant noise in the recovering factor. Indeed, with current data quality, chemical abundances alone are not enough to find structures dispersed in the halo.

\subsection{OPTICS combined with a Siamese neural network}
\begin{figure}
    \centering
    \includegraphics[width=0.9\linewidth]{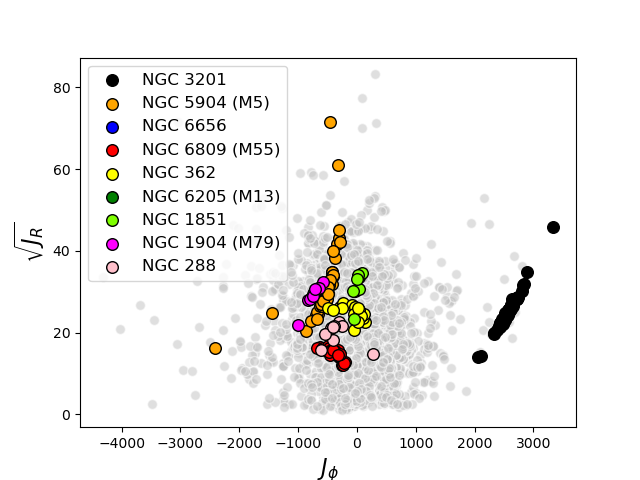}
    \caption{Location of  a few clusters in the $\sqrt{J_R}$ vs $J_{\Phi}$ plane.}
    \label{fig:JrJphiclusters}
\end{figure}

The inability to find clusters based only on the direct application of {\sc OPTICS} leads to the need of a new method that incorporates dynamical information and allows us to recover more clusters based on their chemistry.
We used the orbital properties of our sample of stars to implement a Siamese Neural Network (SNN) using kinematic information.
Although, as we showed in the previous chapter, clusters are well separated in the Toomre diagram, we prefer to avoid to use their current velocities, because they are not invariant and are not good discriminators for identifying mergers or streams. 
We used instead their standardized actions $J_R$, $J_{\phi}$ and $J_Z$,  the three components of angular momentum in cylindrical coordinates. They are exactly analogous to the three peculiar velocities used in the Toomre diagram in their ability to separate the different clusters, as shown in Fig.~\ref{fig:JrJphiclusters}, but they can be used to recover structures that lost their coherence in space and in velocity, as streams, since they should be almost invariant. 

\subsubsection{The implementation of the SNN}

\begin{figure}
    \centering
    \includegraphics[width=0.9\linewidth]{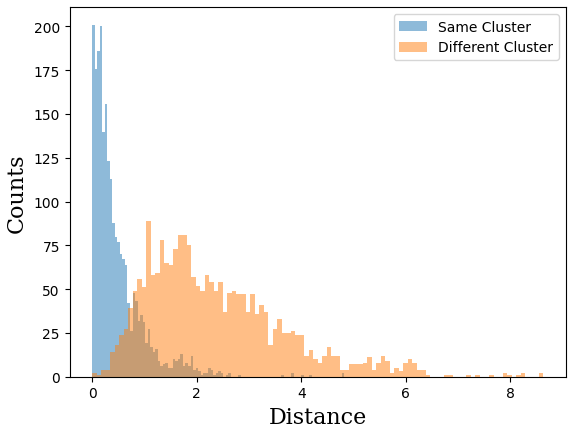}
    \caption{Histogram of the distances for stars belonging to the same cluster (in blue) and stars belonging to different clusters (in yellow).}
    \label{fig:histcluster}
\end{figure}

The process consists in selecting a random number of stars from a GC to create a training set. We select n couples of stars that are in the same clusters and n couples of stars from different clusters. They have to be in the same number as this creates a better and more balanced training sample.
The SNN receives as its input the orbital parameters of the two stars and it is trained to predict whether the couple belong to the same association or not. The SNN used is composed of two branches. Each branch is formed by three hidden layers respectively with 32, 64 and 128 neurons and all with a rectified linear unit (ReLU)  activation function. The training set and the validation set are selected at 80\% and 20\% each. As we saw in Chapter~\ref{chap2}, the two branches of the SNN share the same weights and biases. The outputs of the two branches are then used to calculate a distance metric. That distance is then given as an imput to a sigmoid activation function, which maps that metric into the probability that the two stars belong to the same cluster: it will be 1 or near to 1 if stars are from the same cluster and it will approach 0 if stars are not from the same cluster.\newline
We can now plot the two distributions of distances in a histogram as shown in Fig.~\ref{fig:histcluster} and see that the intra-clusters stars, (light blue in the plot) are compact and generally nearer, whether inter clusters stars (yellow in the plot) are usually far away from each other. \newline
The sigmoid activation function is particularly useful because it allows us to vary the threshold value at which we consider a star as belonging to the same cluster depending on what we aim to observe. We can then consider low probabilities when we look for streams or stars from disrupted clusters or high probabilities when we look for the cores of the clusters.
The stars belonging to the same clusters according to this probability are the ones we consider linked when we create the graph for the next neural network. The threshold value used in this work to consider stars as linked is 0.8.\newline
One important detail to note is that the number of links in a set of points all inter-connected increases as a power law of the number of points. Even though not all points are connected in our graph, the complexity increases rapidly with the amplitude of our dataset. This implies that also the computational power required run creek is expected to grow with the size of the dataset we want to analyze.

\subsubsection{Abundances in Graph Neural Networks (GNN) autoencoder}
The stellar abundances and the links established through the SNN, indeed, will be the inputs of the graph auto-encoder. This latter maps the chemical space into a latent space which also takes into account the kinematic similarity between the stars in our dataset. More specifically, instead of processing the simple chemical abundances, the graph auto-encoder propagates across the whole network the averages chemical abundances (see Section~\ref{GNNs}). Therefore, two stars that have very similar orbital parameters, should also share a similar location within the latent space. That would make stellar associations denser in the latent space than what they are in the simple chemical space, hence also easier to be identified by a clustering algorithm.
The input data need to be standardized. If data are not standardized the machine learning process is slower, as the gradient of the loss function will be steeper on one dimension and flatter on the other: the number of steps required will be higher to adjust one dimension and smaller for another. We then implement a standardisation for every abundance used.
The graph is then passed to the auto-encoder, composed of five hidden layers whose number of neurons was varied. The attempted number of neurons in the hidden layers where 500 and 300 for the outer layer, 100, 200 or 300 for the inner layers and 3 or 5 for the latent space. The final structure is composed of layers of 300, 100 and 5 neurons. The epochs used for this part are 800, even though 500 and 1000 where also tried but deemed worst based on their compilation time and their results. The auto-encoder was supplied with a learning rate scheduler with a patience of 50 and a factor of reduction of 0.5.  The auto-encoder was also provided with an early stopping callback, that acts with a patience of 10. These values were chosen empirically based on the various exploratory tests we carried out.
It is interesting to note that even when the abundances considered were two, an auto-encoder with a five-dimensional latent space was more effective in the recovering of clusters.

\subsubsection{Recovering clusters with OPTICS}
{\sc optics} is then applied to the latent space of this auto-encoder, where information is simplified. Also in this case the latent space was standardized before being used for the clustering analysis. Standardization was important to give all the dimension the same weight.
{\sc optics} produces then a reachability plot, where we can study and observe its dips and compare them with the actual clusters in the dataset.
In the next chapter, we show how we applied the algorithm to the two halo samples. We show the recovering factory on the sample of globular clusters, and then show the limits and potentials in finding new streams and recovering those known in the literature. 

\subsubsection{Computation limits }

The problem we faced is a complex one, which needs appropriate instruments. Ours work has been a first  feasibility study, often dealing with limitations due to machines and computing time. 
The compilation times, indeed,  increase rapidly with the number of stars in our dataset. The computation of orbits itself requires long times, although the code used in the present  work is one of the fastest. The creation of the graph and the use of the auto-encoder also require time and computational power, especially in terms of RAM and GPU, that increases with the dimension of the dataset.
In the future, we aim to ask for computing time in super computers, as the CINECA\footnote{https://www.cineca.it/} ones, in order to extend the application to larger databases. 

%% file: 5-Analysis.tex
\section{Results of the analysis: application of deep clustering to  the APOGEE and {\em Gaia}-ESO datasets}
\label{chap5}

In this chapter we apply the deep clustering algorithm, the {\sc creek}, described in Chapter ~\ref{chap4} to our datasets. We analyse the retrieval of both clusters and streams for APOGEE and {\em Gaia}-ESO. 

\subsection{APOGEE}

\begin{figure}
    \centering
    \includegraphics[width=1\linewidth]{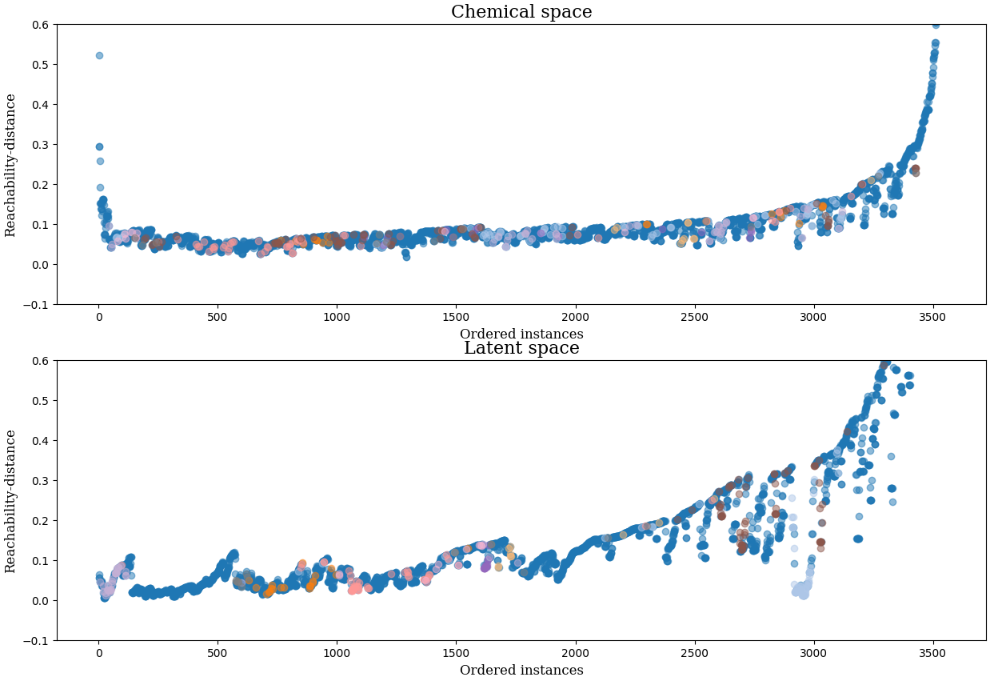}
    \caption{Reachability plots (reachability distances vs ordered instances): In the upper panel {\sc optics} is applied directly to the chemical space, in the lower panel it is applied through the {\sc creek} to the latent space of the auto-encoder. Blue dots are the field stars, and coloured dots are stars belonging to clusters.}
    \label{fig:ExReachPlot}
\end{figure}

The {\sc creek} is first applied to the APOGEE data.
Once the structure of the NNs is fixed, the number of clusters recovered can slightly vary with every iteration depending on the single training. By repeating the process several times, we can find the average number of recovered clusters.  \newline
The {\sc creek}, with {\sc optics}, identifies high-density groups of stars within the latent space. Therefore, we can evaluate the performance of the algorithm by comparing these groups with the real stellar associations. More specifically, we define the metrics \textit{homogeneity} $H$ and \textit{completeness} $C$, that are computed as it follows
\begin{equation}
    H = \frac{Number\;of\;cluster\;stars\;included\;in\;group}{Number\;of\;stars\;in\;group}
\end{equation}
\begin{equation}
    C = \frac{Number\;of\;cluster\;stars\;included\;in\;group}{Number\;of\;stars\;in\;cluster}
\end{equation}
where ``cluster'' is the astrophysical entity defined using a membership analysis based on independent quantities, as proper motions and radial velocities, while ``group'' is the set of stars with similar properties assembled by the {\sc creek}.  
We consider a cluster as recovered if an {\sc optics} group has both homogeneity and completeness over 50\%: $H\geq 0.5$ and $C\geq0.5$. An example of reachability plot is in Fig.~\ref{fig:ExReachPlot}.
In this figure we compare the results of the analysis with the {\sc creek} and with {\sc optics} alone. 
We can see in its upper panel, where {\sc optics} is applied directly to the chemical space (considering as elements $\alpha$-elements and metallicity), {\sc optics} is not able to re-identify clusters. In fact, a stellar cluster must fall within one of the dips of the reachability plot in order to be identified. Therefore, if a cluster is recovered as a group, the corresponding dip would be deep and uniformly coloured with the colour associated to that specific cluster.  
In this case, instead, stars belonging to the same cluster (i.e., stars having the same colour) are spread all over the plot and dips are not deep. The lower panel instead, where {\sc optics} is applied to the latent space of the auto-encoder using the {\sc creek}, shows deeper dips and has coloured stars much more grouped and concentrated into the dips of the reachability plot. 
Even before computing the number of recovered clusters with their homogeneity and completeness, we notice that clusters are better separated and more easily identifiable.

\subsubsection{Globular Clusters in APOGEE}

\begin{figure}
    \centering
    \includegraphics[width=1\linewidth]{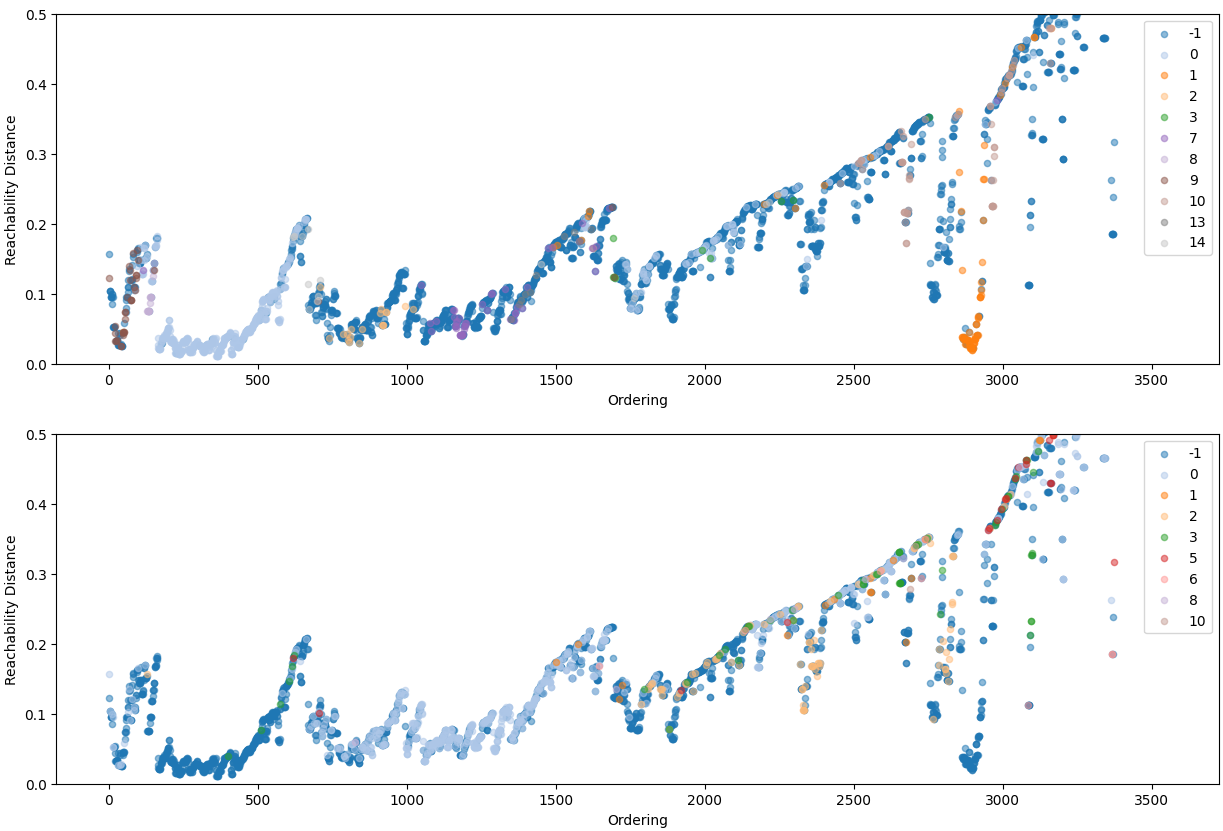}
    \caption{The upper panel shows the reachability plot of the APOGEE dataset colour-coded by clusters, the lower panel is colour-coded by streams as shown in the legend. -1 indicated field stars}
    \label{fig:ReachabilityApogee}
\end{figure}

We applied the {\sc creek} to the  APOGEE dataset to test the recovering of  GCs through the application of {\sc optics} to the latent space of the auto-encoder. Depending on the parameters of {\sc optics} and on the auto-encoder processing we can recover a variable number of clusters. We now consider the realization with {\sc optics} parameters xi = 0.001 and min\_sample = 7. 
A value of  xi = 0.001 indicates that the steepness of a dip can be quite small for it to be considered a cluster and the change in reachability distance between neighboring points is relatively gradual.
The reachability plot obtained for this configuration is shown in Fig.~\ref{fig:ReachabilityApogee}.

{\sc optics} produces a long hierarchy of groups and sub-groups, of which 4 correspond within the $H$ and $C$ criteria established above to real clusters.  The recovered clusters are in Table \ref{tab:Clrec05}: {\sc optics} recovers 4 of the 10 clusters with more than 10 members at 50\%.
If we relax the criteria of  $H$ and $C$ to 30\% instead, we manage to recover 8 out of the 10 clusters selected. 

\begin{table}[H]
    \centering
    \begin{tabular}{|c|c|c|l|} \hline 
         Cluster&  Number& Homogeneity &Completeness\\ \hline 
         Omega Centauri&  0&  0.94&0.65\\ \hline 
         NGC 6656 (M22)&  1&  0.67&0.77\\ \hline 
         NGC 6121 (M4)&  3&  0.5&0.5\\ \hline
         NGC 6544& 8& 0.54&0.93\\ \hline
         NGC 362& 2& 0.32&0.30\\ \hline 
         NGC 6809 (M55)&  9&  0.31& 0.86\\ \hline 
         NGC 5904 (M5)&  7&  0.40& 0.43\\ \hline 
         NGC 3201&  10&  0.56& 0.33\\ \hline
    \end{tabular}
    \caption{Table of recovered clusters. The numbers refers to  Fig.~\ref{fig:ReachabilityApogee}}
    \label{tab:Clrec05}
\end{table}



\begin{figure}
    \centering
    \includegraphics[width=1\linewidth]{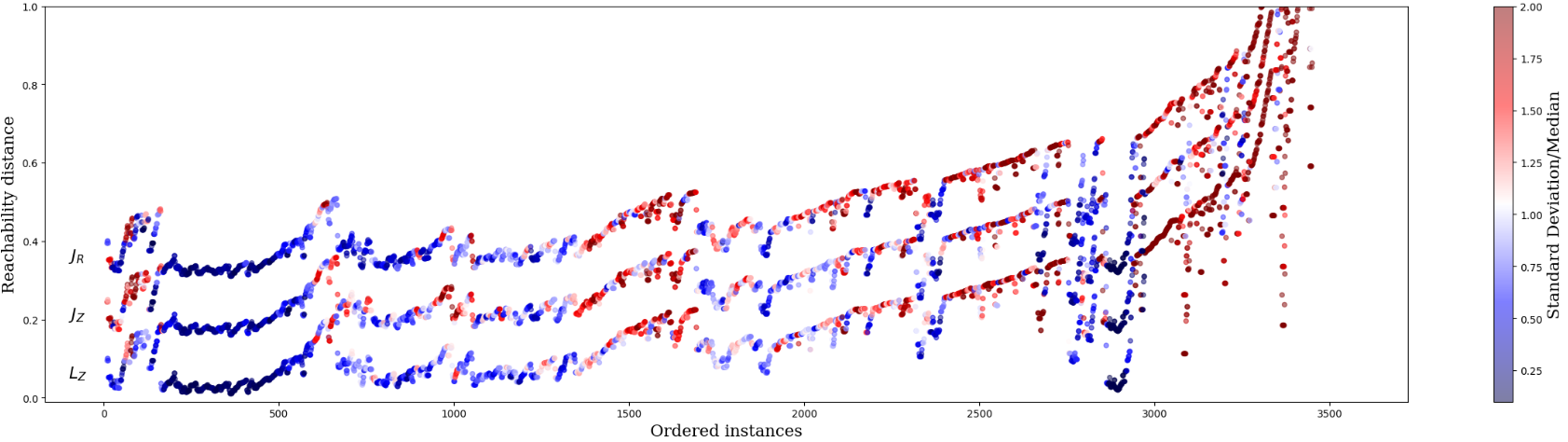}
    \caption{Reachability plot of the APOGEE dataset. The points are 
    colour-coded according to the ratio between the standard deviation and the median over segments of length 10 in $L_Z$, $J_R$ and $J_Z$. Blue (red) stars have a lower (higher) standard deviation. The three variables are plotted with a vertical offset to avoid overlapping.}
    \label{fig:std}
\end{figure}

Another important information can be gathered from Fig.~\ref{fig:std}. In this figure, the colours are attributed to instances based on the ratio between the standard deviation and the median over segments of length 10 in three different parameters: $L_Z$, $J_R$ and $J_Z$. It allows us to see that most dips in the reachability plot have a lower standard deviation in this three variables compared to stars which are located outside the dips. This means that the groups identified by the {\sc creek} correspond to stars with a lower spread in the action space, thus well clustered in that space,  confirming the good retrieval of clusters. On the other hand, not all dips are characterized by a lower standard deviation, which is also expected as there must exist random density fluctuations in the latent space which are not directly associated to groups of stars moving together.

\subsubsection{Stellar streams in Apogee}

\begin{figure}
    \centering
    \includegraphics[width=1\linewidth]{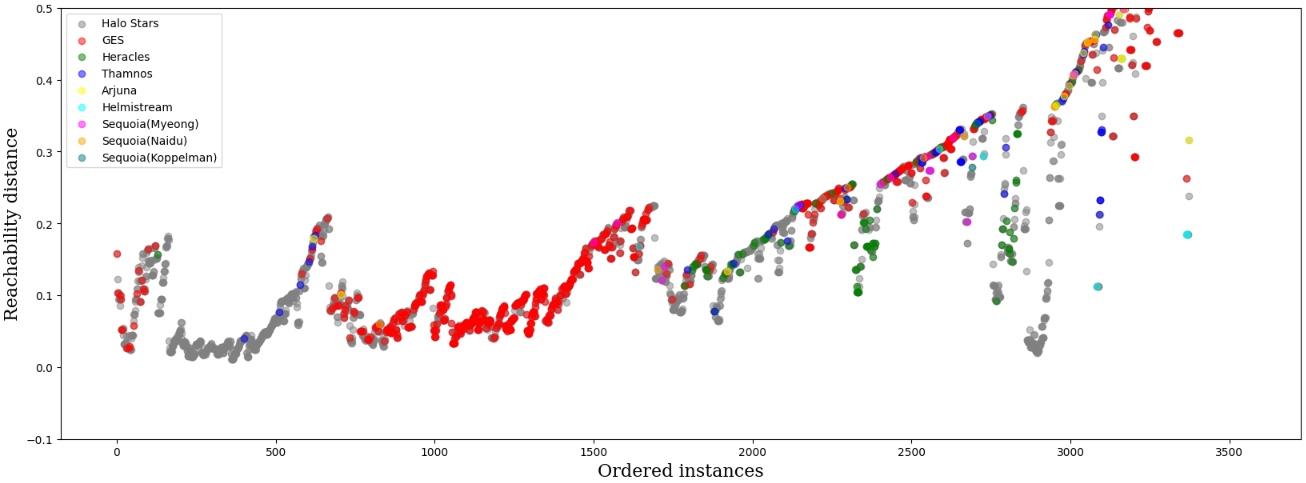}
    \caption{Reachability plot for the stellar streams in the APOGEE dataset. Streams are colour-coded as in the legend, while field stars are in grey.}
    \label{fig:stramreach}
\end{figure}

Since we have shown that we can find 80\% of clusters to at least 30\% in completeness and homogeneity, we can now tackle the search for non-spatially coherent structures, such as stellar streams or merger remnants. 
We can investigate how some well-studied streams are located in the reachability plot and if they fall in its dips.
We recall that streams have been selected according the selection of  \citet{Horta2023}. 
As streams may also not be completely chemically homogeneous, having formed from galaxies or clusters with possibly an extended SFH, instead of computing the effective recovery factors as we did for GCs, we just colour their location in the reachability plot, as shown in Fig.~\ref{fig:stramreach}. Though they are more sparse compared to clusters, streams are still positioned in the dips of the reachability plot and they can be distinguished from GCs. They, indeed, occupy different dips than GCs in the reachability plot. 
In what follows, we will describe some of the most interesting streams found with the {\sc creek}.

\paragraph{Gaia-Enceladus (GES)}

\begin{figure}
    \centering
    \includegraphics[width=1.1\linewidth]{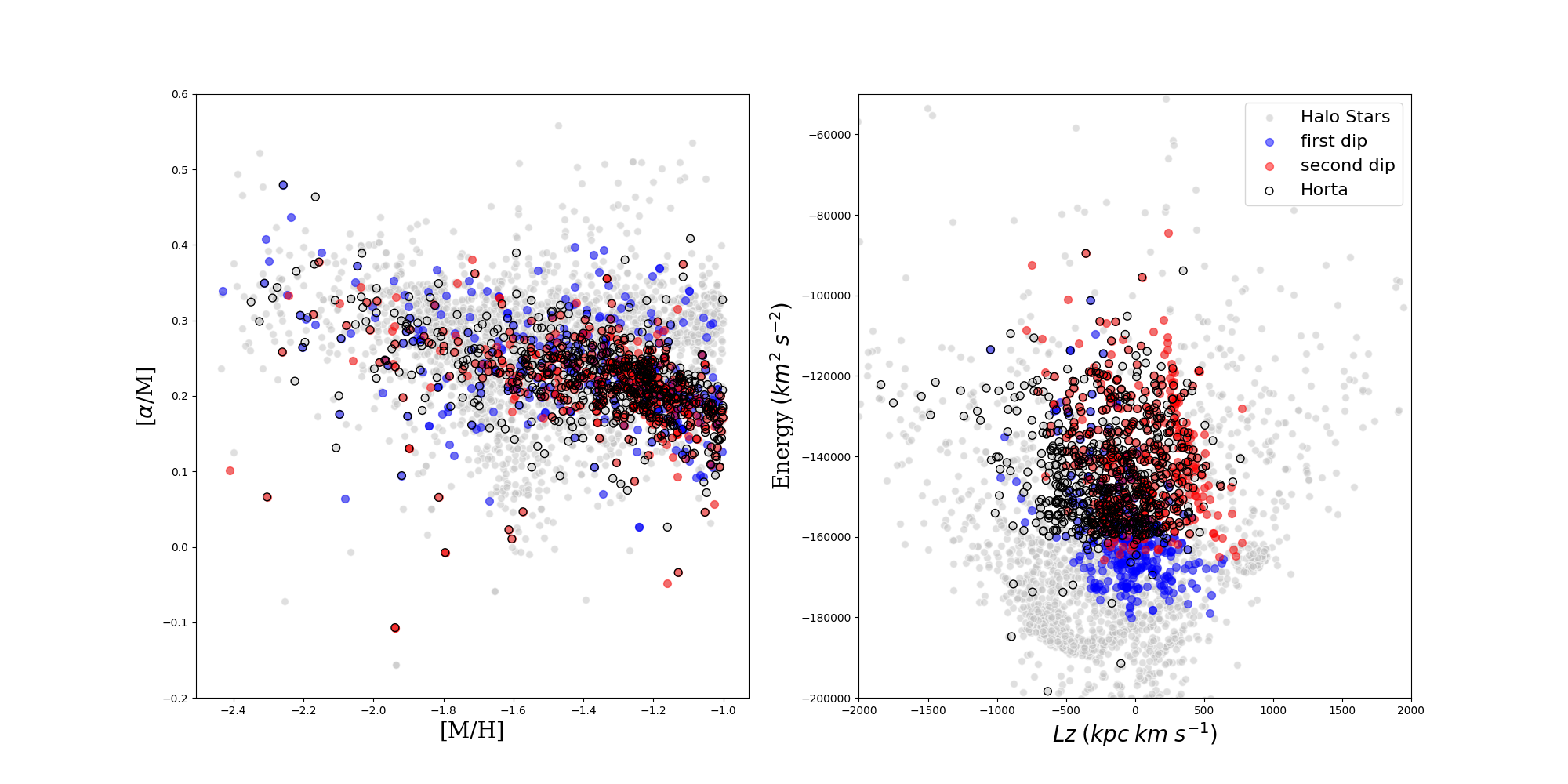}
    \caption{Left panel-- [$\alpha$/M] vs. metallicity and right panel-- Lindblad diagram for the stars in the two dips corresponding to GES. The blue stars in the dip correspond to ordered instances between 667-992, while red are stars in the dip between 993-1438. The black empty circles are the stars belonging to GES according to \citet{Horta2023}. Smaller symbols are halo stars not in GES. }
    \label{fig:dipGES}
\end{figure}

\begin{figure}
    \centering
    \includegraphics[width=0.9\linewidth]{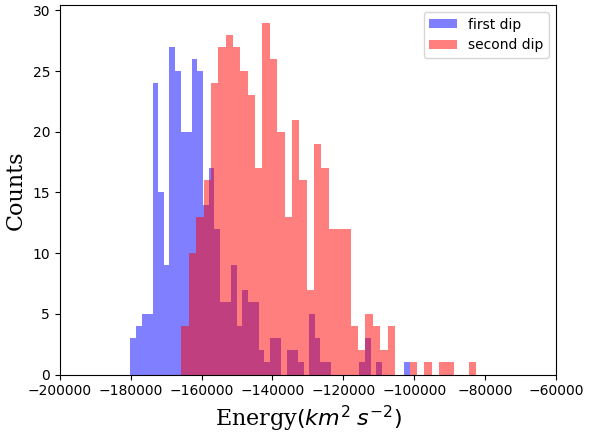}
    \caption{Energy distribution of the two dips of GES, colour-coded as in the legend.}
    \label{fig:Energyhistogram}
\end{figure}

\begin{figure}
    \centering
    \includegraphics[width=0.95\linewidth]{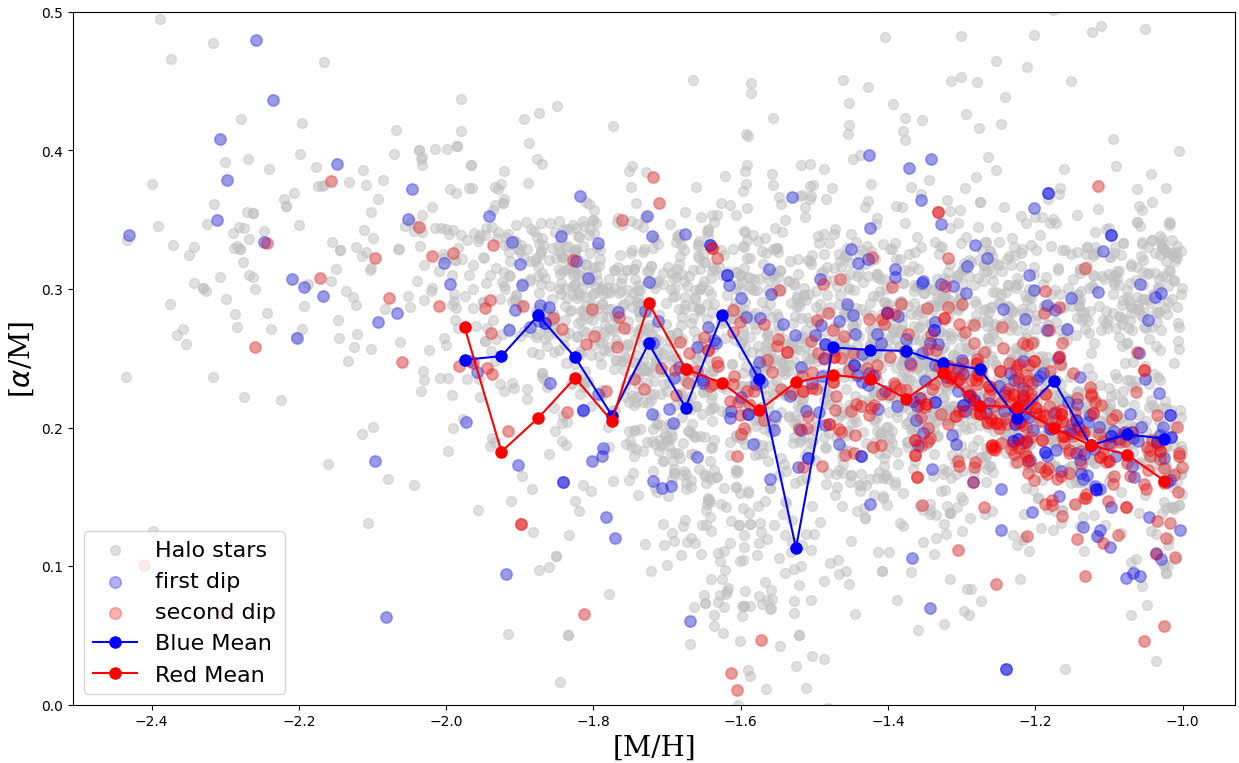}
    \caption{[$\alpha$/M] vs [M/H] in APOGEE. The halo sample is represented with grey circles, the GES first dip population with semi-transparent blue circles and the second dip in red.  Mean abundance of the $\alpha$-elements of the first and the second dip of GES are shown with blue and red circles, connected by continuous lines. }
    \label{fig:AlphaGES}
\end{figure}

\begin{figure}
    \centering
    \includegraphics[width=0.95\linewidth]{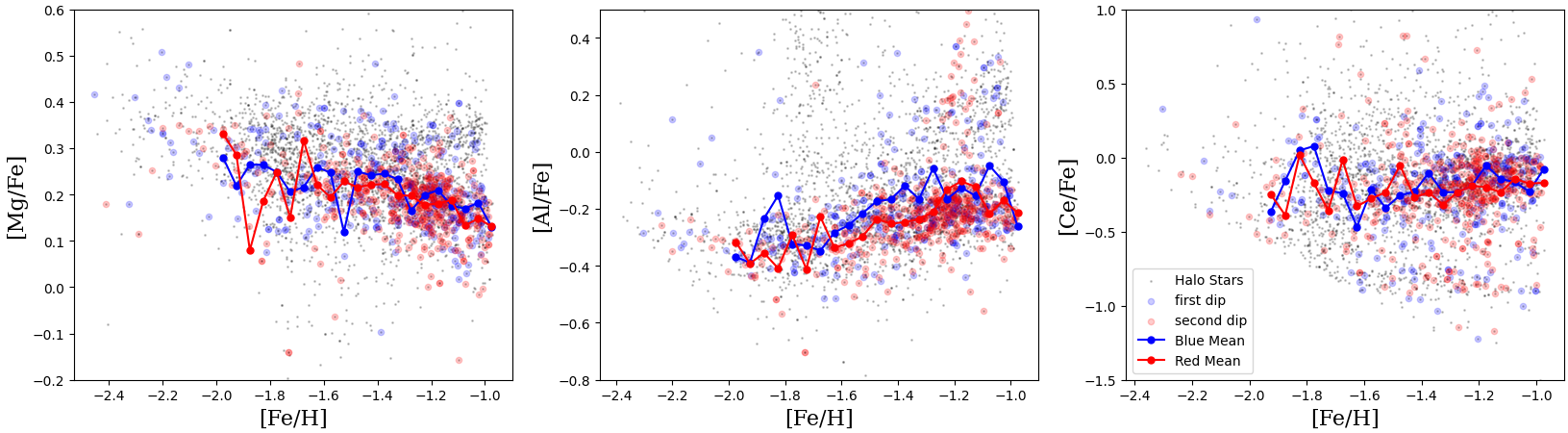}
    \caption{Abundance ratios of Mg, Al and Ce over Fe as a function of [Fe/H]. Stars in the two dips of GES are shown with  semi-transparent blue and red circles. In the same colours, we plot with filled circles the mean binned abundances with continuous lines for  the two dips.}
    \label{fig:ElementsGES}
\end{figure}

\begin{figure}
        \centering
    \includegraphics[width=0.95\linewidth]{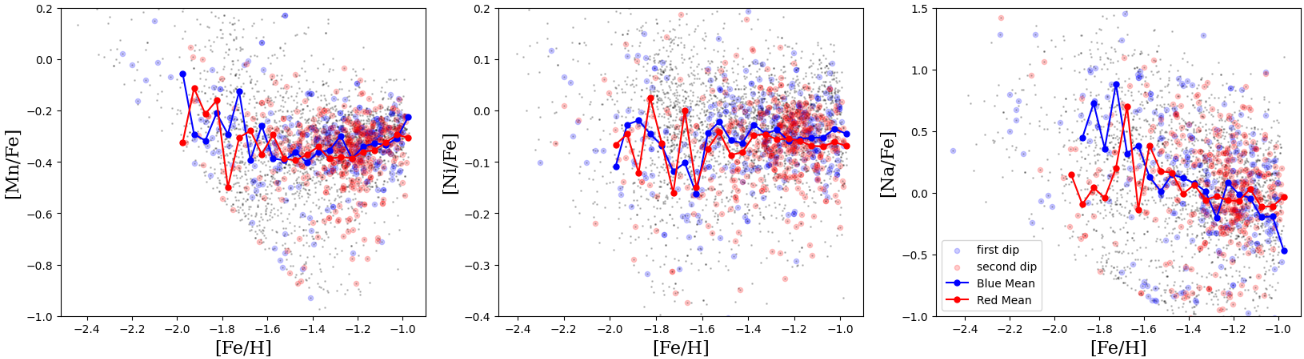}
    \caption{Abundance ratios of Mn, Ni and Na over Fe as a function of [Fe/H]. Stars in the the two dips of GES are shown with  semi-transparent blue and red circles. Mean binned abundances are shown with larger circles connected with continuous lines in blue and red for the two dips of GES for Mn, Ni and Na.}
    \label{fig:Elements2GES}
\end{figure}

GES is particularly well grouped in a wide dip of the reachability plot, and we note that it appears to be composed of two sub-groups according to {\sc optics} (667-992 and 993-1438 in the ordered instances). This is not a single instance as it happens every time we implement the learning algorithm.\newline
With the aim of finding the differences between the two dips of GES, we plot them in both the [$\alpha$/Fe] vs [Fe/H] plane  (left panel) and in the Lindblad diagram (right panel) in Fig.~\ref{fig:dipGES}. We notice that the first dip corresponds to stars with lower energy and includes a significant amount of stars that are cut off from the selection given by \citet{Horta2023}. In Fig.~\ref{fig:Energyhistogram}, we can see that the distributions of energy for the two dips are peaked at different values. In Fig.~\ref{fig:AlphaGES} we see that the mean abundances of $\alpha$-elements are slightly higher for stars in the first dip of the reachability plot.
We then check abundances for other elements in order to understand whether they indicate two completely different chemical  structures. In Fig.~\ref{fig:ElementsGES} and in Fig.~\ref{fig:Elements2GES} we show the abundances of Mg, Al, Ce, Mn, Ni and Na for the two GES substructures. We note that $\alpha$-elements and Al tend to be quite uniformly separated, with the first dip having consistently higher abundances than the second dip. Other elements, instead, appear more homogeneous in the two dips. This probably depends on the fact that the {\sc creek} used $\alpha$-elements and [M/H] to separate the {\sc optics} groups.\newline
The difference between the two sub-groups is also verified through the application of the two-sample Kolmogorov-Smirnov test \citep{Kolmogorovan1933sulla}, a non parametric test used to find whether two samples come from the same unknown distribution. We assume the null hypothesis to be that the two samples, in our case abundances for the two sub-groups, come from the same distribution. The significance level, which is the threshold for determining whether to reject the null hypothesis, is 0.05 and corresponds to a 5\% chance of rejecting the null hypothesis when it is actually true. We find that the two sub-groups can be  considered from two separate distributions for $\alpha$-elements, [M/H], [Mg/Fe], [Ni/Fe], [Ce/Fe], [O/Fe], [Si/Fe], [Al/Fe]. They come from the same distribution for [Mn/Fe], [Na/Fe] and [Cr/Fe].
Therefore, it is likely that what appears as a unique structure in the selection provided by \citet{Horta2023} is actually composed of two separate structures, with different energies and different abundances.


\begin{figure}
    \centering
    \includegraphics[width=1.1\linewidth]{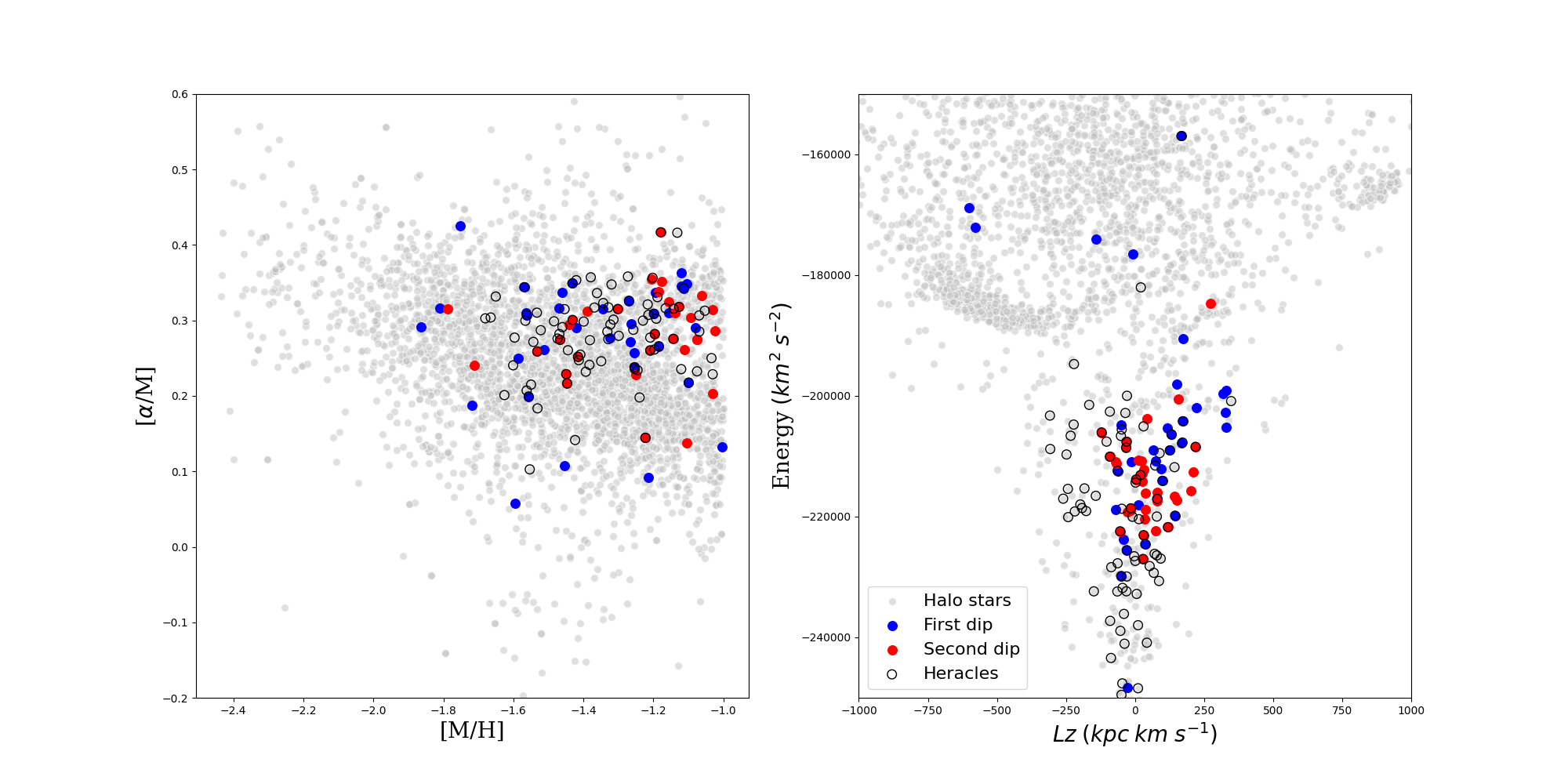}
    \caption{[$\alpha$/M] vs. metallicity (left) and Lindblad diagram (right) for the stars in the two dips corresponding to Heracles. In blue, we show  stars in the dip corresponding to ordered instances between 2318-2354, in red are stars in the dip between 2355-2389. The empty black circles are the stars belonging to Heracles according to \citet{Horta2023}.}
    \label{fig:HeraclesEnergy}
\end{figure}

\paragraph{Heracles}
Another interesting stream is Heracles, in green in Fig.~\ref{fig:stramreach}, situated between the ordered instances 2315-2399.  This stream is well separated into two dips, even though the energy and chemistry seem more intertwined, as we can see in Fig.~\ref{fig:HeraclesEnergy}. 
Additionally it appears that the {\sc creek} identifies a smaller percentage of stars for Heracles compared to GES, with some stars being grouped in the dip located around 2800.
Stars in this group are not enough to compute a mean for every dip at bins of metallicity as we did for GES.

\paragraph{Thamnos, Sequoia, Arjuna, Helmi streams}

These streams are composed of a smaller number of members in  the selection of \citet{Horta2023}. They do not appear clustered in either chemical or dynamic properties. Some of the stars of Thamnos fall in one of the dips around 3200, while the others are in distant positions. 
Some of the stars in Sequoia in the Myeong and Naidu selections fall into a dip around 1800, while others are spread in different dips. 

\paragraph{Other dips in the reachability plot}

\begin{figure}
    \centering
    \includegraphics[width=1\linewidth]{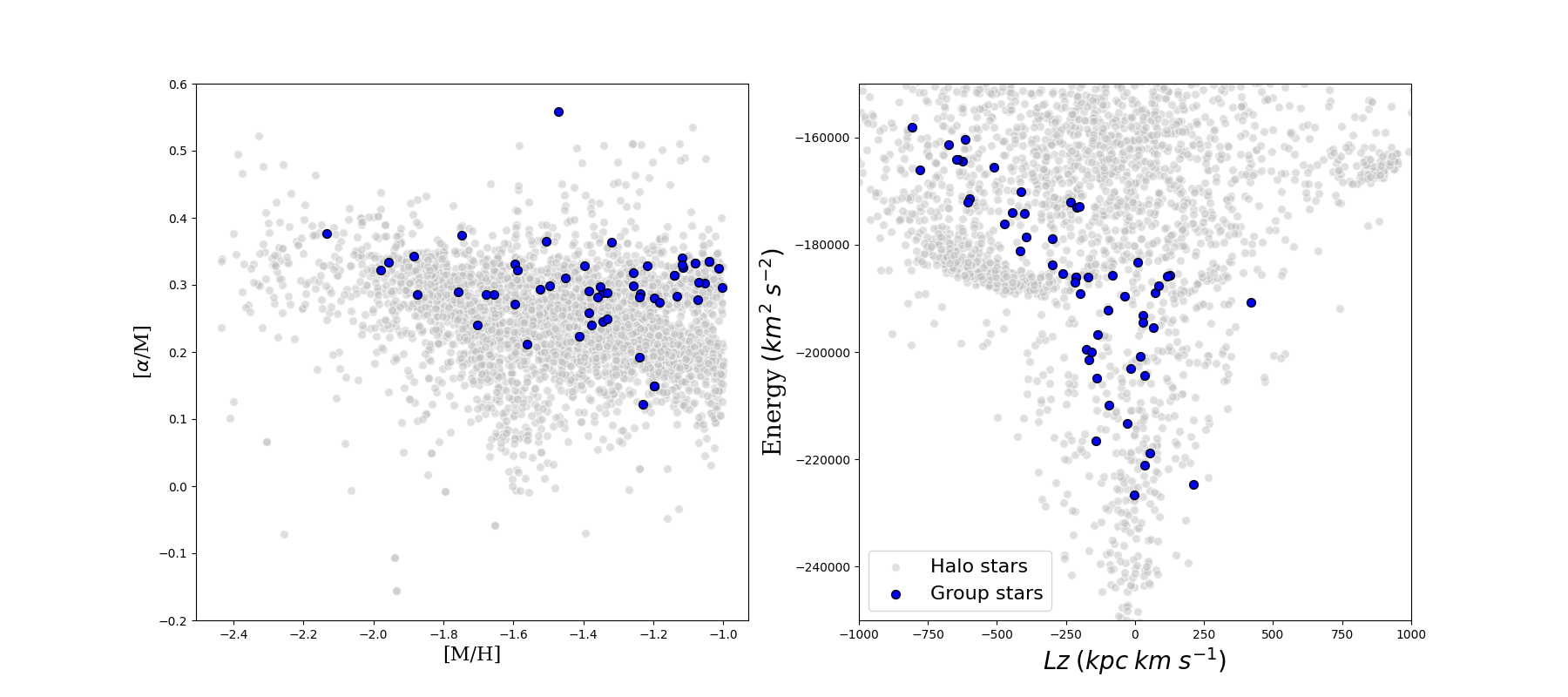}
    \caption{[$\alpha$/M] vs. metallicity (left) and Lindblad diagram (right) for the stars in the newly identified dip. In blue are stars in the dip corresponding to ordered instances between 1874-1931.}
    \label{fig:DipEnergy}
\end{figure}

Observing the most significant dips in the reachability plot, we also identify another group that seems interesting. This group does not show any particular intersection with clusters nor streams, but it contains some of the members of Sequoia.  
It is visible in Fig.~\ref{fig:stramreach} between 1874-1931. This dip shows quite clustered abundances and energies in the central part of the Lindblad diagram, as shown in Fig.~\ref{fig:DipEnergy}. It has a distribution similar to that of the empirical selection of Sequoia members \citep[see, e.g.][]{massari2019origin, Horta2023}, but it was attained in a fully automatic and consistent way taking into account kinematic and chemical properties.  

\subsection{{\em Gaia}-ESO}

The procedure is replicated for {\em Gaia}-ESO data. We first applied the {\sc creek} to globular clusters, and estimated the fraction of retrieved clusters. We then used the {\sc creek} on the halo sample to identify known streams and possibly find new ones. 

\subsubsection{Globular Clusters in {\em Gaia}-ESO}

\begin{figure}. 
    \centering
    \includegraphics[width=1\linewidth]{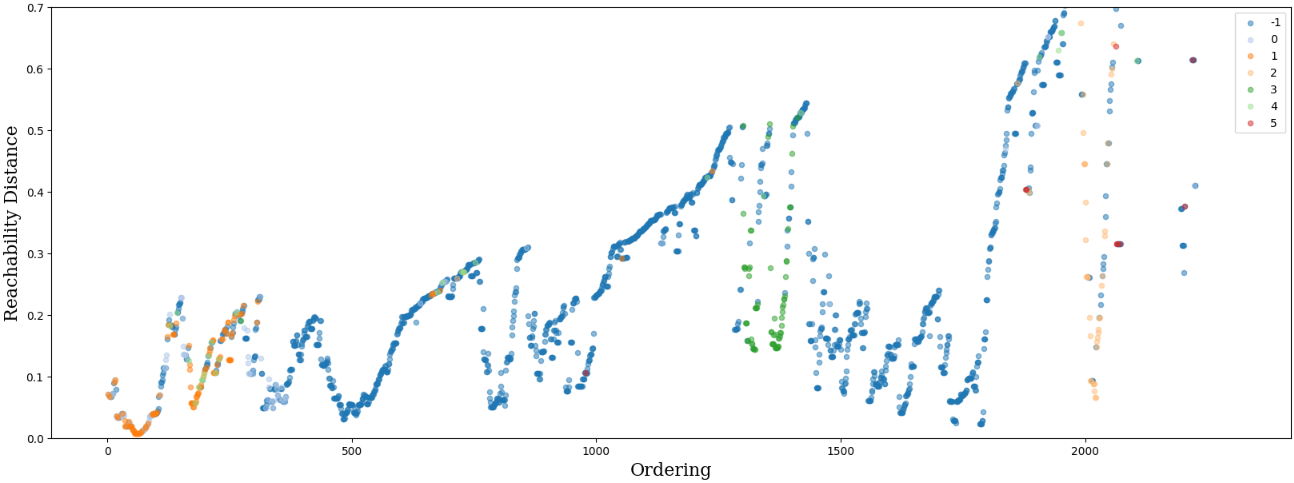}
    \caption{Reachability plot of globular clusters in {\em Gaia}-ESO. Clusters are colour-coded as in the legend, -1 indicates field stars.}
    \label{fig:ReachGaia1}
\end{figure}

\begin{figure}
    \centering
    \includegraphics[width=1\linewidth]{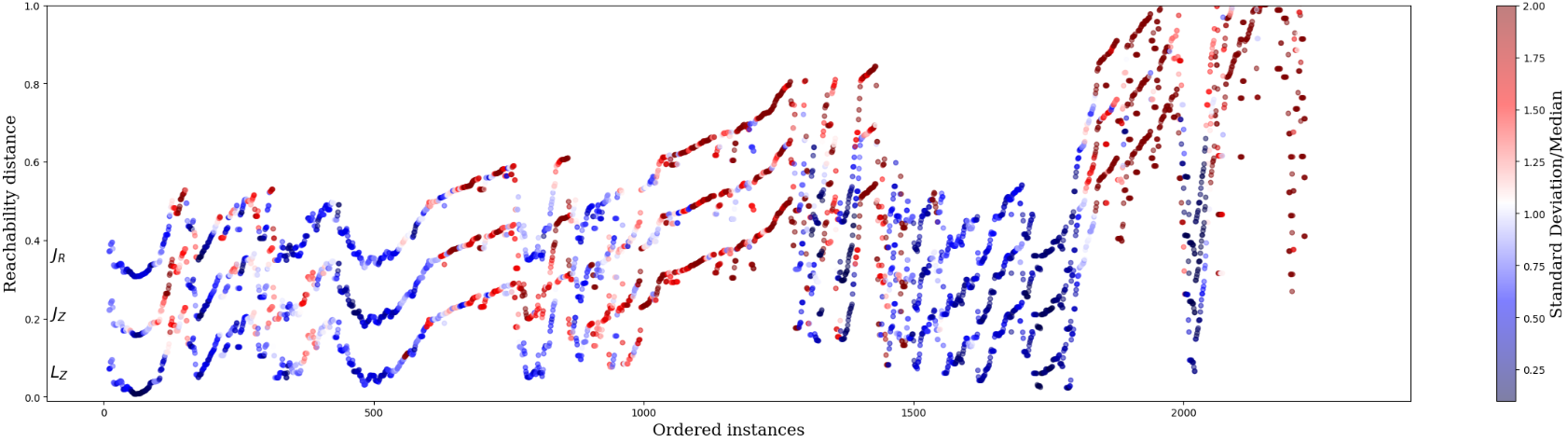}
    \caption{Reachability plot  colour-coded according to the ratio between the standard deviation and the median over a segment of length 10 of $L_Z$, $J_R$ and $J_Z$. Blue (red) stars have a low (high) standard deviation. The three variables are plotted with a vertical offset to avoid overlapping.}
    \label{fig:ReachGaia2}
\end{figure}

We applied the {\sc creek} to the {\em Gaia}-ESO data, implementing {\sc optics} on the latent space. The hyperparameters are still min\_sample = 7 and xi = 0.01 as they allow us to recover the highest number of clusters. 
The obtained reachability plot is shown in Fig.~\ref{fig:ReachGaia1}. In the hierarchy provided by {\sc optics} we recover 2 of the 6 real clusters with homogeneity and completeness over 0.5, as represented in Table~\ref{tab:ClrecGaia1}. If we relax the criteria of $H$ and $C$ to 30\%, we manage to recover 5 of the 6 clusters. These clusters are shown in the same Table~\ref{tab:ClrecGaia1}.\newline
We also show the reachability plot coloured by the ratio between standard deviation and median over segments of length 10 as done for APOGEE in Fig.~\ref{fig:ReachGaia2}. This plot shows that also for {\em Gaia}-ESO the groups identified by {\sc optics} correspond to stars with a lower spread in the action space.

\begin{figure}
    \centering
    \includegraphics[width=1\linewidth]{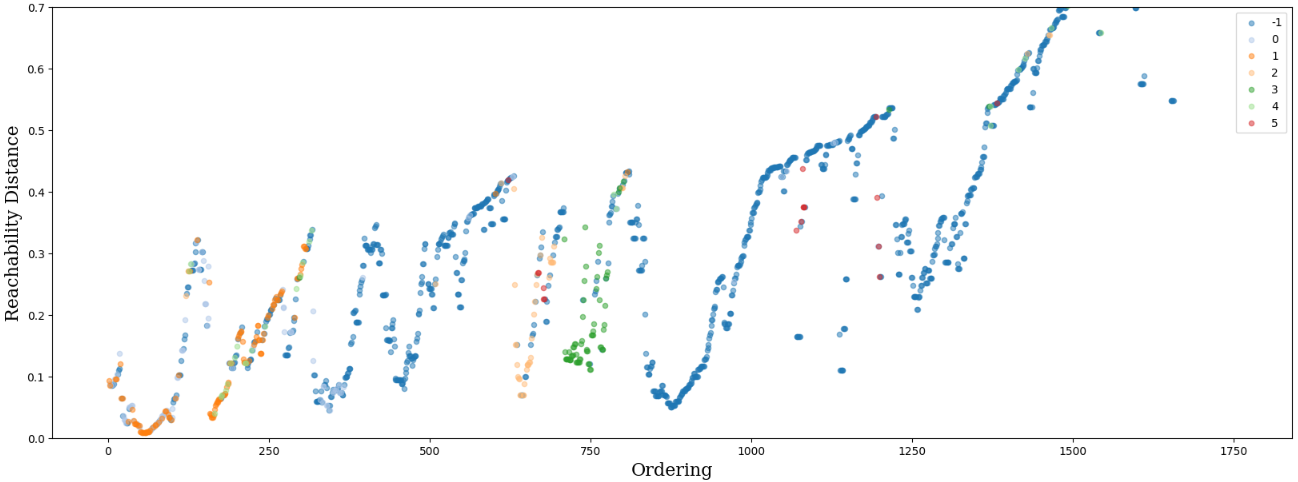}
    \caption{Reachability plot of globular clusters in {\em Gaia}-ESO using Mg, Fe and Cr. Clusters are colour-coded as in the legend, -1 indicates field stars.}
    \label{fig:ReachGaiaCR1}
\end{figure}

\begin{table}
    \centering
    \begin{tabular}{|c|c|c|c|} \hline 
         Cluster&  Number&  Homogeneity& Completeness\\ \hline 
         NGC 4833&  2&  0.66& 0.63\\ \hline 
         NGC 1904 (M 79)&  3&  0.79& 0.55\\ \hline
         NGC 362&  0&  0.41& 0.49\\ \hline 
         NGC 1851&  1&  0.5& 0.46\\ \hline 
         NGC 7078 (M 15)&  5&  0.31& 0.44\\ \hline 
    \end{tabular}
    \caption{Table of  clusters recovered with homogeneity and completeness over 0.3. Numbers refer to Fig.~\ref{fig:ReachGaia1}}
    \label{tab:ClrecGaia1}
\end{table}
We also test the addition of Cr abundance, to Mg and metallicity. In this way, we manage to slightly improve the performance, recovering three clusters with more than 50\%. 
The reachability plot is in Fig.~\ref{fig:ReachGaiaCR1}, where clusters are highlighted against field stars.\newline

\subsubsection{Stellar streams in {\em Gaia}-ESO}

\begin{figure}
    \centering
    \includegraphics[width=1\linewidth]{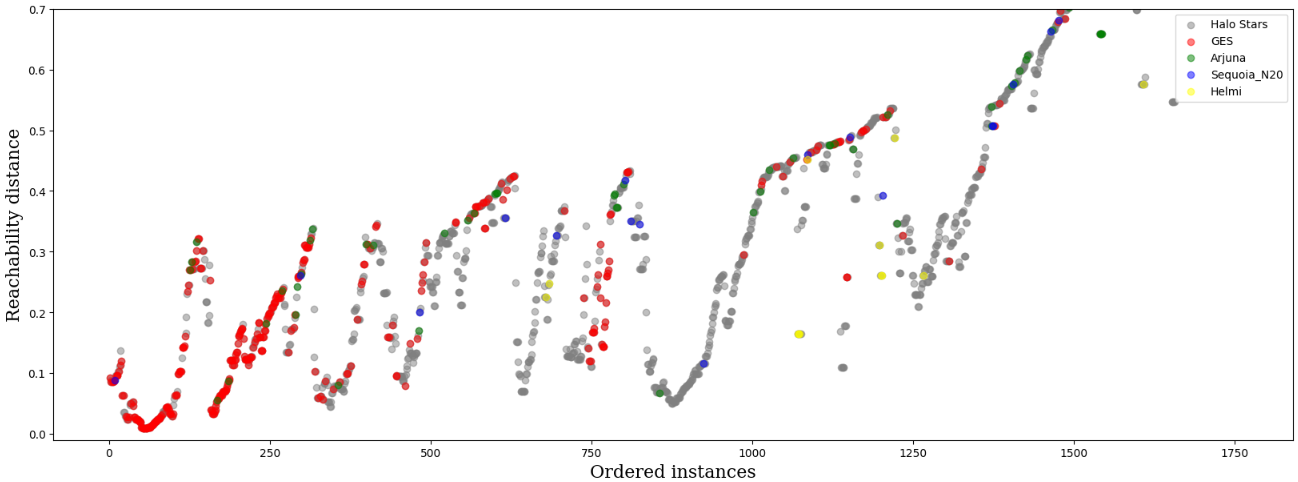}
    \caption{Reachability plot for Mg, Fe and Cr with streams colour-coded as in legend, -1 indicates field stars.}
    \label{fig:ReachGaiaStreamCR}
\end{figure}

As it is done for APOGEE, we observed the location of the streams in the reachability plot. In Fig~\ref{fig:ReachGaiaStreamCR} we show the streams as they are selected in {\em Gaia}-ESO (see Chapter~\ref{chap3} for details). As already discussed, the halo sample of {\em Gaia}-ESO is particularly limited in number and dominated by globular clusters. The most numerous streams and mergers by member number are GES, Arjuna, Sequoia and Helmi. Given such a limited sample, the {\sc creek} is only able to re-detect the member stars of GES, while the members of Helmi fall into various separate dips, as do those of Arjuna and Sequoia. In addition, GES stars falls in the same dip occupied by a GC. 
As the contamination is very high and the sample is dominated by the globular cluster members, we did not analyse the separation of GES into  substructures.

\paragraph{Other dips in the reachability plot}

\begin{figure}
    \centering
    \includegraphics[width=0.75\linewidth]{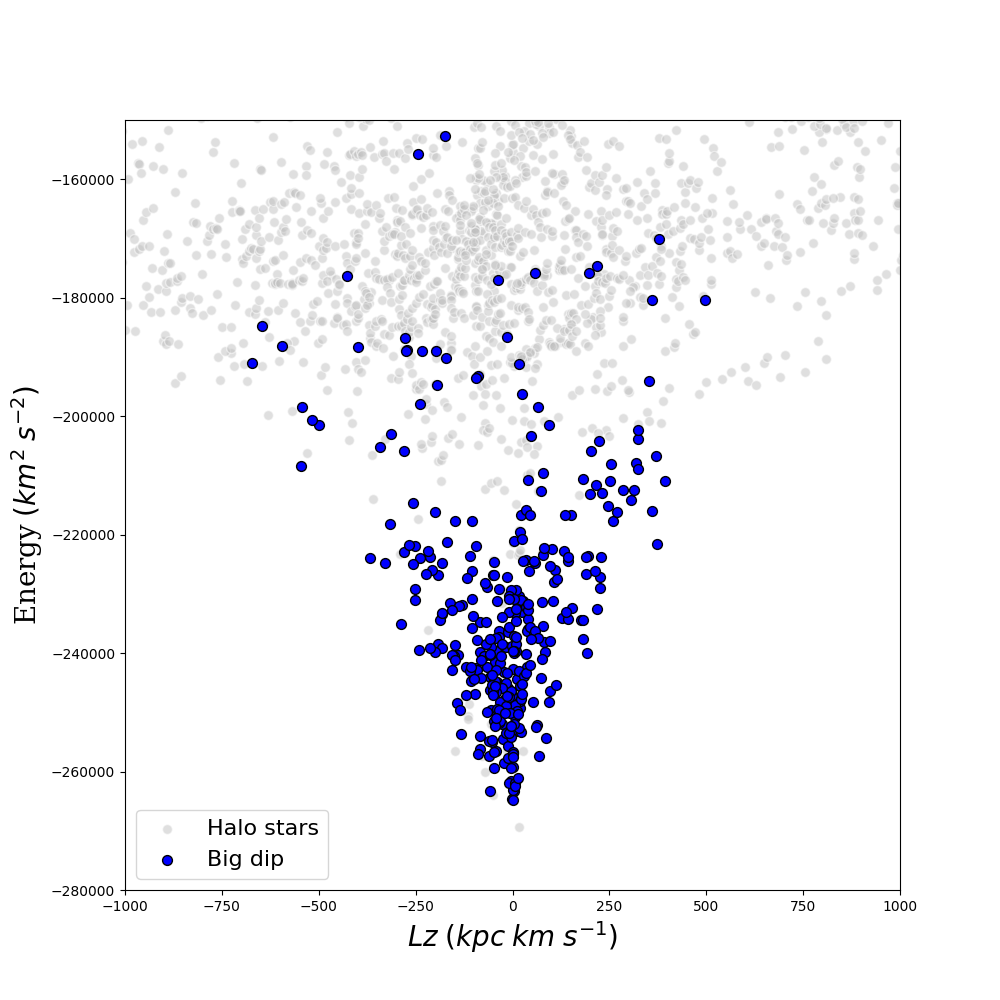}
    \caption{Lindblad diagram of the dip between 1500 and 1800 in the reachability plot in Fig.~\ref{fig:ReachGaia1}.}
    \label{fig:Lindblad}
\end{figure}

An important observation concerns the big dip present in the reachability plot obtained through only Mg and Fe in Fig.~\ref{fig:ReachGaiaCR1} between values of 1500 and 1800 approximately. This dip, once observed more in detail in the Lindblad diagram in Fig.~\ref{fig:Lindblad}, appears like a contamination from disc stars.

\subsection{Our results compared with recent works}

In recent years the research on clusters and streams has expanded significantly as they have emerged as tools to understand the process of formation of the milky way. While many papers have been written on these subject, our study shows some peculiarities.\newline
Firstly, chemical tagging is usually not applied to globular clusters as they cannot be considered completely homogeneous in chemistry. It has been applied to open clusters, \citep[see e.g.][]{blanco2015testing}, and to faint dwarf galaxies \citep[see e.g.][]{aoki2020chemical}, the Milky Way halo has generally been regarded as chemically inaccessible. 
Secondly, many studies focus on identifying clusters and streams are based exclusively on positional data, velocities or orbit invariants \citep[see e.g.][]{borsato2020identifying, lovdal2022substructure}. This allows for the analysis of larger samples of stars (e.g. {\em Gaia} {\sc dr3}) but excludes a priori disrupted clusters and streams.
Some of these works applied directly a clustering on data, others instead, exactly like we did, passed through machine learning algorithms \citep{shih2024via}. Sometimes they were applied with the idea of separating accreted stars and not the single clusters \citep{li2024exploring, ostdiek2020cataloging} but still relying on dynamical information. Chemistry would be especially useful in this.
Our work integrates both chemistry and kinematics, resulting in smaller datasets but allowing to compensate for the inherent imprecisions in abundance measurements with orbital invariants, and not exclude chemistry from our analysis.

%% file: 6-FuturePerspectives.tex
\section{Future perspectives}
\label{chap6}

The overarching aim of this entire project is developing and pioneering a new strategy of data analysis that could efficiently find dispersed stellar associations in the Galactic halo through clustering in the chemical abundance space and also using the information of stellar kinematics. In the previous sections we described the new algorithm called {\sc creek} and the results obtained from its application on two dataset already available such as those of APOGEE and of the {\em Gaia}-ESO Survey. 

There is a broad space for future developments of the present work. In fact, soon much larger datasets will be acquired for halo stars, which will produce the perfect ground for scientific applications of the {\sc creek}. At the same time, the method of analysis itself could be further developed and refined. In this Chapter I am going to summarize future perspectives, applications and developments of this work.

\subsection{The Galactic halo observed at medium spectral resolution}

The coming years will see the advent of numerous medium-resolution spectral surveys in 4 m class telescopes. In particular,  in the northern hemisphere we will have the WEAVE spectrograph mounted on the WHT telescope in the Canary Islands, while in the southern hemisphere, we will have the 4MOST spectrograph operating at  the VISTA telescope in Chile. 
Both instruments are organised in large consortia, among which Galactic Archaeology will play a crucial role. With regard to the study of Galactic stellar populations, the general purpose of the large spectroscopic surveys is to be complementary to {\it Gaia} and thus to provide low resolution spectra with R=5000 for radial velocities at 17$\leq$ G $\leq$ 20 mag and medium resolution spectra with R=20000 for stellar abundances at 12 $\leq$ G $\leq$ 17 mag.

\subsubsection{WEAVE}
WEAVE is the  new wide-field, massively multiplexed spectroscopic survey facility for the William Herschel Telescope, which has seen its first light in late 2022 \citep{Jin2023MNRAS.tmp..715J}, and which will start multi-object spectroscopic observations in 2024.   
It comprises a 2-degree field-of-view prime-focus corrector system, a nearly 1000 multiplex
fibre positioner, 20 individually deployable ‘mini’ integral field units (IFUs), and a single large IFU. These fibre systems feed a dual-beam spectrograph covering the wavelength range 366-959 nm at R=5000, or two shorter ranges at R=20 000. The WEAVE consortium is organized in several programmes, among which we will describe the one dedicated to the study our Galaxy’s origins. 
The Galactic Archaeology programme aims at  completing
the phase-space information of {\it Gaia}, providing metallicities  for about 3 million stars and detailed abundances for about  1.5 million for brighter field and open-cluster stars. 

\begin{figure}
    \centering
    \includegraphics[width=0.99\linewidth]{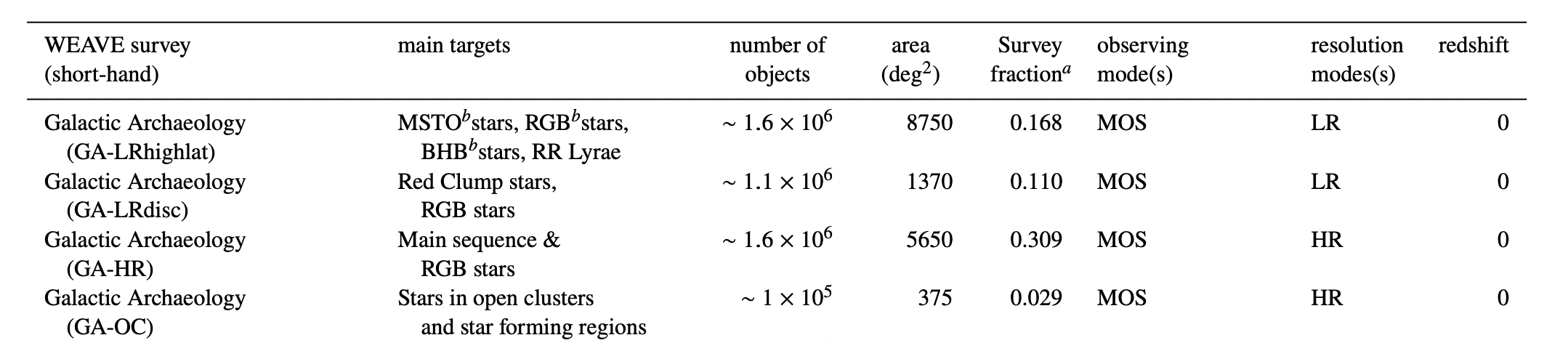}
    \caption{WEAVE survey parameters for the Galactic Archaeology programs from \citet{Jin2023MNRAS.tmp..715J}. }
    \label{table_weave}
\end{figure}

In Table~\ref{table_weave}, we show the survey parameters for the four main stellar surveys that will be carried out with WEAVE. 
The main question to be answered is how our Galaxy and its components were assembled over time. Finding an answer to this central question has general significance in the study of galaxy formation and evolution. 
In particular, the study of halo populations, which can be obtained thanks to WEAVE's large field of view, will provide crucial information on the percentage of the in situ and accreted populations. 
This problem will be addressed by two of WEAVE's sub-surveys: the high-latitude LR survey ('LH-highlat') that will search for traces of accretion events, and characterise its progenitors, including streams and dwarf galaxies, by determining precise stellar parameters, metallicity and radial velocities, and the sub-survey  HR chemo-dynamical survey (‘HR’) that will observe at intermediate and
high Galactic latitudes, providing full chemical
information of the three major Galactic populations:
the thin and thick discs and the halo. 

In addition to the standard stellar parameters and, there will be some spectral analysis pipelines that will produce abundances of other elements. In particular, the group at the Arcetri Astrophysical Observatory has set up two pipelines for high-resolution spectra to measure Li, C, N, and neutron capture elements \citep[see, e.g.][]{Van2019sf2a.conf..121V, Franciosini2022A&A...668A..49F}. 
These abundances will add fundamental information to the study of galactic populations, particularly halo populations for a full characterization of the in situ and accreted populations. 
For example, the abundance ratio [C/N] in giant stars is an excellent tracer of their age, as are the ratios of slow neutron capture elements to $\alpha$ elements. In addition, Li abundances in very metal poor stars are crucial to solve the cosmological Li problem, and to investigate possible differences in Li abundances in accreted and in situ population \citep[see][]{Simpson2021MNRAS.507...43S}. 

\begin{figure}
    \centering
    \includegraphics[width=0.75\linewidth]{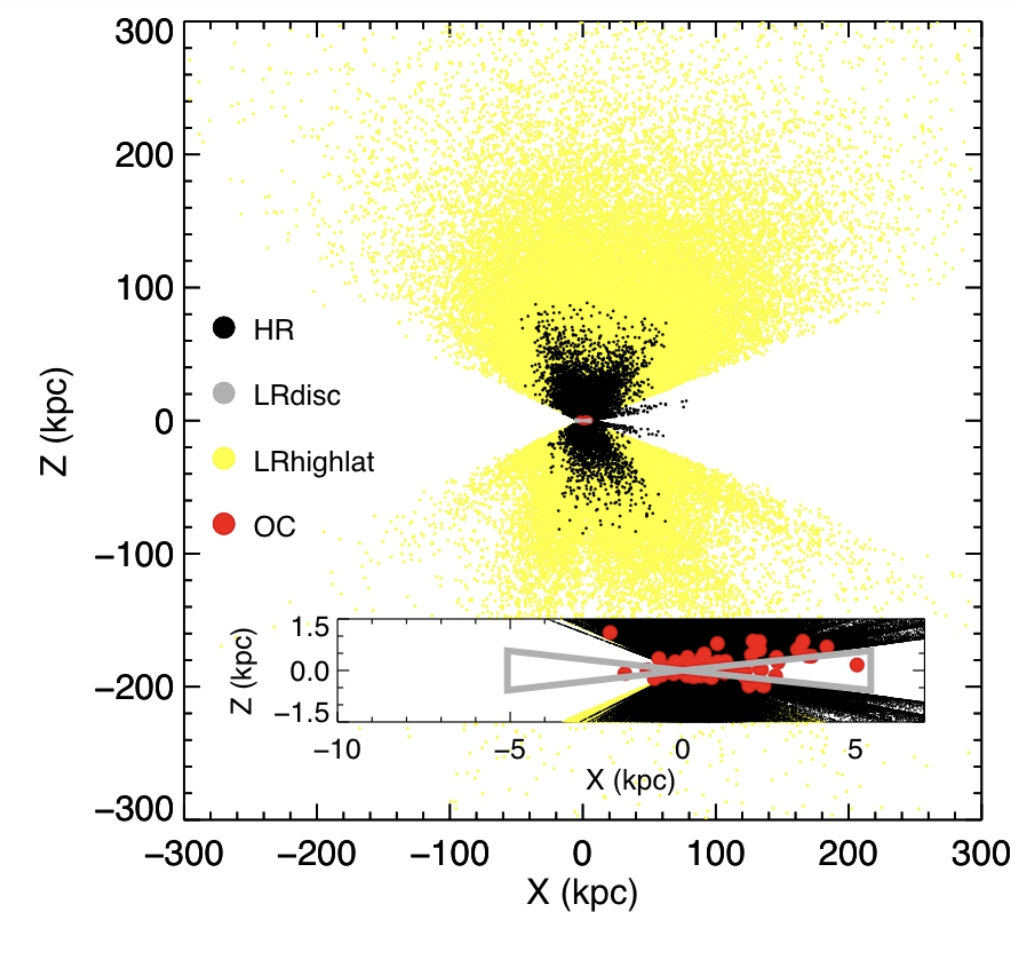}
    \caption{Expected coverage of the WEAVE GA sub-surveys in the X, Y plane (Cartesian coordinates centred on the Sun’s position in the Galaxy) with the Galactic Centre located at (-8.5, 0) kpc from \citet{Jin2023MNRAS.tmp..715J}. The two surveys described in the present Chapter are HR and LRHighlat. Figure adapted from \citet{Jin2023MNRAS.tmp..715J}. }
    \label{table_weave1}
\end{figure}

\subsubsection{4MOST} 
The 4-metre Multi-Object Spectroscopic Telescope project \citep[4MOST][]{deJong2019Msngr.175....3D} is the next ESO spectroscopic survey facility on the VISTA telescope. It is scheduled to start observations in 2025. With its large field-of-view it will be able to simultaneously obtain spectra of $\sim$2400 objects, combining both medium and low resolution spectra. 
 The 4MOST consortium is organised into several surveys Galactic Consortium Surveys
designed to improve our understanding of the Milky Way and how it evolved to its present-day configuration. The key objectives in Galactic Archaeology that the 4MOST surveys want to address are determining the density profile, shape and characteristic parameters of the dark matter halo of the Milky Way; better understanding the current Milky Way disc structure and dynamics (bar, spiral arms, vertical structure, stellar radial migration, merger history); reconstructing the growth history of the Milky Way; finding and characterizing kinematic and chemical patterns within the Magellanic Clouds system; and identifying stars that were accreted to the Galactic halo from globular clusters, and quantifying their contribution to the build-up of the halo.

These, and many more questions, will be addressed using the five Galactic Consortium Surveys to be conducted with 4MOST. In total more than 15 million low-resolution spectra and 3 million high-resolution spectra will sample the different stellar populations, providing a dense and deep map of the stellar properties in the Milky Way and the Magellanic Clouds. In addition, several tailored sub-surveys will study White Dwarfs, compact X-ray binaries, Cepheids and hot dwarf stars.

\begin{figure}
    \centering
    \includegraphics[width=0.75\linewidth]{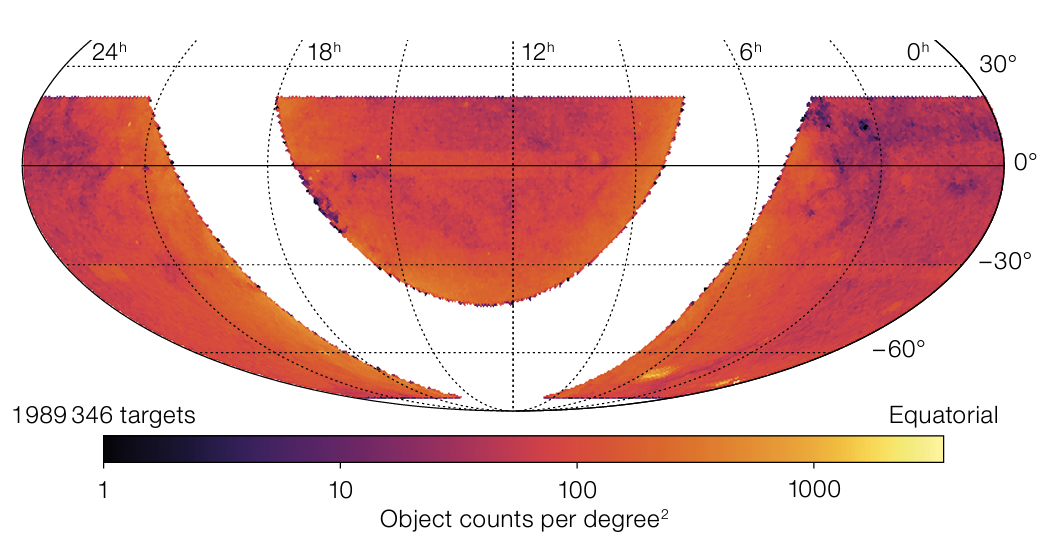}
    \caption{Target density of the three sub-surveys within the 4MOST {\bf Milky Way Halo High Resolution Survey}. Figure adapted  from \citep{Christlieb2019Msngr.175...26C}.  }
    \label{fig_halo_4most}
\end{figure}

In particular, the Milky Way halo will be observed both at low and medium spectral resolution with the 
{\bf Milky Way Halo Low Resolution Survey} \citep{Helmi2019Msngr.175...23H} and the {\bf Milky Way Halo High Resolution Survey} \citep{Christlieb2019Msngr.175...26C}. 
In Fig.~\ref{fig_halo_4most}, we show the target density for the {\bf Milky Way Halo High Resolution Survey} \citep{Christlieb2019Msngr.175...26C} which aims at identifying  and determining the elemental abundance patterns of halo stars belonging to different populations. It aims at separating stars that were formed in situ in the Galaxy, were contributed from a few major accretion events, or were contributed by minor, low-mass accretion events and were accreted to the halo from globular clusters.

\subsubsection{Deep clustering applied to the next generation of medium resolution surveys}
Dealing with such a large amount of data will therefore require automatic methods to separate stellar populations, using both spectroscopic and {\it Gaia} data. In this framework, our algorithm, which has been proven to be effective with samples from the current available spectroscopic surveys, will provide an excellent method of analysis to accurately separate populations belonging to streams,  clusters, merger events from those locally formed. 
In particular, it would be fundamental to have a large number of precise chemical abundances to fully characterize the Galactic Halo populations. 
Elements belonging to the various nucleosynthesis chains, in particular neutron capture elements, have proven to be crucial in separating populations of different origins in the halo. Current spectroscopic surveys, unfortunately, were limited in this respect. APOGEE, operating in the near-infrared, had a limited number of neutron capture elements, while {\it Gaia}-ESO, in the visible, has about 90\% of its sample observed at medium resolution and in a limited spectral range, so providing very precise abundances for only about 6000 stars.

4MOST  and WEAVE are expected  to measure abundance ratios to better than 0.1 dex for many elements, including  Fe, Mg, Si, Ca, Ti, Na, Al, V, Cr, Mn, Co, Ni, Y, Ba, Nd, and Eu, and better than 0.2 dex for Zr, La, and Sr \citep[see, e.g.][]{Feltzing2018IAUS..334..225F, Bensby2019Msngr.175...35B}. 
This precision comes from the number of lines in simulated spectra at different S/N, and exclude systematic uncertainties related to stellar parameters or atomic data.
Both surveys will observe about 1 million stars each at high-resolution in high Galactic latitude fields, thus including high percentages of halo stars. 

Thus, samples such as those provided by 4MOST and WEAVE will be ideal for the application of the {\sc creek}, and will allow us to identify chemo-kinematic structures in the Galactic halo in a fast and accurate way.

\subsection{The Galactic halo observed at high spectral resolution}
\begin{figure}
    \centering
    \includegraphics[width=0.95\linewidth]{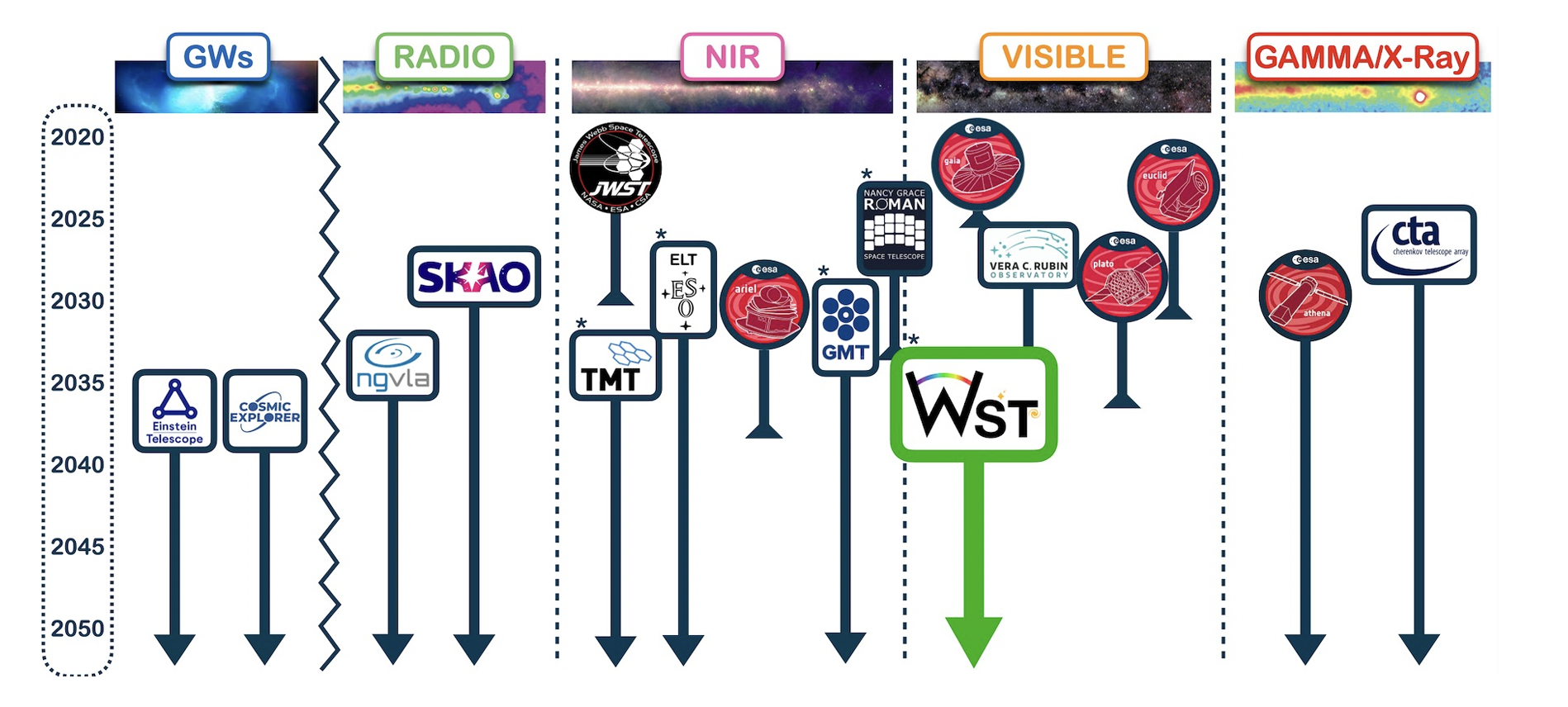}
    \caption{A graphical representation of the current and upcoming major astronomical facilities. The facility marked with an asterisk (e.g., ELT, TMT, GMT) have been listed in the spectral range they cover the most for visual purposes, but they also partly cover other bands. The duration of space missions reflect the publicly available nominal values.  Figure adapted from \citet{WST2024arXiv240305398M}.  }
    \label{fig_wst}
\end{figure}

\subsubsection{WST} 

The next large project that the astronomical community will propose to ESO after the Extremely Large Telescope (ELT) will be the Wide Spectroscopic Telescope (WST), a new concept telescope fully dedicated to spectroscopic observations. 
In Fig.~\ref{fig_wst}, we show a representation of the current and upcoming major astronomical facilities. WST will have strong synergies with many of those, filling in a gap in the current landscape. 
Indeed, by mid of the next decade, on-going and upcoming spectroscopic surveys will provide spectra for about 50$\times 10^6$ stars observed by {\it Gaia}. 
However, the vast majority of them will be at medium spectral resolution, providing essentially radial velocity and global metallicity. 
As shown in the present Thesis, even an accurate knowledge of their kinematics is not sufficient to trace these stars back to their initial position, due to the many unknowns, including the time variation of the Galactic potential, and transient structures such as bar and spiral arms. 
So, it is not possible to reassemble stellar associations now dispersed through the Galaxy with kinematics only. 
On the other hand, it is also well known that the chemical pattern of a star provides the fossil information of the environment where it formed. 
Therefore, chemistry and kinematics together can maximize our capability to search for missing structures in the Galaxy, aiming at reconstructing the star formation history of the Galaxy, as shown in our work. 

WST in its high-resolution R=40 000 MOS mode will provide chemical abundances covering all nucleosynthesis chains with  a high precision ($\sigma <$ 0.05 dex). The combination of R= 40,000 spectral resolution, high-multiplex and large collecting area make WST uniquely positioned to combine the chemistry and kinematics of the stellar populations in the Milky Way and finally understand the full formation history of our Galaxy.
\begin{figure}
    \centering
    \includegraphics[width=0.95\linewidth]{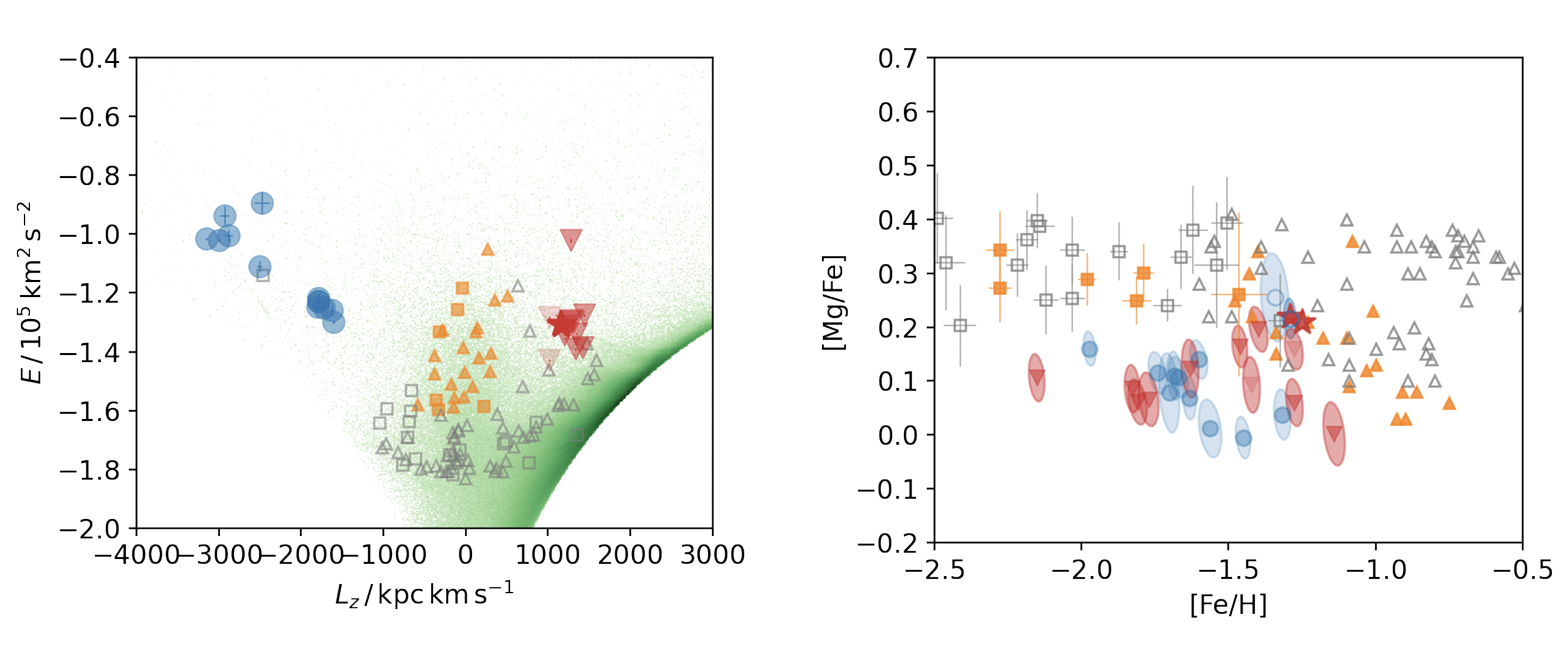}
    \caption{Left: Substructures of the Galactic halo identified through their kinematics, tracing distinct accreted galaxies (red: Helmi streams, blue: Sequoia, orange: Gaia-Enceladus-Sausage). Right: Chemical abundance of stars of different origins, with the same colour code. Figure adapted from \citet{WST2024arXiv240305398M}.  }
    \label{fig_wst_halo}
\end{figure}
Among the numerous science cases proposed for WST \citep[see][for the WST White Paper]{WST2024arXiv240305398M}, there is the identification and characterisation of the assembly and accretion history of the Milky Way. The distinctions in chemical abundance ratios in the Galactic populations will open an opportunity to identify and characterise both large-scale stellar populations, such as thin/thick discs, halo, Bulge, and small-scale stellar populations, including remnants of past encounters or disrupted star clusters, as shown in Fig.~\ref{fig_wst_halo}.

\begin{figure}
    \centering
    \includegraphics[width=0.95\linewidth]{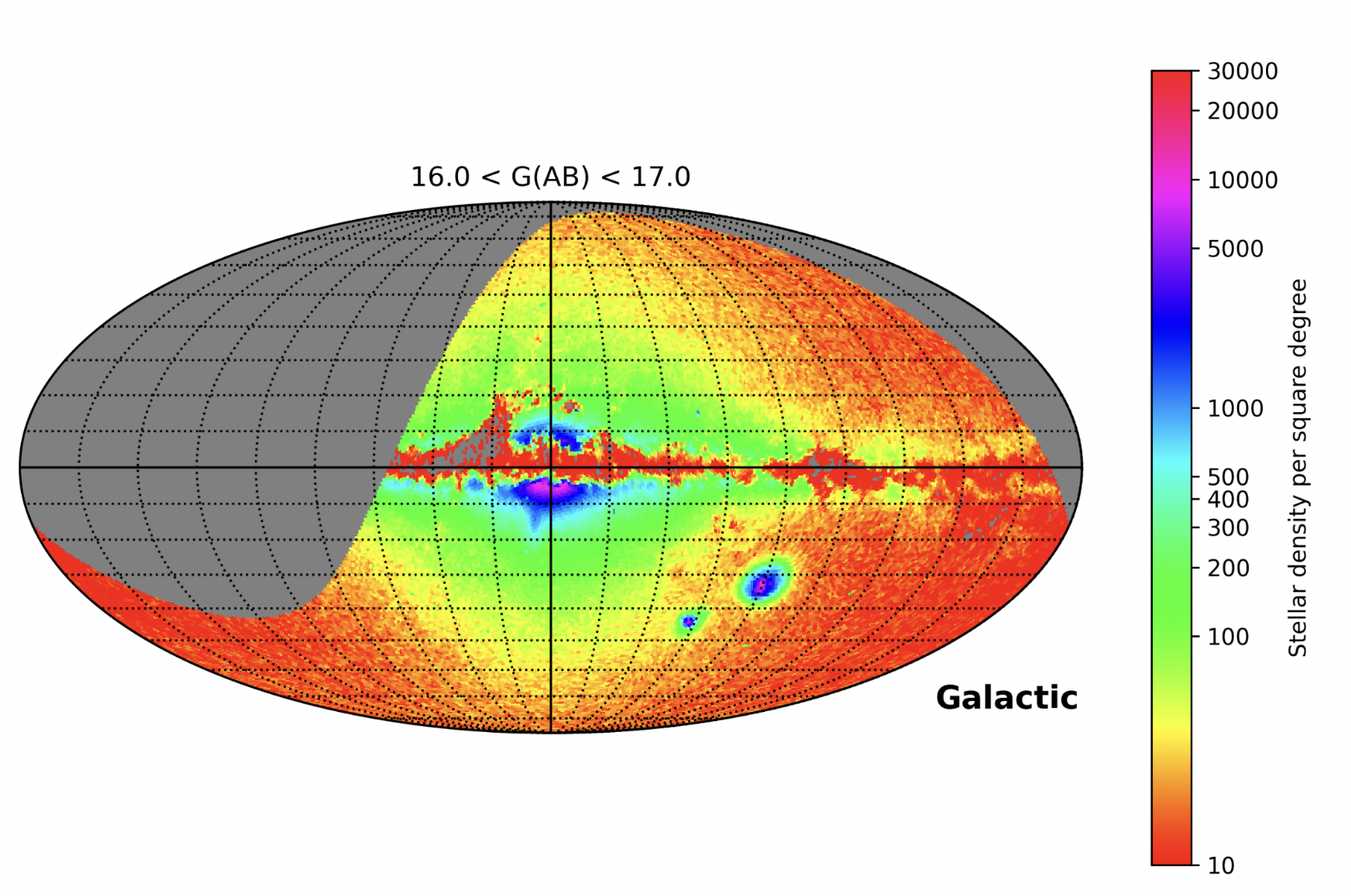}
    \caption{Mollweide view of the faintest targets in the Galactic halo that can be observed with WST, HR MOS mode. Figure adapted from \citet{WST2024arXiv240305398M}.  }
    \label{fig_wst_halo_moll}
\end{figure}

WST will observe few million stars in the MW, among them stars in the Galactic halo, as shown in the Mollweide view of the faintest and most numerous targets that can be observed with WST in its HR MOS mode. As in the case of medium resolution spectroscopy surveys, the combination of kinematics and chemical information will enormously improve our knowledge of the Galaxy, and in particular of the Galactic halo. 
Machine learning techniques, as the Deep clustering algorithm presented in this Thesis, are mandatory to approach the characterization of such large amount of data. 

\subsubsection{HRMOS}

\begin{figure}
    \centering
    \includegraphics[width=0.95\linewidth]{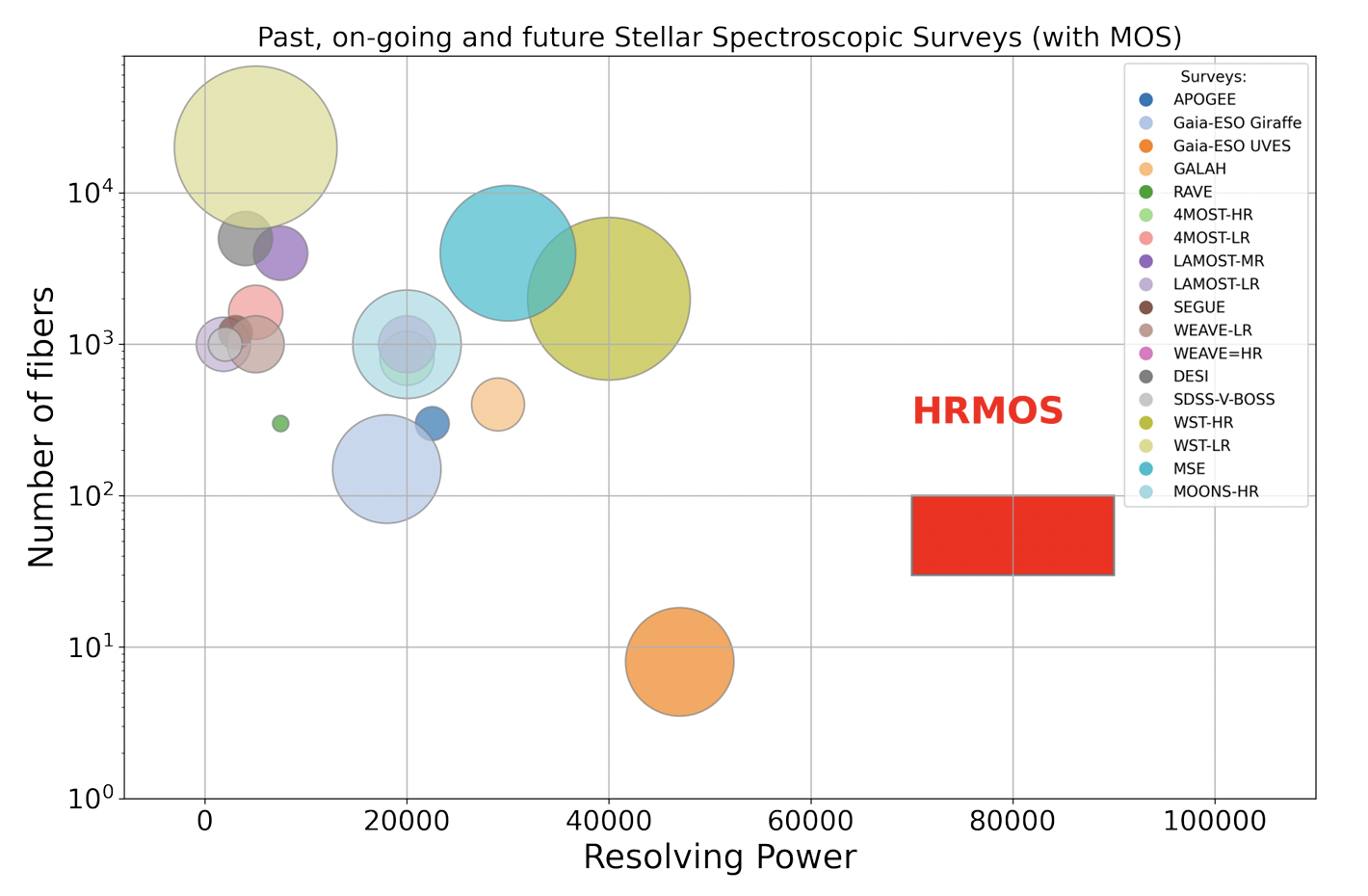}
    \caption{HRMOS in the landscape of spectroscopic surveys with completed, ongoing, and future MOS instrumentation: Number of fibres as a function of the resolving power (R). The size of the symbol is proportional to the collecting area of the telescope. HRMOS is represented with a red rectangle, with R centred on 80000, and the number of fibres from 20 to 100. Figure adapted from \citet{Magrini2023arXiv231208270M}.  }
    \label{fig_hrmos}
\end{figure}

In the framework of the large variety of spectroscopic surveys and multi-object instrumentation that have mapped, are mapping, or are planned to map, the stellar populations of our Galaxy and nearby galaxies, the astronomical community is proposing  e a new instrument for the ESO-VLT in the context of the VLT2030 roadmap. This is planned to be a very high resolution, multi-object spectrograph (HRMOS). 
HRMOS is proposed to have a very high spectral resolution (R $>$ 60 000) than all current, and upcoming, spectrographs dedicated to spectroscopic surveys of Galactic populations \citep{Magrini2023arXiv231208270M}. 
In Fig.~\ref{fig_hrmos}, we show the location of HRMOS in the current and future landscape of  spectroscopic surveys. 

Although HRMOS will be a scaled-down instrument compared to WST, it will achieve precision in chemical abundances never before obtained systematically and in large stellar samples. Using differential spectral analysis techniques, precision on the order of 0.01-0.02 dex is expected. Furthermore, HRMOS will observe in the blue spectral range, including absorption lines of many heavy elements. 
Although HRMOS will not deal with extremely large samples as WST, being able to obtain high-resolution spectra of 10-20 targets at a time, it would revamp our view of the halo, allowing us to address various issues including the characterisation of the first stars and the determination of abundances and isotopic ratios of heavy elements in metal-poor stars. In addition, there will be the unique and fundamental addition of the observation of the lines of radioactive isotope of Th~{\sc II}, which, together with Eu, can give a model-independent estimate of stellar ages. 

In this case, the major contribution of this thesis work, and of the Deep Clustering algorithm presented here, will be to pre-select samples of stars belonging to streams, mergers, and disrupted clusters to be observed with HRMOS. 
The deep spectra of HRMOS will provide extremely precise abundances, isotopic abundance ratios and ages from radioactive isotopic ratios.

\subsection{A promising algorithm to select targets for future surveys and to analyse large datasets}

\begin{figure}
    \centering
    \includegraphics[width=0.95\linewidth]{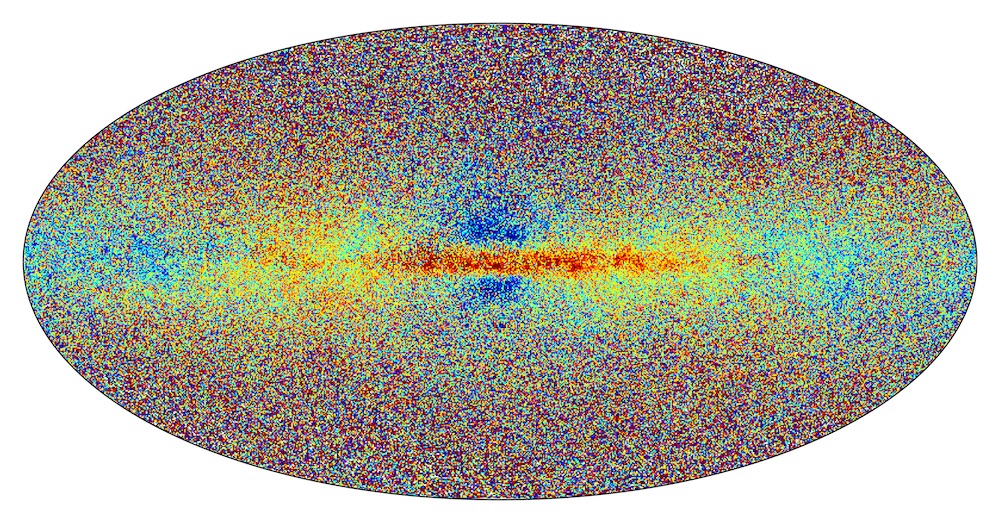}
    \caption{All sky view in Galactic coordinates (HEALPix map) showing the stars in the Gaia DR3 GSP-Spec database (Gaia Collaboration, Recio-Blanco et al. 2022). The colour indicates the stellar metallicity, [M/H], that is the mean abundance of all chemical elements except hydrogen and helium. Redder stars are richer in metals. Figure adapted from \citet{Gaia2023A&A...674A..38G}.  }
    \label{fig_gaia_abundances}
\end{figure}

As potential developments of this project will rely on {\em Gaia} DR3 data, we attempted to apply the {\sc creek} to this dataset. In {\em Gaia} DR3 clusters are sparse: we manage to select 115476 halo stars considering a total velocity over 180 km s$^{-1}$ and excluding stars from Andromeda and quasars. Among these stars, only 569 were found in clusters, of which only 223 had measured metallicity. This is a emphasizes the necessity of integrating {\em Gaia} data with future surveys.
The missing metallicities are a clear indication of which stars would be interesting to observe and analyse in order to get a more complete sample of stars.
The combination of {\em Gaia} data with WST and later WEAVE, will allow to reach much fainter stars that the ones available today. In the wait for these surveys we can request observing time and analyse the data, providing metallicities and abundances in order to enhance the application of our algorithm.

\subsection{Future developments of the {\sc creek}}

The analysis method described in Chapter~\ref{chap4} can be further refined in two fundamental aspects:

\begin{itemize}
\item conditioning the latent space for improved clustering analysis;
\item learning the strength of the links between stars.

\end{itemize}

\subsubsection{Conditioning the latent space}

In order to map the original complex data to a feature space that is easy to cluster, many methods of analysis first focus on feature extraction, feature transformation or dimensionality reduction. The use of autoencoders precisely serves this purpose, which is to map the original dataset into a latent space that is a better representation for subsequent tasks, such as clustering, but also classification, or anomaly detection. However, the problem with traditional autoencoders in generating a latent space is that they often do not enforce any specific structure on the latent space. This can lead to a disorganized and sparse representation, making it challenging to perform the subsequent tasks such as clustering. Specifically, the latent space can have discontinuities and regions that do not correspond to any meaningful data. In order to solve this issue it is possible to add to the standard reconstruction loss of the autoencoder, another term that is calculated based on the distribution of points in the latent space and which encourages this latter to form distinct clusters. All that is achieved by applying a clustering algorithm (such as k-means) directly to the latent representations and calculating a loss based on how well the latent points fit the desired cluster distribution \citep[e.g.][]{Ren22}. Hence, during training the autoencoder optimizes both the reconstruction loss and the clustering loss simultaneously. The reconstruction loss ensures that the latent representations retain the essential features of the input data, while the clustering loss arranges these representations into meaningful clusters. This combined approach results in a latent space that is not only effective for data reconstruction but also well-structured for clustering tasks. On the other hand, this approach requires the setting of an additional hyper-parameter, which is the number of clusters to form in the latent space. However, also this additional hyper-parameter can be tuned just like the others hyper-parameters of the autoencoder (e.g., number of neurons, number of layers, etc...).

\subsubsection{The strength of the links as a learnable parameter}

The method of clustering analysis developed in this work makes use of links between stars which are established between stars with similar orbital actions. In fact, the graph autoencoder, which maps stars from the chemical abundance space into the latent space where clustering analysis is performed, aggregates information from neighboring nodes (i.e., stars connected by a link) through multiple layers. By doing so, two strongly connected stars (i.e., stars with many common neighbors) should also be mapped into very adjacent regions of the latent space even though they may have slightly dissimilar abundances. The graph autoencoder I used in this work runs giving the same weights to all the links between stars. This approach is based on the assumption that stars with very similar kinematics must have a similar origin, hence a similar chemical composition, and that stars with a similar origin must have similar kinematics. However, in practice this assumption is not true as stars can undergo severe changes of their orbital actions due to interactions with non-axisymmetric potentials of the Galaxy (e.g., giant molecular clouds, Galactic bar, spirals). Thus, a further development of this method of analysis relied in the possibility of breaking the links that are less significant in mapping stars into the latent space. All that could be done though the use of a more advanced family of neural networks called Graph Attention Networks \citep{Vaswani17}, which handle the strength of the links as a learnable parameter. This further development is expected to give less importance to links between stars that have migrated and that therefore should not be close neighbors in the latent space.

%% file: 7-Summary_Conclusions.tex
\section{Summary and conclusions}

\label{chap7}

\begin{figure}[ht]
    \centering
    \includegraphics[width=1\linewidth]{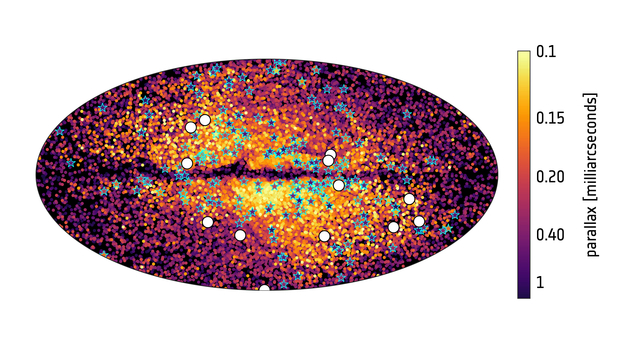}
    \caption{All-sky distribution of stars identifies as part of the   debris of a galaxy that merged with our Milky Way during its early formation stages, 10 billion years ago, and named this galaxy Gaia-Enceladus. The stars of Gaia-Enceladus are represented with different colours depending on their parallax with purple hues indicating the most nearby stars and yellow the most distant ones. White circles indicate globular clusters that were observed to follow similar trajectories as the stars from Gaia-Enceladus, indicating that they were originally part of that system; cyan star symbols indicate variable stars that are also associated as Gaia-Enceladus debris. Figure adapted from \citep{helmi2018merger}.}
    \label{fig_gaia_enc}
\end{figure}

Strong chemical tagging rests on the assumption that stars born from the collapse of the same cloud share the same photospheric chemical composition and that stars born from different molecular clouds have unique chemical patterns that make them distinguishable. 
However, recent work has shown that incorporating additional data, such as kinematics or ages, can broaden the applicability of chemical tagging to resolved populations in our Galaxy. 
Among the several populations of the Galaxy, the halo is the one that presents the most favourable conditions for attempting to recover destroyed structures. In fact, the stellar density is much lower compared to the density of the disc and bulge, and traces of past mergers and streams generated by the destruction of star clusters or galaxies, as in the case of Sagittarius dwarf galaxy, can be identified.\\
Recovering disrupted mergers and clusters in the Milky Way halo is essential for unraveling the complex merging history, for understanding the origin of the current structure of our Galaxy and shedding light on fundamental questions about the formation and evolution of galaxies. 
The identification of stellar streams is crucial since it provides valuable insights into the past interactions of our Galaxy with dwarf galaxies. Their chemical composition can be used to infer the properties of the disrupted objects, such as their mass and star formation history. Moreover, the Milky Way halo is a fundamental environment for studying and testing cosmological models of hierarchical structure formation. Therefore, our study contributes to a broad spectrum of research in this area.
However, despite the ongoing efforts to identify large-scale streams that is being conducted using astrometry and kinematics \citep[see, e.g.][]{shih2024via}, the characterisation of nearby streams has proved even more complicated. Often, as in the case of Gaia-Enceladus and as shown in Fig.~\ref{fig_gaia_enc}, nearby streams or debris compose a large part of the local stellar halo, and surround the position where we are. We can therefore identify them only by the combined use of some of their invariant properties, such as orbital energy and angular momentum, or actions, and chemical composition.\newline
In this work, we have presented a thoughtful attempt to make use of both the orbital and chemical properties of a selected sample of halo stars observed by two of the largest public spectroscopic surveys {\it Gaia}-ESO and APOGEE. 

After framing the main topics of Galactic archaeology, we described the fundamental ML techniques that are currently indispensable for analysing large astronomical databases. 
We then outlined our ML algorithm, the {\sc creek}, which allows us to take into account both orbital parameters  and chemical composition to recover star clusters, and to identify or re-identify stellar streams, mergers, or debris.

\paragraph{A new concept algorithm, the {\sc \textbf{creek}} }

In this work, we have designed a new concept algorithm, the {\sc creek}, that considers at the same time the orbital and chemical properties of stars to provide a deep clustering analysis of the halo stellar population. 
The name {\sc creek} is chosen to recall the streams and debris that populate the halo of our Galaxy. 

The {\sc creek} operates as follows:
\begin{itemize}
\item {\bf Data selection:} within the Milky Way stellar population, we selected halo stars from the datasets of APOGEE and {\em Gaia}-ESO, based both on their velocities and metallicities. We then computed their orbital parameters.
\item {\bf Using orbital properties}: the selected data were passed to a Siamese Neural Network (SNN) that associated stars from the same cluster based on their orbital parameters; stars with probability over 80\% to be in the same cluster according to the SNN were connected through a link and the resulting graph was passed to a Graph Neural Network (GNN). 
\item {\bf Using chemical properties:} 
The GNN autoencoder took as inputs the selected abundances, which are [$\alpha$/M] and [M/H] for APOGEE and [Mg/Fe], [Fe/H] and [Cr/Fe] for {\em Gaia}-ESO. The abundances were chosen to maximize homogeneity within stars from the same cluster while ensuring distinctiveness between stars from different clusters. Additionally, we prioritised elements with smallest errors. The GNN autoencoder computed a mean of the abundances of all connected stars, weighted on the number of links of each star. It subsequently analysed them in order to map the chemical space into a more efficient representation in the latent space. This step aimed to include the chemical similarity among the stars in our dataset, facilitating the delineation of their structures and relationships.
\item {\bf Clustering with {\sc \textbf{optics}} in the latent space: } Finally, {\sc OPTICS} was applied to the latent space, providing groups based on the chemical similarities of the stars, but also considering their kinematic links. A flowchart representative of the {\sc CREEK} algorithm is depicted in Fig.~\ref{fig:chart} in Chapter~\ref{chap4}.
\end{itemize}
Recent works predominantly rely on kinematics to identify clusters and streams. However, with the {\sc creek}, we were able to tackle clustering in the Milky Way halo integrating both chemistry and kinematics, obtaining promising results where a conventional chemical tagging, i.e. using only chemical abundances,  would bear no outcome.

\paragraph{The application of the {\sc creek} to APOGEE and {\em Gaia}-ESO surveys}

The process of analysis was conducted on both APOGEE and {\em Gaia}-ESO. The recovery of groups was quite effective once compared to real globular clusters, with approximately 80\% of clusters recovered with a 30\% threshold of completeness and homogeneity.
When comparing the obtained groups with real streams, we obtained different results depending on the dataset. In APOGEE we were able to recover some streams, such as Gaia-Enceladus and Heracles. {\em Gaia}-ESO, instead, proved less suitable for our purposes, likely due to the lower number of stars in the halo with measured abundances, leading to difficulties in separating streams from clusters.
Nonetheless, the results were satisfying, considering that the definition of streams in literature often depends on the authors and is not unambiguous. With our method, we established a repeatable, objective, and automatic technique for defining and identifying streams.
The limitations of the current databases, related to the low number of halo stars, and the large uncertainties on the chemical abundances, currently limit the results of our analysis.

\paragraph{Scientific results}

The {\sc creek} method proved to be successful from both a technical point of view and in yielding interesting scientific results. A notable result was obtained for our selection of the stream Gaia-Enceladus in APOGEE. Compared to the literature selection \citep[see, e.g.][]{Horta2023}, the {\sc creek} was able to separate two substructures within GES, different both kinematically and chemically.
The two substructure are indeed separated, as verified through a Kolmogorov-Smirnov test, in Energy and the most relevant chemical abundances, such as $\alpha$-elements, metallicity and [Ce/Fe].

In addition, we have, for the first time, exploited {\em Gaia}-ESO data to chemically characterize stream stars. We confirm with such data their belonging to the low sequence in the [Mg/Fe] vs [Fe/H] plane, which corresponds to the accreted population. We find a strong similarity and partial overlapping between GES and Arjuna that could possibly suggest a common origin or even a belonging to the same structure. We also find, in some streams, the interesting underabundance of [Al/Fe], typical of accreted populations.

\paragraph{Future perspective and new application}

In the coming years, we anticipate the advent of various medium-resolution spectral surveys operated through 4-meter class telescopes, notably WEAVE and 4MOST. These two surveys will be fundamental for our study as they will provide abundance ratios for C, N and neutron capture elements, which can be used as chemical clocks, and thus adding the age information to the problem. WEAVE will provide metallicities for 3 million stars and 4MOST will deliver high-resolution data for a similar number of stars, alongside an additional 15 million stars at lower resolution.\\
The Wide Spectroscopic Telescope (WST) will also be fundamental for our study, with its high-resolution of R = 40 000 and its ability to provide abundances for all nucleosynthetic chains at high precision.
Looking further ahead, HRMOS is poised to stand out among these instruments. It will have a spectral resolution over 60 000 and will yield abundances with precisions of 0.01-0.02 dex.
The {\sc creek} will prove fundamental for the analysis of these surveys, particularly in separating clusters and streams, and delineating populations with different origins in general. The possibility to analyse chemical abundances for all nucleosynthetic chains of these new surveys will allow the {\sc creek} to distinguish and separate different halo populations and will facilitate the tracing of stars back to their birth positions.\\
An interesting aspect of the {\sc creek} is that it is not only able to separate stars within a dataset, but it can also pre-select samples of stars belonging to accreted structures in order to be observed with very high resolution instruments, such as HRMOS.

From a technical point of view, the algorithm can be refined by conditioning the latent space for improved clustering analysis and by learning the strength of the links between stars.

%% file: Aknowledgements.tex
\section*{Acknowledgements}

First and foremost, I would like to express my gratitude to Davide Massari for supplying the code for orbital computation. Additionally, I extend my sincere thanks to Daniel Horta for providing his stream dataset from APOGEE and for his prompt and kind response, and to Rodolfo Smiljanic for his timely and helpful replies.